%% file: VecchiaDL_main.tex
\documentclass[11pt]{article}
\usepackage{fullpage,setspace}
\usepackage{amssymb, amsthm, amsmath}
\usepackage{graphicx}
\usepackage[authoryear]{natbib}
\usepackage{bm}
\usepackage{lineno}
\usepackage{comment}
\usepackage{caption}
\usepackage{subcaption}
\usepackage{mathtools}

\usepackage[T1]{fontenc}
\usepackage[sc]{mathpazo}
\usepackage{hyperref}
\usepackage{algorithm}
\usepackage{algpseudocode}
\usepackage{booktabs}
\input{commands}

\usepackage{color}
\newcommand{\edit}[1]{{\color{blue} #1}}

\begin{document} % \linenumbers

\begin{center}
{\Large  Modeling Extremal Streamflow using Deep Learning Approximations and a Flexible Spatial Process}\\\vspace{6pt}
{\large Reetam Majumder\footnote[1]{North Carolina State University}, Brian J. Reich$^1$ and Benjamin A. Shaby\footnote[2]{Colorado State University}}\\
\today
\end{center}

\begin{abstract}\begin{singlespace}\noindent
Quantifying changes in the probability and magnitude of extreme flooding events is key to mitigating their impacts. While hydrodynamic data are inherently spatially dependent, traditional spatial models such as Gaussian processes are poorly suited for modeling extreme events. Spatial extreme value  models with more realistic tail dependence characteristics are under active development. They are theoretically justified, but give intractable likelihoods, making computation challenging for small datasets and prohibitive for continental-scale studies. We propose a process mixture model (PMM) which specifies spatial dependence in extreme values as a convex combination of a Gaussian process and a max-stable process, yielding desirable tail dependence properties but intractable likelihoods.  To address this, we employ a unique computational strategy where a feed-forward neural network is embedded in a density regression model to approximate the conditional distribution at one spatial location given a set of neighbors. We then use this univariate density function to approximate the joint likelihood for all locations by way of a Vecchia approximation. The PMM is used to analyze changes in annual maximum streamflow within the US over the last 50 years, and is able to detect areas which show increases in extreme streamflow over time.

%\textbf{Alternate title ideas:} A flexible spatial process with deep learning approximations for extremal streamflow data; Modeling extremal streamflow using a flexible spatial process with deep learning approximations; Spatial process modeling of annual maximum streamflow using deep learning approximations;  Modeling Extremal Streamflow using Deep Learning Approximations of a Flexible Spatial Process
\vspace{12pt}
\end{singlespace}\end{abstract}

\newpage
\singlespacing

\section{Introduction}\label{s:intro}
The Intergovernmental Panel on Climate Change released its Sixth Assessment in 2021 and projected an increased frequency of hydroclimatic extremes. 
%To obtain these climate change projections, Global Climate Models (GCMs) are forced under Shared Socio-Economic Pathways (SSPs), which are represented by CO$_2$ emission and mitigation scenarios. 
In addition to changes in the mean of climate variables, the impact of climate change is more severe with changes in the frequency and magnitude of hydroclimatic extremes. 
%Climate change is often described in terms of the mean, but will be mostly experienced through extremes.  
%In order to properly account for spatial dependence while modeling rare event probabilities, we use spatial extreme value analysis (EVA).  EVA poses several challenges not present in traditional mean value analysis: (i) data for extreme events are by definition sparse, and therefore parametric models must be carefully chosen based on extremal theory to estimate small probabilities; (ii) climate data are inherently spatially-dependent, but standard measures of dependence such as correlation and spatial models such as Gaussian processes do not adequately model  extreme events; (iii) theoretically-justified EVA models (e.g., the max-stable process) give intractable likelihoods, making computation challenging for small datasets and prohibitive for continental-scale studies; and  (iv) GCMs, which are the linchpin of climate change studies, exhibit bias for predicting extreme quantiles because they are evaluated on spatial resolutions that are too coarse to reliably produce smaller-spatial scale (e.g., watershed) extreme events.
Floods are responsible for huge economic and human costs \citep{hirabayashi2013global,winsemius2018disaster}, and this cost is projected to increase due to sea level rise and extreme precipitation events brought about by our changing climate \citep{winsemius2018disaster}. Effective prediction of future flooding events is required for water infrastructure design, but is challenging due to the complexity of flooding events and uncertain climate predictions \citep{merz2014floods,condon2015climate,kundzewicz2017differences,franccois2019design}. Extensive research has been conducted looking at changing climate signals in historical extreme rainfall  \citep{knox1993large,kunkel2020precipitation} and in flooding \citep{franks2002identification, sharma2018if,bloschl2019changing,milly2005global,
 walter2010increasing,
 meehl2000trends,
 hirsch2011perspective,
 vogel2011nonstationarity,
 archfield2016fragmented}.  For example, \citet{hirsch2012has}
 found a significant change in annual maximum streamflow (a key measure of flood risk) at 48 of 200 US Geological Survey (USGS) gauges and spatial clustering in the direction and magnitude of the changes.  As a result, there is a need to account for spatial and temporal variability (i.e., nonstationarity) in flood frequency patterns when assessing current and future risk
 \citep{merz2014floods,kundzewicz2014flood,milly2008stationarity,milly2015critiques,vogel2011nonstationarity,salas2014revisiting}.

One approach to projecting flood risk on the basis of extreme streamflow involves the statistical extrapolation of the spatiotemporal trends observed in the historical record.  Of particular interest is estimating the joint probability of extremal streamflow at multiple locations, which is useful for understanding regional flood impacts and assessment to support federal and state emergency management agencies. For example, 
\citet{vsraj2016influence,dawdy2012regional,lima2016hierarchical} use extreme value analysis (EVA) methods to model nonstationarity with regressions or hierarchical models for the relationship between flooding and watershed characteristics and weather. 
Classic non-spatial EVA \citep{coles2001introduction} begins by isolating the extreme events of interest.  This is done systematically by either selecting all exceedances over a threshold or computing the block maximum, e.g., the annual maximum of daily streamflow.  
%Formally, if $Y_1(\bs),...,Y_n(\bs)$ are $n$ independent and identically distributed  observations at spatial location $\bs$, then under certain conditions the distribution of the exceedance above threshold $u$, $Y_t(\bs) | Y_t(\bs)>u$, converges to a generalized Pareto distribution for large $u$.  For block maxima, under certain conditions, a properly scaled version of $Y(\bs) = \max\{Y_1(\bs),...,Y_n(\bs)\}$ converges to a Generalized Extreme Value (GEV) distribution as $n$ increases. The fitted GEV model gives estimates of the probability of rare events and the return level, i.e., the threshold that is exceeded on average once every $m$ years and thus flood frequency curves of return level by return period.
A spatial EVA analyzes exceedances or pointwise maxima (i.e., computed separately at each spatial location) as a stochastic process over space. 
%Applying spatial EVA methods to flood frequency curve estimation has several advantages over previous region flood frequency analysis procedures \citep{hosking1993some,sankarasubramanian1999investigation}.  
 Modeling spatial dependence allows for predictions at ungauged locations and the estimation of the joint probability of extremes at multiple locations.
 It also facilitates the borrowing of information across locations to estimate the marginal distribution at each location, which is particularly useful for EVA where data are sparse and low-probability events are of interest, and  gives valid statistical inference for model parameters by properly accounting for spatial dependence.  

In this study, we consider extreme streamflow data from the United States Geological Survey's Hydro-Climatic Data Network (HCDN) \citep{lins2012usgs}. Our primary objective is to identify regions within the US where the distribution of extreme streamflow has changed over time. The HCDN has a long historical record, and consists of locations that are minimally impacted by anthropogenic activity while excluding sites where human activities affect the flow of the watercourse. 
%This makes the HCDN data suitable for studying the effects of changing climate on streamflow, often in conjunction with other climate variables \citep{Sankaretal2001,OhSankar-2012,Awasthietal2022}.
 We focus on the modeling of block maxima of streamflow with the help of the max-stable process (MSP) \citep{de2006extreme}.  MSPs are a limiting class of models for spatial extremes, featuring strong forms of tail dependence \citep{Smith-1990, Tawn, Schlather, Kabluchko-Schlather-deHaan, Wadsworth-Tawn-2012,Reich-Shaby}. They are a natural asymptotic model for block maxima, but can also be applied to peaks over a threshold using a censored likelihood \cite[e.g.,][]{Reich-Shaby-Cooley,Huser-Davison-2014}.
 
 In practice, MSPs pose two challenges. First, the analytic forms of (censored) MSP densities are computationally intractable for all but a small number of spatial locations \citep{Schlather, Kabluchko-Schlather-deHaan, Wadsworth-Tawn-2012,wadsworth-2014a,wadsworth-2015a}. For general MSPs, \citet{Castruccio-etal} stated that full likelihood inference seemed limited to $n=13$ locations. These low dimensional results have led to the use of composite likelihood (CL) approximations \citep{Padoan-Ribatet-Sisson}. However, CL suffers from  statistical inefficiency for large $n$ \citep{Huser-Davison-Genton}, finite-sample bias when using all pairs of observations \citep{Sang-Genton, wadsworth-2015a, Castruccio-etal}, and computational challenges posed by computing likelihoods at all $O(n^2)$ pairs. More recently, \citet{Huser-etal} proposed an expectation-maximization algorithm for full likelihood inference, with computation time of 19.8 hours with $n=20$.  %This has recently been implemented for non-Gaussian spatial data specified as directed acyclic graphs by \citet{Zheng-et-al-2022}. 
 \citet{Huser-Stein-Zhong-2022} have also applied the Vecchia approximation that requires only moderate-dimensional (say 10 or 15) joint distribution functions, which are available for some MSPs. More recently, deep learning has been used to estimate parameters in spatial models within a  framework of simulation-based inference. They leverage a likelihood-free approach by simulating datasets with different parameter values and using deep learning to identify features of the simulated data that are indicative of particular parameter values. \citet{gerber2021fast} used it to estimate covariance parameters for spatial Gaussian process (GP) models, by training convolutional neural networks (CNNs) to take moderate size spatial fields or variograms as input and return the range and noise-to-signal covariance parameters as output. \citet{lenzi2021neural} used simulated data as input and trained CNNs to learn the parameters of an MSP. Finally, \citet{sainsburydale2023fast} have used permutation invariant neural networks for large spatial extremes datasets in a Bayesian setting for estimating parameters from independent replicates. However, it is difficult to extend them to problems with large numbers of parameters; for example, a crucial assumption in our application is that the marginal distributions have spatio-temporally varying coefficients (STVC) which substantially expands the parameter space. Bayesian approaches have also been proposed since they provide stability by incorporating  prior information as available and are often preferred for uncertainty quantification. But they are restricted to either small $n$ \citep{Ribatet-Cooley-Davison} or very specific models \citep{Reich-Shaby, Morris-Reich-Thibaud, bopp2021hierarchical}. For lower dimensional problems, approximate Bayesian computation (ABC) can replace likelihood evaluation with repeated simulation from the MSP model \citep{erhardt2012approximate}. For general intractable likelihood estimation problems, neural networks can be leveraged for conditional density estimation. For example, \cite{pmlr-v97-greenberg19a} use a sequential neural posterior estimation method for simulation-based inference. A related method is normalizing flows \citep{Papamakariosetal2021,Kobyzevetal2021}, where a simple density is pushed through a series of transformations, often involving neural networks, to obtain more complex densities. 
%This has led to the predominant use of the composite likelihood (CL) \citep{Lindsay, Varin-Reid-Firth}. CL constructs marginal likelihoods on subsets of data and integrates them using working independence assumptions. Pairwise CL is widely used and approximates the likelihood by the product of bivariate likelihoods. The seminal work of \citet{Padoan-Ribatet-Sisson}, among others \citep{Genton-Ma-Sang, Huser-Davison-2013, Sang-Genton, Castruccio-etal}, has cemented the pairwise CL as a leading method for inference with MSPs. The pairwise CL offers a trade-off between statistical efficiency and computational speed and gives a consistent and asymptotically normal estimator \citep{Padoan-Ribatet-Sisson}. However, it suffers from  statistical inefficiency for large $n$ \citep{Huser-Davison-Genton}, finite-sample bias when using all pairs of observations \citep{Sang-Genton, Wadsworth-2015, Castruccio-etal} and computational challenges posed by computing likelihoods at all $O(n^2)$ pairs. Also, spatially-varying coefficient (SVC) models are essential for modeling the spatial distribution of extremes, and while SVC model-fitting tools are available in {\tt R} packages, to our knowledge these have not been investigated in the literature, presumably due to the computational burden of estimating a large number of parameters with pairwise CL. 

A second challenge posed by MSPs is that they are restrictive in the class of dependence types they can incorporate. Environmental data often have weakening spatial dependence with increasing levels of extreme quantiles as we go farther out into the tails of the distributions; however, MSPs are unable to accommodate this behavior. \citet{Wadsworth-Tawn-2012} addressed this with a max-mixture model that took an MSP and incorporated asymptotic independence at the boundary point of the parameter space using a mixing parameter. A more general approach was taken in \citet{Huser-Wadsworth} which combined a Pareto random variable with a GP resulting in a hybrid model which interpolates between perfect dependence and asymptotic independence, indexed similarly by a mixing parameter. This flexible model can establish asymptotic dependence or asymptotic independence from the data without needing a prior assumption. A limitation of this model is that the Pareto random variable is shared by the spatial locations, inducing dependence between distant sites. This might be unrealistic for an analysis over a large spatial domain. Finally, \citet{Hazra-Huser-Bolin} consider a mixture of a GP with a stochastic scale process; it can capture a range of extremal dependence structures, but does not employ a mixing parameter and therefore assumes equal contribution from both its constituent processes.

In this paper, we propose a spatial EVA model and an associated computational algorithm to address the aforementioned limitations of the MSP and related approaches. The EVA model is specified as a convex combination of an MSP and a GP for residual dependency, and has GEV margins with STVC. We refer to it as the process mixture model (PMM). From a modeling perspective, the mixture of the two spatial processes allows asymptotic dependence or independence for locations separated by  distance $h$, with independence as $h\to \infty$ (long-range independence). Furthermore, the STVC can account for temporal nonstationarity which is key for large-scale climate studies. This flexibility comes at a computational cost: the model has hundreds of parameters and even bivariate PDFs do not have a closed form to the best of our knowledge. Therefore we develop a new computational algorithm that uses a feed-forward neural network (FFNN) embedded in a density regression model \citep{xu-reich-2021-biometrics} to approximate the conditional distribution at one spatial location given a set of neighbors. Following this, the univariate density functions are used to approximate the joint likelihood for all locations by means of a Vecchia approximation \citep{vecchia1988estimation}.  This specification partitions the parameter space into a low dimensional vector of spatial dependence parameters and a higher dimensional vector of marginal parameters, and decouples the likelihood approximation from parameter estimation. The FFNN is trained on synthetic data generated from a design distribution using different parameter values; this allows us to avoid data scarcity issues and accommodate a range of marginal densities. Parameter estimation is carried out using MCMC. This computational framework is quite general. Unlike many of the approaches mentioned above, it can be applied to virtually any spatial process (e.g., GP, MSP, and mixtures), can accommodate high-dimensional STVC margins, as well as missing and censored data. We use the PMM to analyze changes in annual maximum streamflow within the US over the past 50 years.

The rest of this paper is organized as follows. Section 2 provides background on the construction and dependence measures for MSPs. Section 3 introduces the PMM for block maxima. Section 4 describes inference for the PMM which employs a deep learning Vecchia approximated density regression approach. Section 5 consists of a detailed simulation study demonstrating the method. Section 6 analyzes annual streamflow maxima data for HCDN stations across the US and identifies changes in their behavior over the past 50 years. Section 7 concludes with a discussion. Additional theoretical details, simulation studies, and results from our application are provided in a supplement.
%\section{Description of the data}\label{s:data}

\section{Background}\label{s:MSP-BR}
\subsection{The max-stable process}

A random process $\{R(\bs):\bs\in \mathcal{S} \subset\mathbb{R}^d\}$ indexed by spatial locations $\bs$ is called max-stable if there exists a sequence $\{X_i(\bs):i\in \mathbb{N}\}$ of independent copies of the process $\{X(\bs):\bs\in \mathcal{S}\}$, and normalizing functions $a_n(\bs)>0, b_n(\bs)\in \mathbb R$ such that
\begin{align*}
    R(\bs) \stackrel{d}{=} \frac{\max_{i=1:n}X_i(\bs)-b_n(\bs)}{a_n(\bs)}.
\end{align*}
Further, it can be shown that if there exist continuous functions $c_n(\bs) >0, d_n(\bs) \in \mathbb R$, such that as $n\to \infty$,
\begin{align*}
    \frac{\max_{i = 1:n} X_i(\bs) - d_n(\bs)}{c_n(\bs)} \rightarrow R(\bs),
\end{align*}
then $R(\bs)$ is either degenerate, or an MSP \citep{deHaan}. If $R(\bs)$ is non-degenerate, the pointwise distributions of $R(\bs)$ are in the GEV family \citep{deHaan-Ferreira}. 

Max-stable processes arise as the pointwise maxima taken over an infinite number of appropriately rescaled stochastic processes, and are therefore widely applied for studying spatial extremes, in particular block maxima (with a block size of $n$). 
MSPs can be constructed through a spectral representation \citep{deHaan,Penrose1992}. Let $\{Z_i: i\in \mathbb N\}$ be the points of a Poisson process on $(0,\infty)$ with intensity $1/z^2$. Then there exists a non-negative stochastic process $W(\bs)$ with continuous sample paths and with $\mathbb E W(\bs) = 1 $ for all $\bs\in \mathcal{S}$, such that
\begin{align*}%\label{e:MSPspectral}
    R(\bs) \stackrel{d}{=} \max_{i \geq 1} Z_i W_i(\bs),
\end{align*}
where $W_i(\bs)$ are independent copies of $W(\bs)$. Common parametric sub-classes of MSPs include mixed moving maxima processes \citep{wang_stoev_2010}, the Schlather processes \citep{Schlather}, and Brown-Resnick processes \citep{Brown-Resnick,Kabluchko-Schlather-deHaan}. 

The finite dimensional distribution of an MSP $R(\bs)$ at a set of locations $(\bs_1,\ldots,\bs_k) \in \mathcal{S}$ has the form $Pr\{R(\bs_j)<r_j, j=1:k\} = \exp\{-V(r_1,\ldots, r_k)\}$, where $V$ is known as the exponent function and is given by:
\begin{align*}
    V(r_1,\ldots,r_k) = \mathbb{E}\biggl [ \max_{j=1:k} \frac{W(\bs_j)}{r_j} \biggr].
\end{align*}

%Defining $\vartheta(\bs_1,\ldots,\bs_k) := \mathbb{E}[\max_{j=1:k}W(\bs_j)]$, we can write:

%   \begin{align*}
%       V(r,\ldots,r) = \frac{\vartheta(\bs_1,\ldots,\bs_k)}{r},
%\  end{align*}
%   where $\vartheta(\bs_1,\ldots,\bs_k)$ is known as the extremal coefficient.

\subsection{Dependence properties}
%The extremal coefficient $\vartheta(\bs_1,\ldots,\bs_k)$ summarizes the dependence among the $k$ elements of the random vector $R(\bs)$. The full distribution is intractable for large $k$, but explicit expressions are available for specific cases. For example, for spatial data, it is more convenient to restrict our attention to the extremal coefficient function \citep{SchlatherTawn2003} for the bivariate case (i.e., $k=2$), defined as
%\begin{align*}
%    \vartheta(\bs_1,\bs_2) = - r\log Pr\bigl[R(\bs_1)\leq r, R(\bs_2)\leq r\bigr] = \mathbb E\bigl [\max\{W(\bs_1),W(\bs_2)\}\bigr ].
%\end{align*}
%The extremal coefficient function $\vartheta(\bs_1,\bs_2)$ takes values in $[1,2]$; the lower and upper bounds correspond to complete dependence and complete independence, respectively.

Let $F_1$ and $F_2$ be the CDFs of $R(\bs_1)$ and $R(\bs_2)$, and let $U(\bs_i) = F_i (R(\bs_i))$ for $i=1,2$. The joint tail behavior of the two random variables $U(\bs_1)$ and $U(\bs_2)$ with uniform marginals can be studied in terms of the conditional exceedance probability, given by:
\begin{align*}\label{e:chi_defn}
    \chi_u(\bs_1,\bs_2) := Pr[U(\bs_1) > u | U(\bs_2) > u] \in (0,1),
\end{align*}
where $u\in(0,1)$ is a threshold. A commonly used measure of extremal dependence is the upper-tail coefficient \citep{Joe}, defined as:
\begin{align*}
    \chi(\bs_1,\bs_2) = \lim_{u\rightarrow 1}\chi_u(\bs_1,\bs_2).
\end{align*}
The random variables $U(\bs_1)$ and $U(\bs_2)$ are considered \textit{asymptotically dependent} if 
the upper-tail coefficient is strictly positive, and \textit{asymptotically independent} if it is zero.

We note that asymptotic (in)dependence is different from complete (in)dependence, since asymptotic (in)dependence is determined specifically by the joint behavior of the tails of the distribution as $u\to 1$. Asymptotic independence is also different from long-range independence which is determined by the asymptotic behavior of $\chi_u(\bs_1,\bs_2)$ as $||\bs_1-\bs_2||\to \infty$.

In the asymptotic independence scenario, the coefficient of tail dependence proposed by \citet{Ledford-Tawn-1996,Ledford-Tawn-1997} is useful to study the joint tail behavior of the process. Consider $R(\bs)$ with unit Fr\'{e}chet margins. The joint survivor function of $R(\bs_1)$ and $R(\bs_2)$ can be expressed as:
\begin{align*}
    \Bar{F}(r,r) := Pr[R(\bs_1)>r,R(\bs_2)>r] \sim \mathcal{L}(r)r^{-1/\eta}, \mbox{ as } r\to \infty,
\end{align*}
where $\mathcal{L}$ is a slowly varying function that satisfies $\mathcal{L}(tr)/\mathcal{L}(r)\to 1$ as $r\to \infty$ for all fixed $t>0$, and $\eta \in (0,1]$ is a constant that effectively determines the decay rate of $\Bar{F}(r,r)$ for large $r$. The parameter $\eta$ is known as the coefficient of tail dependence. A value of $\eta = 1/2$ indicates independent marginal variables; values lower and higher than $1/2$ correspond to a negative and a positive association respectively between the pair of variables.

\subsection{The Brown-Resnick process}
Consider $W(s) = \exp\{\epsilon(\bs) - \gamma(\bs)\}$ in the spectral representation of an MSP, where $\epsilon(\bs)$ is an intrinsically stationary Gaussian process with semivariogram $\gamma(\cdot)$, and $\epsilon(0) = 0$ almost surely. $W(\bs)$ is continuous and non-negative. Then $R(\bs)$ is a strictly stationary MSP known as the Brown-Resnick process, whose distribution depends only on $\gamma(\cdot)$. 
%\begin{comment}
The Brown-Resnick process is attractive since a wide range of variograms can be used with them, and they are relatively easily simulated. The exponent function that defines the joint distribution for the pair $R(\bs_1)$ and $R(\bs_2)$ is
\begin{align*}
    V(r_1,r_2) = \frac{1}{r_1}\Phi\biggl\{\frac{a}{2} - \frac{1}{a}\log \biggl(\frac{r_1}{r_2}\biggr)\biggr\} + \frac{1}{r_2}\Phi\biggl\{\frac{a}{2} - \frac{1}{a}\log \biggl(\frac{r_2}{r_1}\biggr)\biggr\},
\end{align*}
where $a = \{2\gamma(\bs_1 - \bs_2)\}^{1/2}$, and $\Phi(\cdot)$ denotes the standard normal distribution function.
%\end{comment}

%\edit{
\begin{comment}
 Its pairwise exponential measure takes the form
\begin{align*}%\label{e:extremal_coeff}
    V(r,r) = \frac{2\Phi}{r} \biggl\{ \biggl(\frac{\gamma(\bs_1 - \bs_2)}{2}\biggr)^{1/2} \biggr \}.
\end{align*}
\end{comment}
%}

\section{A Process Mixture Model for Spatial Extremes}\label{s:model}
Let $\{Y(\bs);\bs \in \mathcal{S}\}$ be a spatial extremes process indexed by the set $\mathcal{S}\subset \mathbb{R}^2$. In this section we consider $Y(\bs)$ to be defined as a block maximum, but the methods can be extended to peaks over a threshold. We assume a potentially different marginal distribution for each spatial location $\bs$ and denote $F_{\bs}$ as the marginal cumulative distribution function (CDF) for site $\bs$. For example, we assume that $F_{\bs}$ is the generalized extreme value (GEV) distribution with location $\mu(\bs) \in \mathbb{R}$, scale $\sigma(\bs)>0$, and shape $\xi(\bs) \in \mathbb{R}$, so that marginally
$$
  Y(\bs) \sim \mbox{GEV}\{\mu(\bs),\sigma(\bs),\xi(\bs)\}.
$$
Its  CDF $F_{\bf s}(y|\mu(\bs),\sigma(\bs),\xi(\bs)) := Pr[Y(\bs)<y]$ is
\begin{equation}
    F_{\bf s}(y|\mu(\bs),\sigma(\bs),\xi(\bs)) = \begin{cases}
        \exp \biggl[-\left\{1+\xi(\bf s)\left(\frac{y-\mu(\bf s)}{\sigma(\bf s)}\right)\right\}_+^{-1/\xi(\bf s)}\biggr], & \xi(\bs)\neq 0,\\
        \exp\left\{-\exp\bigl(\frac{y-\mu(\bf s)}{\sigma(\bf s)}\bigl)\right\}, & \xi(\bs) = 0,
    \end{cases}
\end{equation}
with $\{y\}_+ := \max(0,y)$, and support over the set $\bigr\{y:1+\xi(\bs)(y-\mu(\bs))/\sigma(\bs)>0\bigr\}$ for the CDF. The shape parameter $\xi(\bs)$ controls the lower and upper bounds of the distribution; the GEV distribution is bounded above for $\xi(\bs) <0$, and bounded below for $\xi(\bs)>0$. Therefore, the transformed variables
\begin{equation}
    U(\bs) = F_{\bf s}\{Y(\bs)\}
\end{equation}
share common uniform marginal distributions across the spatial domain. This transformation separates residual spatial dependence in $U(\bs)$ from the spatial dependence induced by spatial variation in the GEV parameters, which we model using GP priors over $\bs$. We note that although we describe the marginal and residual models separately, we fit a joint hierarchical model to simultaneously estimate all model parameters.

We define our spatial dependence model on the residual model $U(\bs)$ by taking $U(\bs) = G\{V(\bs)\}$, such that 
\begin{equation} 
  \label{eq:copula_pareto-02}
  V(\bs) = \delta R(\bs) + (1-\delta)W(\bs),
\end{equation}
where $R(\bs)$ and $W(\bs)$ are an MSP and a GP respectively, both transformed to have standard exponential margins, and $\delta\in[0,1]$ is the weight parameter to control relative contribution of the two spatial processes. Mixing the asymptotically dependent MSP with the asymptotically independent GP provides a rich model for spatial dependence. This generalizes the extremal process of \citet{Huser-Wadsworth}, who assumed a  standard exponential random variable $R$ common to all locations, by replacing it with an MSP. Since \eqref{eq:copula_pareto-02} mixes two processes, we refer to it as the process mixture model (PMM). Further details regarding the transformations required to obtain standard exponential margins, and the Huser-Wadsworth model are provided in the Supplementary Material \citep[][Appendix A.2--A.3]{VecchiaDL_supplement}.

Let $G(\cdot)$ denote the CDF of the marginal distribution of $V(\bs)$. By construction, $V(\bs)$ marginally follows the two-parameter hypoexponential distribution, and its CDF has the following functional form:
\begin{equation}
    \label{eq:hypo}
    G(v) = 1 - \frac{1-\delta}{1-2\delta}\exp\left\{-\frac{v}{1-\delta}\right \} + \frac{\delta}{1-2\delta}\exp\left\{-\frac{v}{\delta} \right \},
\end{equation}
where $\delta \in (0,1), \delta \neq 1/2$, and $v>0$.
  Although other options are possible, we model the correlation of $W(\bs)$ using the isotropic powered-exponential correlation function $\mbox{Cor}\{W(\bs_1,\bs_2)\} = \exp\{-(h/\rho_W)^{\alpha_W}\}$ with distance $h=||\bs_1-\bs_2||$ measured as the $L_2$ norm, smoothness $\alpha_W\in(0,2]$, and range $\rho_W>0$. $R(\bs)$ is assumed to be an isotropic Brown-Resnick process with the variogram $\gamma(h) = (h/\rho_R)^{\alpha_R}$, for smoothness $\alpha_R\in(0,2]$ and range $\rho_R>0$. 
  We also incorporate a nugget into the process mixture. Denoting the proportion of the variance explained by the spatial process by $r$, we construct $W(\bs)$ and $R(\bs)$ to satisfy:
\begin{align*}
    \mbox{Cor}\bigl(W(\bs_1),W(\bs_2)\bigr) &= r\cdot\exp\{-(h/\rho_W)^{\alpha_W}\}\\
    R(\bs) &= \max\{r\cdot R_{1}(\bs),(1-r)\cdot R_{2}(\bs)\},
\end{align*}
where $R_{1}(\bs)$ is an MSP, and $R_{2}(\bs) \iid \mbox{GEV}(1,1,1)$ distributed independently of $R_{1}(\bs)$. Since $R(\bs)$ and $W(\bs)$ are assumed to be isotropic processes, going forward, we can rewrite $\chi_u(\bs_1,\bs_2)$ and $\chi(\bs_1,\bs_2)$ as functions of the distance between locations, $\chi_u(h)$ and $\chi(h)$, respectively. 

\begin{figure}
    \centering
    \begin{subfigure}[b]{0.48\linewidth}
    \includegraphics[width=\linewidth]{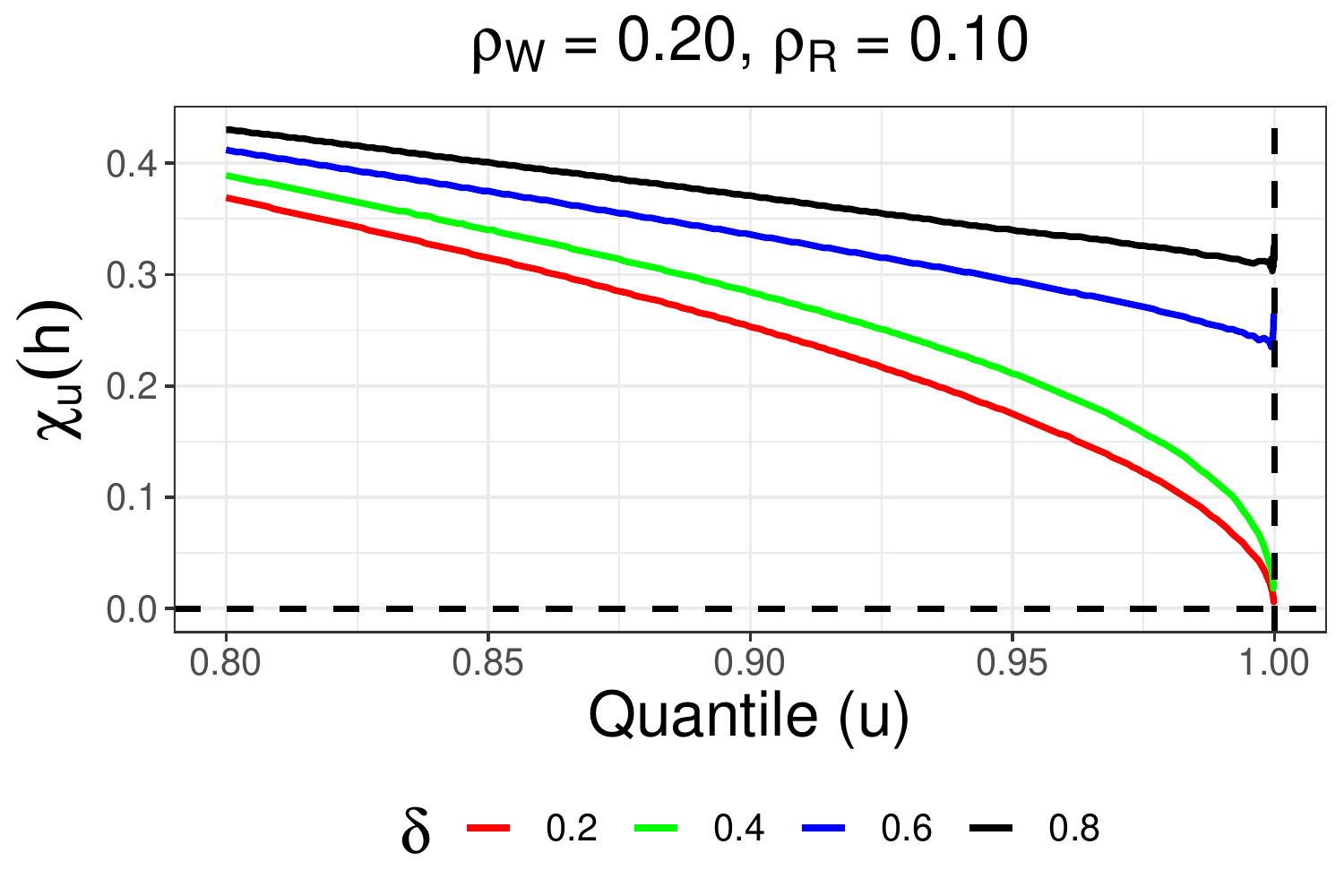}
\caption{$\chi_u(h)$ as a function of $u$ and $\delta$, at distance $h = 0.22$.}
    \label{fig:empericalchi_a}
    \end{subfigure}
    \hfill
    \begin{subfigure}[b]{0.48\linewidth}
    \includegraphics[width=\linewidth]{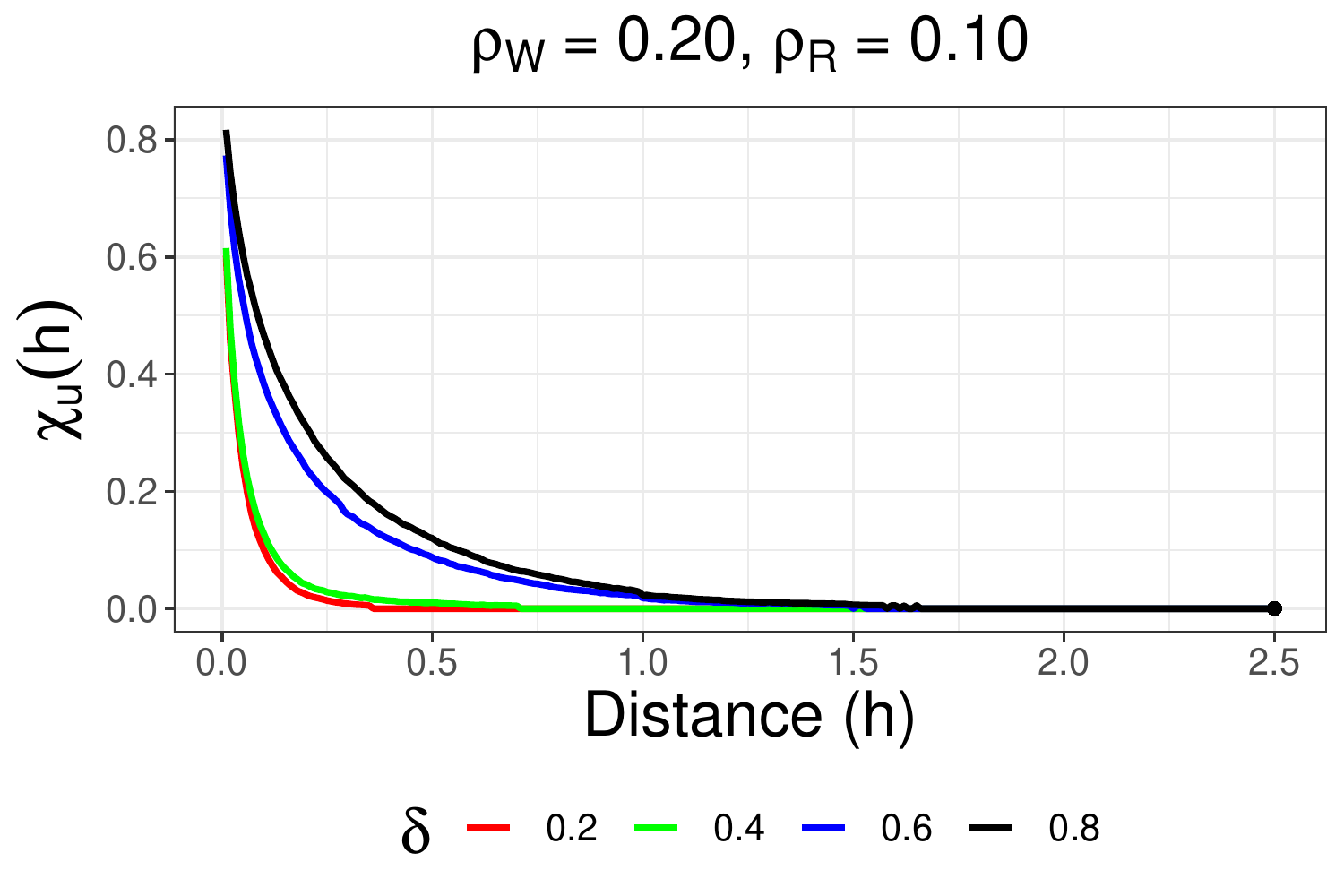}
\caption{$\chi_u(h)$ as a function of $h$ and $\delta$, for threshold $u = 0.999$.}
    \label{fig:empericalchi_b}
    \end{subfigure}
    \caption{{\bf Behavior of the empirical conditional exceedance:} Approximate $\chi_u(h)$ for the PMM plotted as a function of threshold $u$, distance $h$, and asymptotic dependence parameter $\delta$. Smoothness parameters $\alpha_W = \alpha_R = 1$, GP range $\rho_W = 0.2$, and MSP range $\rho_R = 0.1$ are fixed for both plots.}
    \label{fig:empericalchi}
\end{figure}

Figure \ref{fig:empericalchi} plots Monte Carlo approximations of $\chi_u(h)$ as a function of $u$ and $h$ for the PMM.  For these plots we fix $\rho_R=0.1$, $\rho_W=0.2$, $\alpha_R = \alpha_W = 1$, and $\delta\in\{0.2,0.4,0.6,0.8\}$. Figure \ref{fig:empericalchi_a} sets the correlation to $0.4$ by fixing $h=0.22$, and plots $\chi_u(h)$ as a function of the threshold $u$. The limit is zero for $\delta<0.5$ and positive for $\delta>0.5$. For small values of $h$, $R(\bs)$ is approximately the same for both sites (i.e., $R(\bs_1) \approx R(\bs_2)=R$), and thus the univariate $R$ result of \citet{Huser-Wadsworth} that the process is asymptotically dependent if and only if $\delta>0.5$ emerges. An analytical expression for $\chi_u(h)$ in this special case is provided in \citet[][Appendix A.4]{VecchiaDL_supplement}. From Figure \ref{fig:empericalchi_b}, we see that as the distance $h$ increases, $\chi_u(h)$ converges to zero for all $\delta$ because both $R(\bs)$ and $W(\bs)$ have diminishing spatial dependence for long distances. The rate of convergence of $\chi_u(h)$ to zero also depends on the value of $\delta$, with much slower convergence when $\delta>0.5$. We note that $\chi_u(h)$ does not converge to zero for large $h$ under the common $R$ model of \citet{Huser-Wadsworth}, which is unrealistic for studies on a large spatial domain. Additional plots for different values of $\rho_R$ and $\rho_W$ are provided in \citet[][Appendix A.5]{VecchiaDL_supplement}.

\section{Deep Learning Vecchia Approximation for the Process Mixture Model}\label{s:vecchia}

Fitting the PMM introduced in Section \ref{s:model} poses computational challenges, especially for large datasets.  The joint distribution for $W(\bs)$ is available in closed form but is cumbersome for large datasets; the joint distribution of $R(\bs)$ is available only for a moderate number of spatial locations, and the joint distribution of the mixture model is more complicated that either of its components.  An alternative is to build a surrogate likelihood for Bayesian computation
\citep[e.g.,][]{rasmussen2003gaussian,  %jabot2014comparison,  
wilkinson2014accelerating,  
%gutmann2016bayesian,
price2018bayesian,
%drovandi2018accelerating, 
wang2018adaptive, 
%acerbi2018variational, 
%jarvenpaa2019efficient, 
jarvenpaa2021parallel,li2019neural}.  Below we develop a surrogate likelihood based on a Vecchia decomposition \citep{vecchia1988estimation} and deep learning density regression.

Assume the process is observed at $n$ locations $\bs_1,...,\bs_n$.  Partition the parameters into those that affect the marginal distributions, denoted $\btheta^{MARG}$, and those that affect the spatial dependence, denoted $\btheta^{SPAT}$. For the model in Section \ref{s:model}, $\btheta^{MARG}$ includes the GEV parameters $\btheta^{MARG} = \{\mu(\bs_i),\sigma(\bs_i),\xi(\bs_i); i=1,...,n\}$ and $\btheta^{SPAT}=\{\delta,\rho_R,\alpha_R,\rho_W,\alpha_W\}$. Let $Y(\bs_i)\equiv Y_{i}$ and $U_{i}=F(Y_{i};\btheta^{MARG})$ be the transformation of the response so that the distribution of $U_{i}\in[0,1]$ does not depend on $\btheta^{MARG}$.  We approximate the spatial model on this scale and use the standard change of variables formula to define the joint likelihood on the original scale \begin{equation}\label{e:changeofvariables}
  f_y(y_{1},...,y_{n};\btheta^{MARG},\btheta^{SPAT}) = f_u(u_{1},...,u_{n};\btheta^{SPAT})\prod_{i=1}^n\left|\frac{dF(y_{i};\btheta^{MARG})}{dy_{i}}\right|,
\end{equation}
where $f_y(\cdot)$ and $f_u(\cdot)$ are the joint density functions of $Y_1\ldots, Y_n$ and $U_1, \ldots, U_n$ respectively.

We approximate the joint likelihood in \eqref{e:changeofvariables} using a Vecchia approximation \citep{vecchia1988estimation,stein2004approximating,datta2016hierarchical,katzfuss2021general},
\begin{equation}\label{e:vecchia}
    f_u(u_{1},...,u_{n};\btheta^{SPAT}) = \prod_{i=1}^n f(u_{i}|\btheta^{SPAT},u_{1}, ...,u_{i-1})
    \approx
    \prod_{i=1}^n f_i(u_{i}|\btheta^{SPAT},u_{(i)})
\end{equation}
for $u_{(i)} = \{u_j; j\in \calN_i\}$ and ${\cal N}_i\subseteq\{1,...,i-1\}$, e.g., the $m$ locations in ${\cal N}_i$ that are closest to $\bs_i$. The set of locations $\bs_{(i)}$ are analogously defined as $\bs_{(i)} = \{\bs_j; j\in \calN_i\}$; the set is referred to as the Vecchia neighboring set, and its members as the Vecchia neighbors of location $\bs_i$. Of course, not all locations that are dependent on location $\bs_i$ need be included in $\calN_i$ because distant observations may be approximately independent after conditioning on more local observations. The approximation therefore entails truncating the dependence that $u_i$ has on all its previous $i-1$ ordered sites to instead consider dependence on only up to $m$ sites, i.e., $|\calN_i| \leq m$. The first term on the right hand side of \eqref{e:vecchia} is the marginal density $f_1(u_1)$. Different choices are possible for ordering the locations prior to the Vecchia approximation \citep{Guinness2018}. In our work, the spatial locations are scaled to be on the unit square, and ordered by their distance from the origin.

The conditional distributions for the PMM do not have closed-form expressions. We consider two related approximations - the local approximation where individual conditional density functions $f_i(\cdot)$ are fit for each location, and a global approximation where a single conditional density function $f(\cdot)$ is estimated for all locations $\bs_i,i = 2,\ldots,n$. We primarily focus on the local SPQR and present it in this section. Details of the global SPQR approach can be found in the Supplementary Material \citep[][Appendix A.6]{VecchiaDL_supplement}.

For the local SPQR approximation at location $\bs_i$, we fit a density regression viewing $u_{(i)}$ and $\btheta^{SPAT}$ as features (covariates), denoted by $\bx_i$. We approximate the univariate conditional densities  for density regression using the model introduced in \citet{xu-reich-2021-biometrics}:
\begin{equation}\label{e:1}
    f_i(u_{i}|\bx_{i},\mathcal{W})= \sum_{k=1}^{K}\pi_{ik}(\bx_{i},\mathcal{W}_i)B_{k}(u_{i}),
\end{equation}
for $i = 2,\ldots, n$, where $\pi_{ik}(\bx_i,\mathcal{W}_i)\ge 0$ are probability weights with $\sum_{k=1}^K\pi_{ik}(\bx_i)=1$ that depend on the parameters $\mathcal{W}_i$ and $B_k(u_i)\ge0$ are M-spline basis functions that, by definition, satisfy $\int B_k(u)du=1$ for all $k$.  By increasing the number of basis functions $K$ and appropriately selecting the weights $\pi_{ik}(\bx_i)$, this mixture distribution can approximate any continuous density function \citep[e.g.,][]{chui1980,abrahamowicz1992}. 
The weights are modeled using a feed-forward neural network (FFNN) with $H$ hidden layers with $N_l$ neurons in hidden layer $l$ and multinomial logistic weights, equivalent to the softmax activation function. The model is
\begin{eqnarray}\label{e:ffnn}
\pi_{ik}(\bx_i,\mathcal{W}_i) &=& \frac{\exp\{\gamma_{Hk}(\bx_i,\mathcal{W}_i)\}}{\sum_{j=1}^K\exp\{\gamma_{Hj}(\bx_i,\mathcal{W}_i)\}},\\
\gamma_{lk}(\bx_i,\mathcal{W}_i) &=&  W_{ilk0} + \sum_{j=1}^{N_l}W_{ilkj}\psi\left\{\gamma_{l-1,j}(\bx_i,\mathcal{W}_i)\right\} \mbox{\ \ \ for\ \ \ } l \in\{1,...,H\},\nonumber\\
\gamma_{0k}(\bx_i,\mathcal{W}_i) &=&  W_{i0k0} + \sum_{j=1}^pW_{i0kj}x_{ij},\nonumber
\end{eqnarray}
where $\bx_i=(x_{i1},...,x_{ip})$, $\mathcal{W}_i = \{W_{ilkj}\}$ are parameters to be estimated and $\psi$ is the activation function. Activation functions are non-linear transformations applied to each output element of a layer, and is a key feature of neural networks that allows them to learn complex, non-linear dependencies in the data. The SPQR methodology of \citet{xu-reich-2021-biometrics} admits most of the commonly used activation functions, and we use the rectified linear unit (ReLU) \citep{NairHinton2010}. FFNNs use optimization to obtain optimum values of $\pi_{ik}(\bx_i,\mathcal{W}_i)$. In SPQR, the FFNN minimizes the negative log-likelihood loss associated with the density in \eqref{e:1}, using the process values evaluated at locations $\bs_i$ as the response.
Building on the universal approximation theorem for FFNNs \citep{hornik1989multilayer}, \citet{xu-reich-2021-biometrics} argue that with $H=1$ and large $K$ and $N_1$, the model in (\ref{e:ffnn}) can approximate any conditional density function that is smooth in its arguments.  
%To avoid over-fitting, we select prior distributions $W_{abc}|\tau^2\iid \mbox{Normal}(0,\tau^2)$ and the hyperparameter has uninformative half-Cauchy prior distribution $\tau\sim\mbox{t}_+(0,1)$.

Within this framework, approximating the conditional distributions is equivalent to estimating the weights $\mathcal{W}$. Unlike a typical statistical learning problem, observational data are not used to estimate $\mathcal{W}$. Rather, the weights are learned from training data generated from the PMM with parameters $\btheta^{SPAT} \sim p^*$, and then a realization from the process over sites $\bs_i$ and $\bs_{(i)}$ from the model conditioned on $\btheta^{SPAT}$. Specifically, we generate data at the observed spatial location with the same neighbor sets to be used in the analysis.  We select the design distribution $p^*$ with support covering the range of plausible values for $\btheta^{SPAT}$.  Given these values, we generate $U(\bs)$ at $\bs\in\{\bs_i,\bs_{(i)}\}$. The feature set $\bx_i$ for modeling $u_i$ at location $\bs_i$ thus contains the spatial parameters $\btheta^{SPAT}$ and the process values $U(\bs_{(i)})$ at the neighboring locations.

Therefore, all that is required to build the approximation is the ability to generate small datasets from the model. The size of the training data is effectively unlimited, meaning the approximation can be arbitrarily accurate.  Once the weights have been learned, applying the FFNN to the approximate likelihood is straightforward, and the Vecchia approximation ensures that the computational burden increases linearly in the number of spatial locations. The proposed estimation approach is in the same vein as recent simulation based neural inference methods \citep{gerber2021fast,lenzi2021neural,sainsburydale2023fast}; an important distinction is that instead of estimating model parameters using a neural network, our approach estimates a set of conditional densities which approximates the full model likelihood.

The weights in \eqref{e:1} are estimated separately for each location. That is, each component $f_i(u_{i}|\btheta^{SPAT},u_{(i)})$ in \eqref{e:vecchia} is modeled using its own FFNN, for $i = 2, \ldots, n$. The model is fit using the \texttt{R} package \texttt{SPQR} \citep{SPQR_R} and the fitting process is consequently referred to as the SPQR approximation. The \texttt{SPQR} package supports hardware acceleration for systems with a CUDA-compatible NVIDIA graphical processing unit (GPU), which was used for all SPQR models in this paper and provided significant speedups for computation times. All computations were carried out on a mobile workstation with 11th Gen Intel Core i7-11800H processors (8 cores, 16 logical processors), 64 GB of RAM, and an NVIDIA T600 laptop GPU with CUDA support. Algorithm \ref{a:local} outlines the local SPQR procedure.

\begin{algorithm}
\caption{Local SPQR approximation}\label{a:local}
\begin{algorithmic}
\Require Locations $\bs_1, \ldots, \bs_n$ and corresponding sets of neighboring locations $\bs_{(1)}, \ldots, \bs_{(n)}$
\Require Design distribution $p^*$, training sample size $N$
\State $i \gets 2$
\While{$i \leq n$}
\State $k \gets 1$
\While{$k \leq N$}
    \State Draw values of $\mathbf{\btheta}_{k}^{SPAT} \sim p^*$
    \State Generate $U_k(\bs)$ at $\bs \in \{\bs_i, \bs_{(i)}\}$ given $\mathbf{\btheta}_{2k}$ using  \eqref{eq:copula_pareto-02}
    \State Define features $\bx_{ik} = ( \mathbf{\btheta}_{k}^{SPAT},u_{(i)k})$, where $u_{(i)k} = \{U_k(\bs); \bs\in\bs_{(i)}\}$
    \State $k \gets k + 1$
    \EndWhile
\State solve $\hat{{\cal W}_i} \gets \underset{{\cal W}}{\operatorname{argmax}}  \prod_{k=1}^N f_i(u_{ik}|\bx_{ik},{\cal W})$ for $f_i(u_i|\bx_i,{\cal W}_i)$ defined in \eqref{e:1} using {\tt SPQR}
    \State $i \gets i + 1$
\EndWhile
\end{algorithmic}
\end{algorithm}

Given the approximate model in (\ref{e:changeofvariables}) for $f_y$ with an SPQR approximation for $f_u$, a Bayesian analysis using Markov Chain Monte Carlo (MCMC) methods is straightforward.  We use Metropolis updates for both $\btheta^{MARG}$ and $\btheta^{SPAT}$.  For a spatially-varying coefficient model with local GEV coefficients for location $\bs_i$, we update parameters $\{\mu(\bs_i), \sigma(\bs_i), \xi(\bs_i)\}$ as a block sequentially by site, and exploit the Vecchia approximation to use only terms in the likelihood corresponding to sites $j \mbox{ such that } j \in \mathcal{N}_j$, i.e. sites for which site $i$ is included in the neighboring set.  All Metropolis updates are tuned to give acceptance probability 0.4, and convergence is diagnosed based on the visual inspection of the trace plots.  Additional computational details are given for specific analyses below, and MCMC code is provided in a GitHub repository - \url{https://github.com/reetamm/SPQR-for-spatial-extremes}.

\section{Simulation Study}\label{s:sim}
\begin{figure}
    \begin{subfigure}[b]{0.45\linewidth}
    \centering
        \ \includegraphics[width=\linewidth]{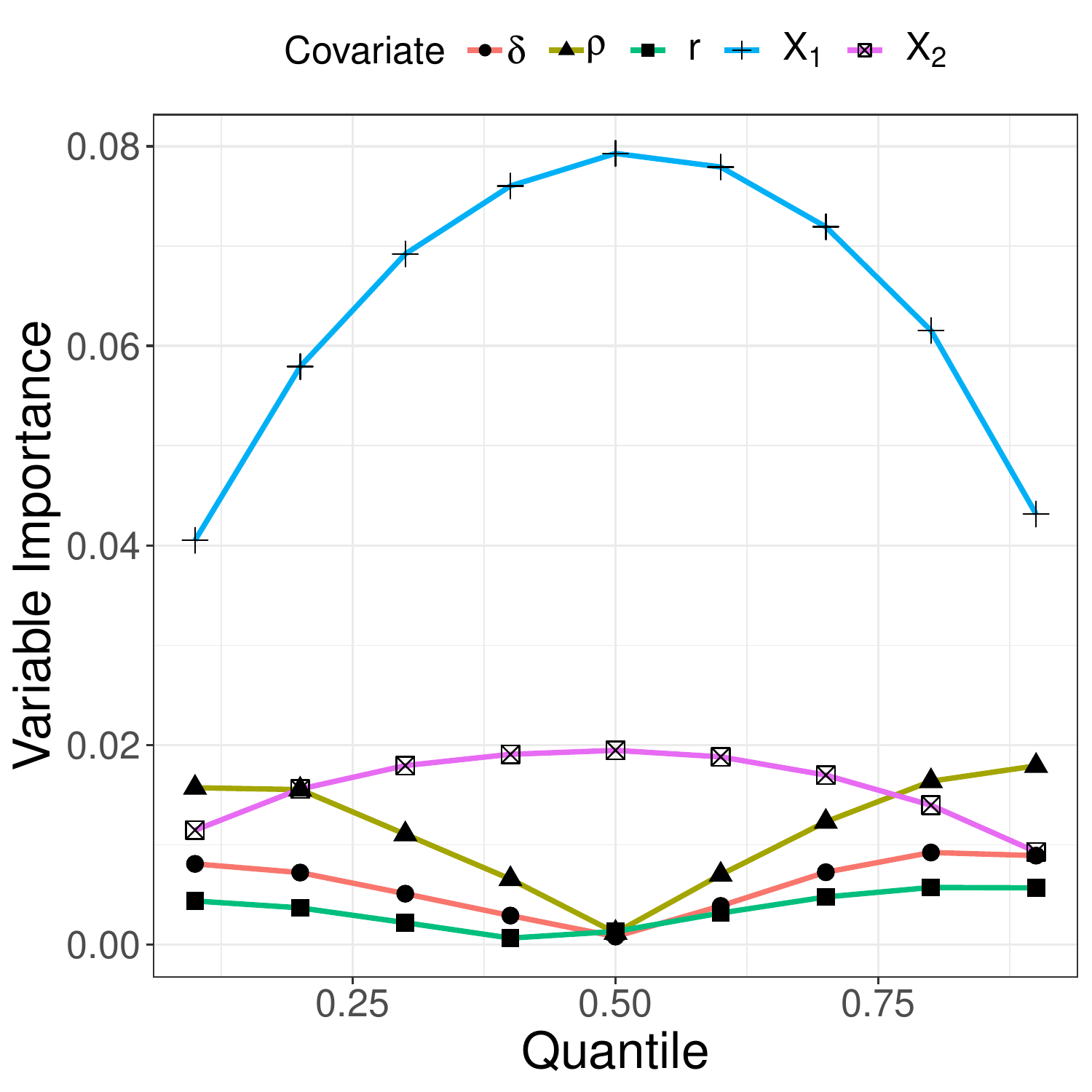}
\caption{Variable importance of $\delta$, $\rho$, $r$, and the two nearest neighbors.}
    \label{fig:EVP_VI}
    \end{subfigure}
    \hfill
        \begin{subfigure}[b]{0.45\linewidth}
    \centering
     \includegraphics[width=\linewidth]{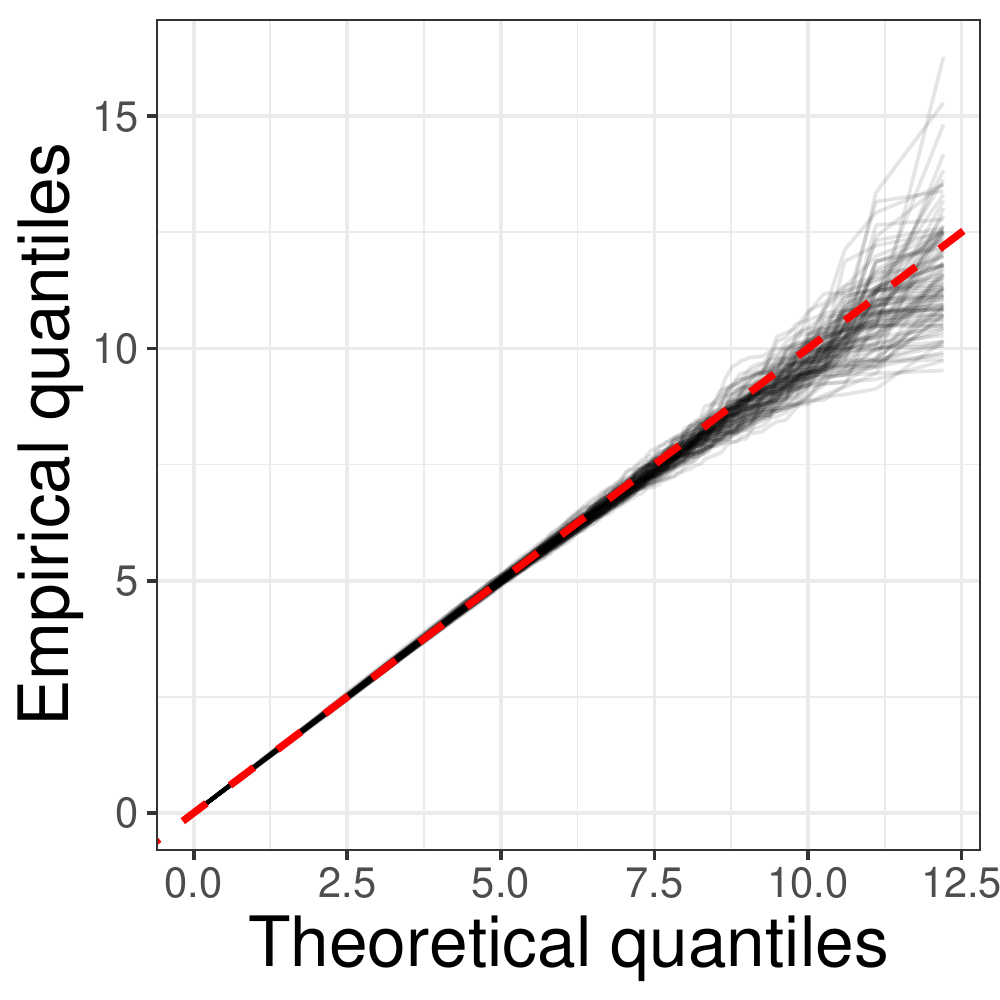}
\caption{Q-Q plots on the exponential scale for checking goodness-of-fit.}
    \label{fig:EVP-qqplot}
    \end{subfigure}
        \caption{{\bf Model diagnostics for local SPQR fit on PMM data}: VI  plot of the 5 most important variables (left) and Q-Q plot on the exponential scale (right) for the local SPQR at site 25. The gray lines correspond to Q-Q plots of fits to different simulated datasets.}
    \label{fig:diag_VI}
\end{figure}

We illustrate the potential of the proposed algorithm with a simulation study to evaluate the performance of the method on extreme value data in terms of both density estimation and parameter estimation. We consider the PMM as the underlying spatial process, and an STVC model for the marginal GEV parameters. The goal of the study is to assess whether our approach can simultaneously model the marginal GEV distributions and the underlying extremal spatial process.
We simulated data from the PMM defined in Section~\ref{s:model} at $n=50$ spatial locations distributed randomly on the unit square.
To put the MSPs and GPs on the same scale, we assume common smoothness parameter $\alpha_R=\alpha_W=\alpha=1$ and parameterize $\rho_W$ and $\rho_R$ to give the same effective range, i.e., the distance at which the correlation of the GP reaches 0.05 ($h=\rho_W\log(20)$) and the $\chi$-coefficient of the MSP reaches 0.05 ($h=\rho_R4\Phi^{-1}(1-0.05/2)^2$ where $\Phi$ is the standard normal distribution function). This results in $\rho=\rho_W$ and $\rho_R=0.19\rho$. We also assume the absence of a nugget term, i.e., $r = 1$.

The locations are ordered by their distance from their origin. For the $i^{th}$ location $\bs_i, \mbox{ with }i>1$ , the Vecchia neighbor set $\bs_{(i)}$ consists of the $m$ nearest neighbors of $\bs_{i}$ among the previous $i-1$ locations. Up to 15 conditioning points are used in the Vecchia neighbor set, i.e., $m = \min(i-1,15)$, where $m = |\mathcal{N}_i|$. We use the local SPQR model outlined in Algorithm~\ref{a:local} to model the conditional densities at each location, which employs stochastic gradient descent with the adaptive moment estimation (Adam) optimizer \citep{kingma2014adam}. We compared multiple SPQR models as part of the density estimation process. Models were compared on the basis of the log-score and the Kullback-Leibler (KL) divergence between the estimated and true densities. Architectures with the lowest validation loss were chosen in each case. Our NN architecture for each SPQR model consists of two hidden layers with 30 and 15 neurons, 15 output nodes, a learning rate 0.001, batch size 100, and 50 epochs. We train the SPQR model with design distribution $p^*$, generating samples uniformly on $\rho\in(0.0,0.5)$ and $\delta\in(0,1)$. 

We first evaluate the SPQR fits of the PMM full conditional distributions. Figure \ref{fig:EVP_VI} plots variable importances for the local SPQR model at site 25. The VI plot identifies the features that are most important for explaining different aspects of the conditional distribution; the spatial parameters are found to be more important at extreme quantiles, while the process realizations at the Vecchia neighbor locations are more important closer to the median. The plots indicate that the conditional distributions at each location are sensitive to the spatial parameters; details of the VI metric used by SPQR are presented in the Supplementary Material \citep[][Appendix A.7]{VecchiaDL_supplement}. To assess goodness of fit, we repeat the process of fitting the local SPQR model at site 25 for 100 independent datasets simulated from the PMM. Figure \ref{fig:EVP-qqplot} is a Q-Q plot based on true and fitted values from the SPQR models, where each line corresponds to one of the 100 datasets. The Q-Q plot is presented on the exponential scale \citep{Heffernan2001ExtremeVA} to verify whether the model can adequately capture tail behavior.
The values fall along the $Y=X$ line, suggesting a good model fit. Computation time for local SPQR at sites with all 15 neighbors is approximately 22 minutes. The \texttt{doParallel} package in \texttt{R} was used to parallelize SPQR model fits and improve computation times.

\begin{table}
\centering
\caption{Coverage (in $\%$) for marginal GEV parameters under 2 scenarios based on MCMC simulations over 100 datasets. The 3 values represent the minimum, mean, and maximum coverage across the 50 study locations.}
\label{t:coverage}
\begin{tabular}{ccccc}
\toprule
 & $\mu_0$ & $\mu_1$ & $\sigma$ & $\xi$ \\\midrule
$\delta = 0.2$ & (86, 92, 96) & (91, 96, 99) & \multicolumn{1}{c}{(83, 91, 98)} & (92, 96, 100) \\
$\delta = 0.8$ & (85, 93, 100) & (90, 95, 100) & (86, 93, 98) & (90, 96, 100)\\\bottomrule
\end{tabular}
\end{table}

\begin{figure}
    \centering
    \includegraphics[width=.4\linewidth]{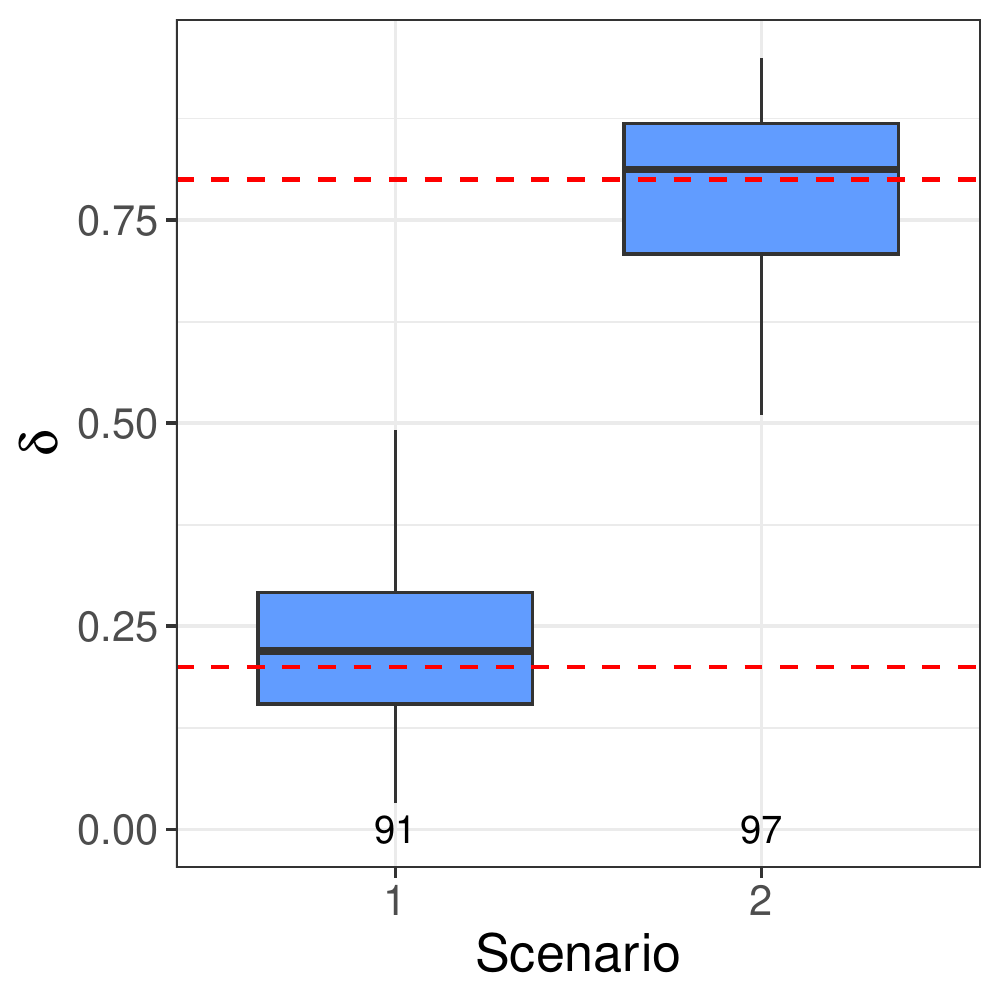}
\caption{Sampling distribution of the posterior mean for the asymptotic dependence parameter $\delta$ for two simulation scenarios. The horizontal dashed lines are true values and the numbers along the bottom give the empirical coverage of the 95\% intervals. }
    \label{fig:sim:pmmsvc_coverage}
\end{figure}

We conduct a simulation study to explore how density-estimation errors propagate to parameter-estimation errors for spatially varying GEV parameters. Two scenarios are considered for the simulation studies, corresponding to $\delta = \{0.2,0.8\}$. For both scenarios, \edit{we assume the true values of} $\rho = 0.15$ and $r=0.80$, and simulate 100 datasets at the 50 locations. Each dataset consists of 50 independent (time) replicates. For spatial coordinates $\bs = (s_1,s_2)$, the marginal GEV parameters are:
\begin{align*}
    \mu_0(\bs) &= \exp\{2+\cos(2\pi s_1) + \cos(2\pi s_2)\}\\
    \mu_1(\bs) &= 1\\
    \sigma(\bs) &= \exp\{\cos(2\pi s_2)\}\\
    \xi(\bs) &= \frac{1}{2}\sin(\frac{\pi}{2}s_1).
\end{align*}
We model the marginal GEV parameters using an STVC model. At each location, we assume the data to be GEV with parameters
\begin{align*}
    Y_t(\bs) \sim GEV\bigl(\mu_0(\bs) +\mu_1(\bs)X^*_t, \sigma(\bs),\xi(\bs)\bigr),
\end{align*}
with $X^*_t = (t-25.5)/10$, $t=1:50$. The variable $X_t^*$ represents changes in the location parameter over time. 

The intercept process $\mu_0(\bs)$ is assigned a GP prior with a Mat\'ern covariance function and nugget effects allow local heterogeneity:
\begin{align*}
      \mu_0(\bs) &= \tilde{\mu}_0(\bs) + e_0(\bs) \\
      e_0(\bs)&\iid \mbox{Normal}(0,v_{\mu_0})\\
      \tilde{\mu_0}(\bs)&\sim \mbox{GP}\bigl(\beta_{\mu_0},\tau^2_{\mu_0}K(\bs,\bs';\rho_{\mu_0},\kappa_{\mu_0})\bigr)\\
      \beta_{\mu_0} &\sim \mbox{Normal}(0,10^2), \tau^2_{\mu_0}, v^2_{\mu_0}\iid \mbox{IG}(0.1,0.1)\\
      \log \rho_{\mu_0} &\sim \mbox{Normal}(-1,1), \log \kappa_{\mu_0} \sim \mbox{Normal}(-2,1),
  \end{align*}
where $K(\bs,\bs';\rho_{\mu_0},\kappa_{\mu_0})$ is the Mat\'ern correlation function with spatial range $\rho_{\mu_0}$ and smoothness parameter $\kappa_{\mu_0}$, and IG$(\cdot,\cdot)$ is the inverse-Gamma distribution. The slope $\mu_1(\bs)$, the log-scale $\log \sigma(\bs)$, and the shape $\xi(\bs)$ are modeled similarly using GPs. The spatial parameters have priors $\delta\sim\mbox{Uniform}(0,1)$, $\rho\sim\mbox{Uniform}(0.0,0.5)$, and $r\sim \mbox{Uniform(0,1)}$. Runtimes were approximately 1 minute for 1,000 iterations of the MCMC.

Table \ref{t:coverage} details coverage of the empirical $95\%$ intervals for the posterior distribution of the marginal GEV parameters. Mean coverage across locations is near or at nominal level across different parameters and scenarios. Figure \ref{fig:sim:pmmsvc_coverage} plots the sampling distribution of the posterior mean estimator of $\delta$ for the 2 scenarios, and provides empirical coverage of the $95\%$ posterior interval. Both scenarios show low bias; coverage is $91\%$ for the asymptotic independence scenario, and $97\%$ for the asymptotic dependence scenario. Overall, the SPQR approach is able to distinguish between the two asymptotic regimes in the presence of spatially varying marginals.

Additional simulation studies are presented in the Supplementary Material \citep[][Appendix B]{VecchiaDL_supplement} which involve special cases of the PMM. The first considers a GP as the spatial process, and evaluates both the global and local SPQR methods. The second study is also for a PMM but in a non-STVC setting, where were explore a few additional scenarios including negative shape parameters, and missing and censored data. The final demonstrates the need of non-linear NN layers by fitting a SPQR model without any hidden NN layers.

\section{Analysis of Extreme Streamflow in the US}\label{s:app}
\subsection{Data description and exploratory analysis}\label{s:app:data}

\begin{figure}[t]
    \centering
    \includegraphics[width=0.8\linewidth]{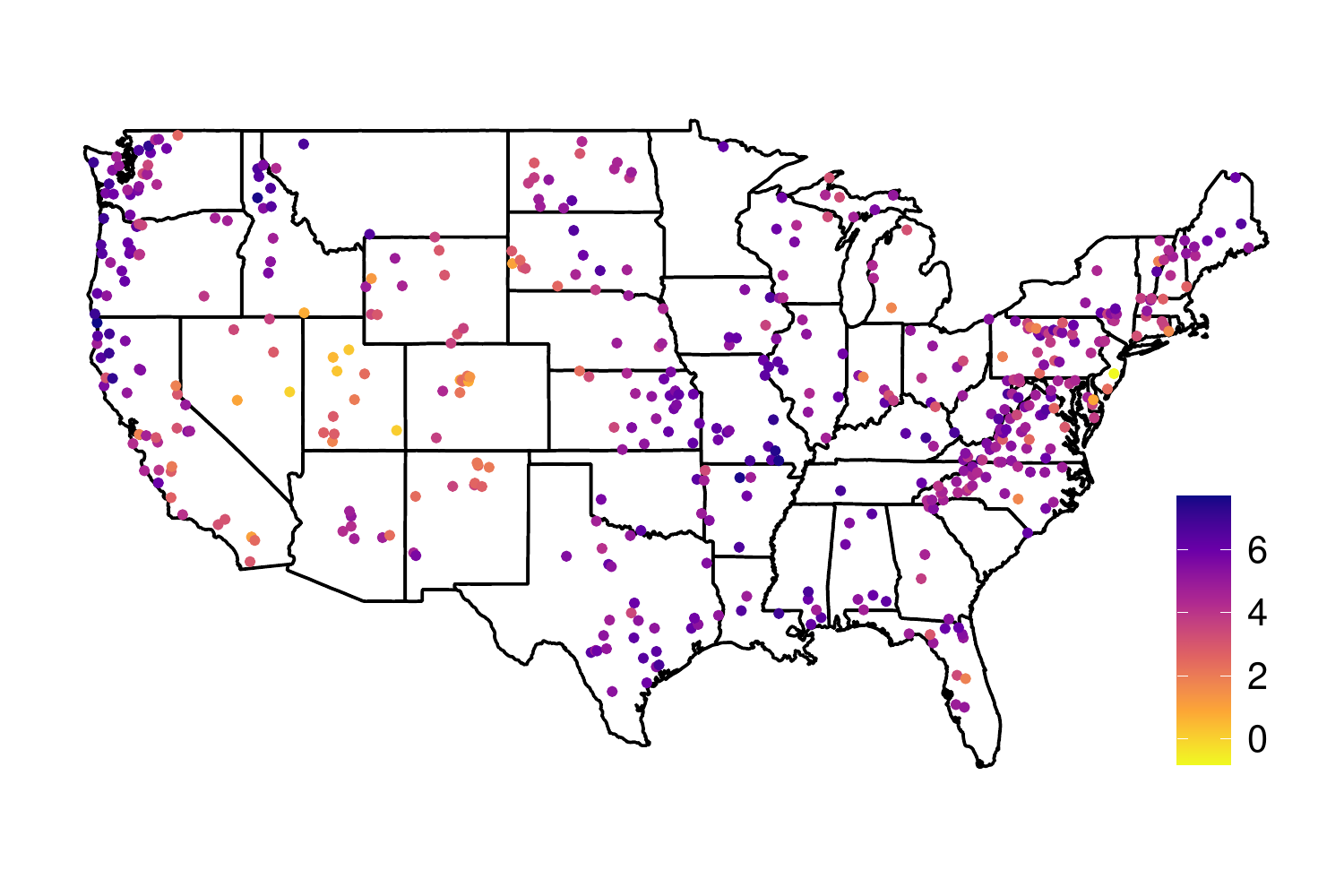}
\caption{{\bf HCDN annual maxima}: Sample 0.9 quantile of the log annual streamflow maxima $Y_t(\bs)$ (in m$^3/$s) at each of the 487 gauges.}
    \label{fig:HCDN_quantiles}
\end{figure}

We apply the proposed methods to model extreme streamflow from 1972--2021 at 487 stations across the US with complete data.  These locations are part of the USGS Hydro-Climatic Data Network (HCDN) 2009 \citep{lins2012usgs} and are relatively unaffected by human activities. The data is downloaded using the \texttt{dataRetrieval} package in \texttt{R} \citep{dataRetrieval}, and the code is made available in our GitHub repository\footnote{\url{https://github.com/reetamm/SPQR-for-spatial-extremes}}. Our goal is to identify regions within the US where the distribution of extreme streamflow is changing over time. The annual maximum of daily streamflow is measured in m$^3/$s, and for each of the $T=50$ years and $n=487$ stations, the response $Y_t(\bs)$ is taken to be the logarithm of the annual maximum. The log transformation was chosen as a Box-Cox transformation parameter after comparing parameter values between -2 and 2 on the basis of goodness of fit, profile likelihood values, and the stability of initial MLE estimates at the locations. Figure \ref{fig:HCDN_quantiles} plots the sample 0.9 quantile of the $T=50$~observations at each station, which show considerable spatial variation.

\begin{figure}
    \begin{subfigure}[b]{0.4\linewidth}
         \centering
    \includegraphics[width=\linewidth]{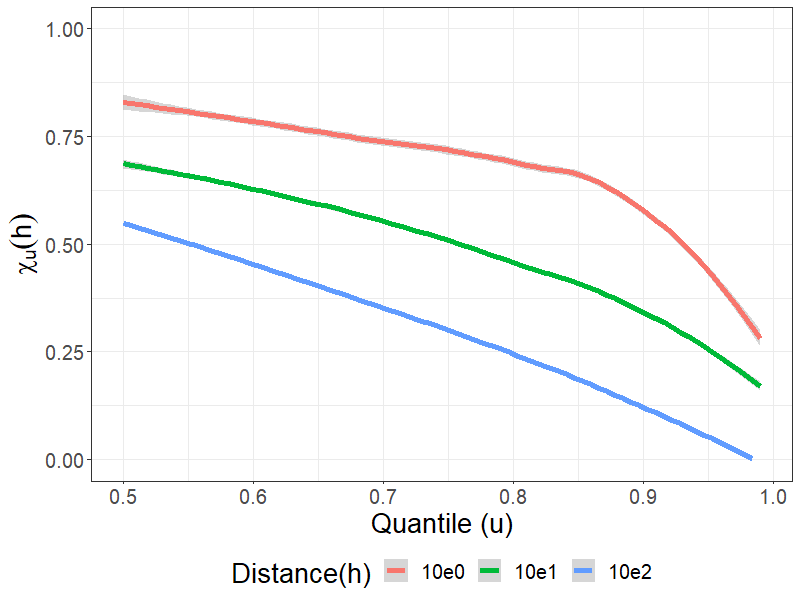}
\caption{Empirical conditional exceedance $\chi_u(h)$ for log annual maximum streamflow at different distances.}
    \label{fig:chi_h}
    \end{subfigure}
    \hfill
    \begin{subfigure}[b]{0.4\linewidth}
        \centering
    \ \includegraphics[width=\linewidth]{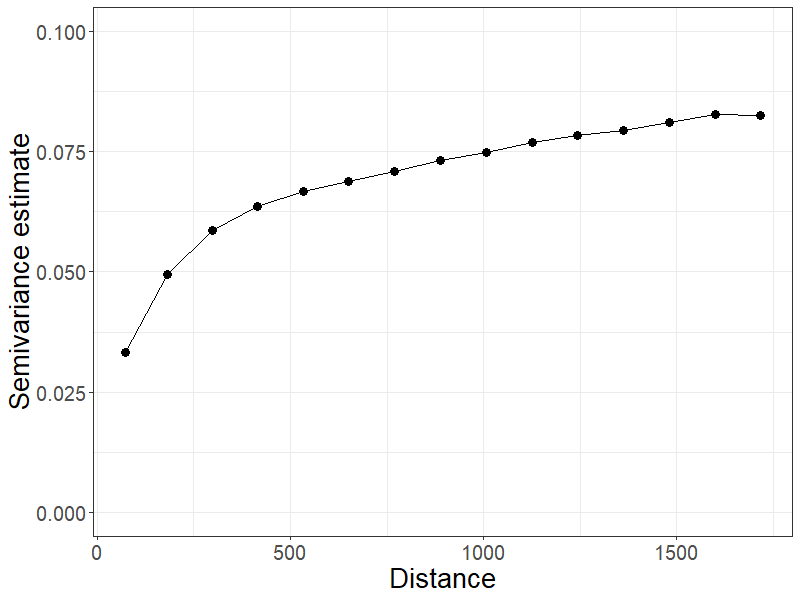}
\caption{Sample variogram for log annual maximum streamflow, averaged over 50 years of data.}
    \label{fig:avg_variogram}
    \end{subfigure}
        \caption{Spatial behavior of log annual maximum streamflow in terms of the conditional exceedance and the variogram, units of km.}
    \label{fig:spatial_HUC02}
\end{figure}

In order to study the dependence structure of the process, especially at its extremes, we consider the conditional exceedance probability $\chi_u(h)$ of maximum streamflow at pairs of stations separated by a distance $h = ||\bs_j - \bs_k||$ in kilometers (km).
Figure \ref{fig:chi_h} plots $\chi_u(h)$ for rank-standardized streamflow data as a function of $u$ for different values of $h$. The rank standardization ensures a Uniform$(0,1)$ marginal distribution at each location. Stations farther away from each other can be seen to have less extremal dependence, with tail dependence approaching 0 for stations 1000 km apart.
Figure \ref{fig:avg_variogram} plots the mean of the annual variograms of the streamflow data. It shows a range of over 1500 km, as well as the presence of a nugget effect. Both plots suggest that extremal streamflow is spatially dependent at distances of 1000~km or more, even after accounting for the spatial differences in the marginal distributions.

\begin{figure}
    \centering
    \includegraphics[width=0.24\linewidth]{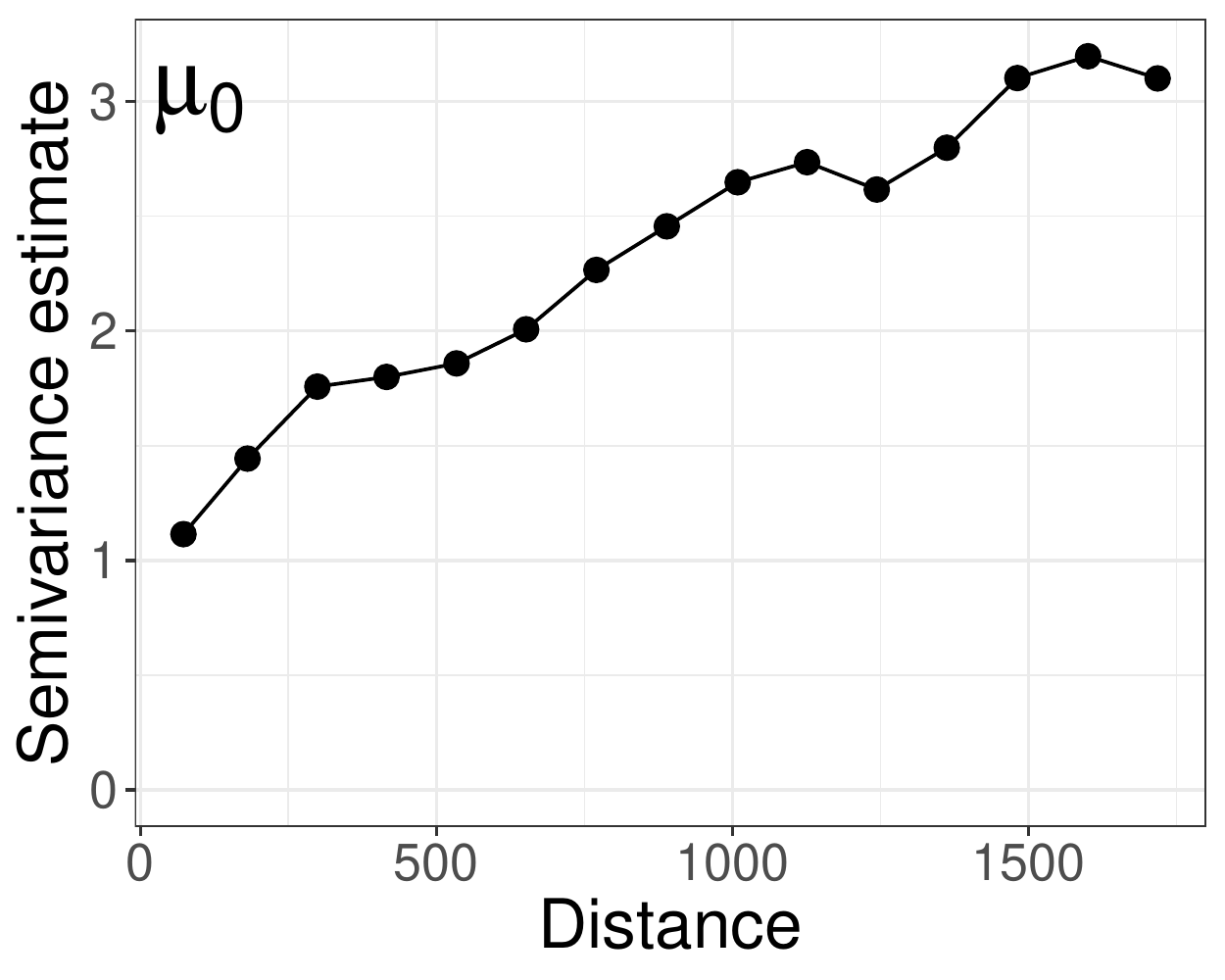}
    \includegraphics[width=0.24\linewidth]{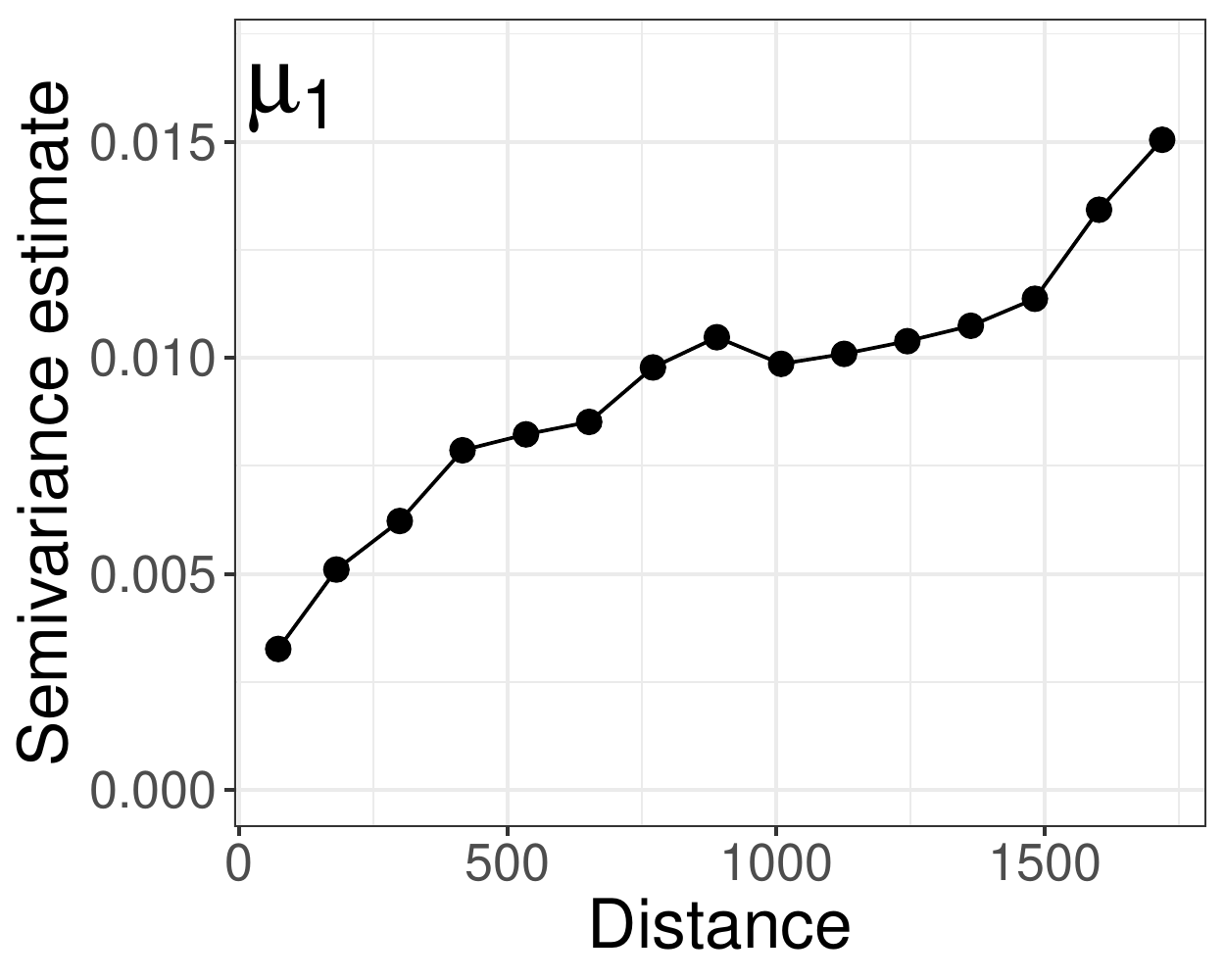}
    \includegraphics[width=0.24\linewidth]{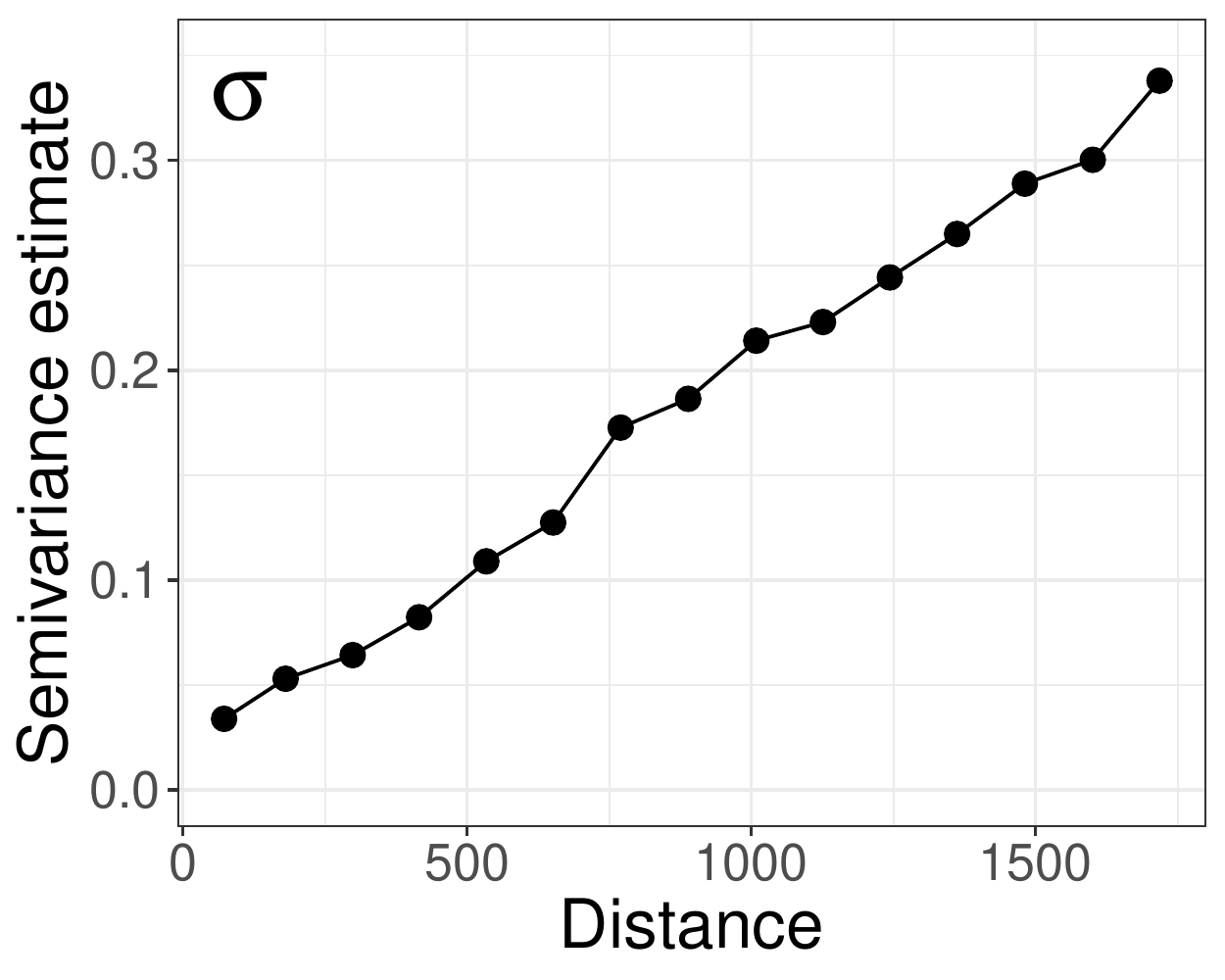}
    \includegraphics[width=0.24\linewidth]{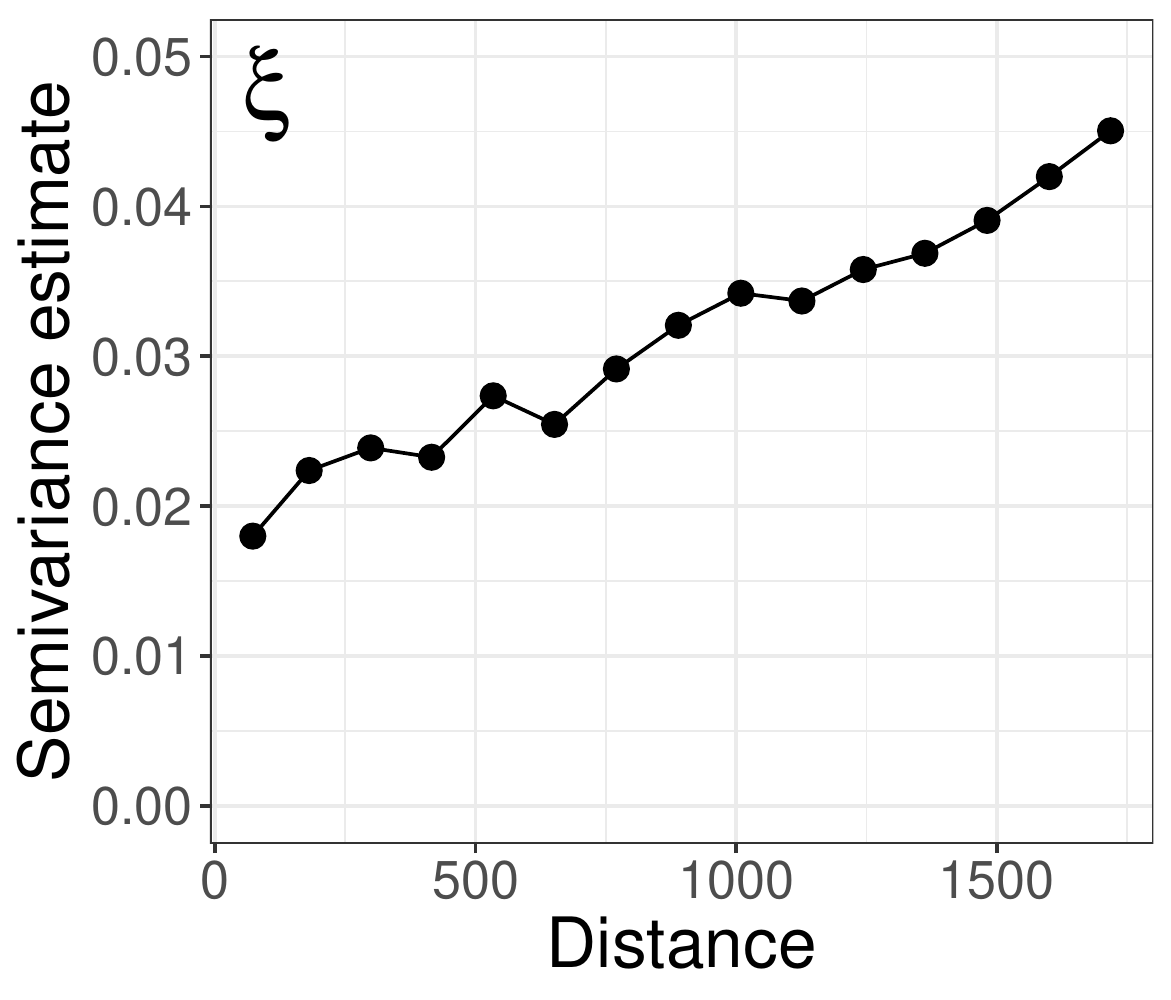}
    \caption{\small Sample variogram plots for the MLE of GEV parameters for extreme streamflow at each location. Parameters are labeled on the top-left of each panel.}
    \label{fig:variograms_HUC02}
\end{figure}

For the marginals at each location, we assume GEV distributions with STVC, 
\begin{equation}\label{e:streamflow_GEV}
   Y_t(\bs) \sim \mbox{GEV}\left[\mu_0(\bs)+ \mu_1(\bs) X_t,\sigma(\bs),\xi(\bs)\right],
\end{equation}

where $X_t = (\text{year}_t-1996.5)/10$ for $\text{year}_t = 1972 + t-1$. This parameterization attempts to capture changes in the location parameter in the past 50 years due to changing climate; positive values of $\mu_1(\bs)$ would suggest an increase in the magnitude of the annual extremal streamflow. Figure \ref{fig:variograms_HUC02} plots variograms for MLE estimates (estimated separately by location) of the GEV parameters at each location. All 4 GEV parameters show spatial dependency, which motivates the STVC specification.

%\subsection{STVC marginal model specification}
The marginal GEV parameters for each location are assigned GP priors with a Mat\'ern correlation functions and nugget effects to allow for local heterogeneity. The prior specification is similar to the one used in Section \ref{s:sim}, except with the additional assumption of a common Mat\'ern smoothness parameter was made for all four GP priors to improve MCMC convergence, i.e., $\kappa_{\mu_0} = \kappa_{\mu_1} = \kappa_{\sigma} = \kappa_{\xi} = \kappa$.  For the residual model, we use the PMM in Section \ref{s:model} for spatial dependence and assume independence across years. The specification includes a nugget based on the mean variogram (Figure \ref{fig:avg_variogram}).
 As priors for the joint parameters $\btheta^{SPAT}$, we set $\delta,r \sim \mbox{Uniform}(0,1)$, and $\rho \sim \mbox{Uniform}(0,6251)$ measured in km. 

\subsection{Results}\label{s:app:results}
The local SPQR approximation for the log of annual streamflow maxima is thus based around $\btheta^{SPAT} = (\delta,\rho,r)$, and models are fit using 200,000 synthetic observations at each of the 486 locations with neighbors.
Once the local SPQR models have been fit, we run two MCMC chains for 30,000 iterations each, with two different starting values of $\delta$. The first 10,000 iterations from each chain are discarded as burn-in; additional results are provided in the Supplementary Material \citep[][Appendix C.2]{VecchiaDL_supplement}.

\subsubsection{Parameter estimates:}

The posterior means (standard deviations) of the spatial parameters are ${\hat \delta} = 0.45\, (0.02)$, ${\hat \rho} = 807\, (45)$ km, and ${\hat r} = 0.92\, (0.004)$. The posterior of $\delta$ has a $95\%$ interval of $(0.40,0.49)$ which puts the process in the asymptotic independence regime with high probability. The GEV Mat\'ern smoothness parameter estimate is $\hat{\kappa} = 0.60\, (0.03)$, and the four range parameters (in km) are $\hat{\rho}_{\mu_0} = 12435\, (10645)$, $\hat{\rho}_{\mu_1} = 27605\, (10689)$, $\hat{\rho}_{\sigma} = 20311\, (11232)$, and $\hat{\rho}_{\xi} = 20320\, (11481)$. The STVC parameters are therefore much smoother over space than the year-to-year variation captured by $\rho$. The intercept is the most variable GEV parameter across space and the slope is the least, having the smallest and largest range parameter estimates, respectively. It is likely that the intercept varies the most because the magnitude of streamflow at a station is dependent on very local features. Conversely, the slope parameter may vary smoothly because the drivers of change are regional rather than local in nature.

\begin{figure}
    \begin{subfigure}[b]{0.49\linewidth}
        \centering
     \includegraphics[width=0.85\linewidth]{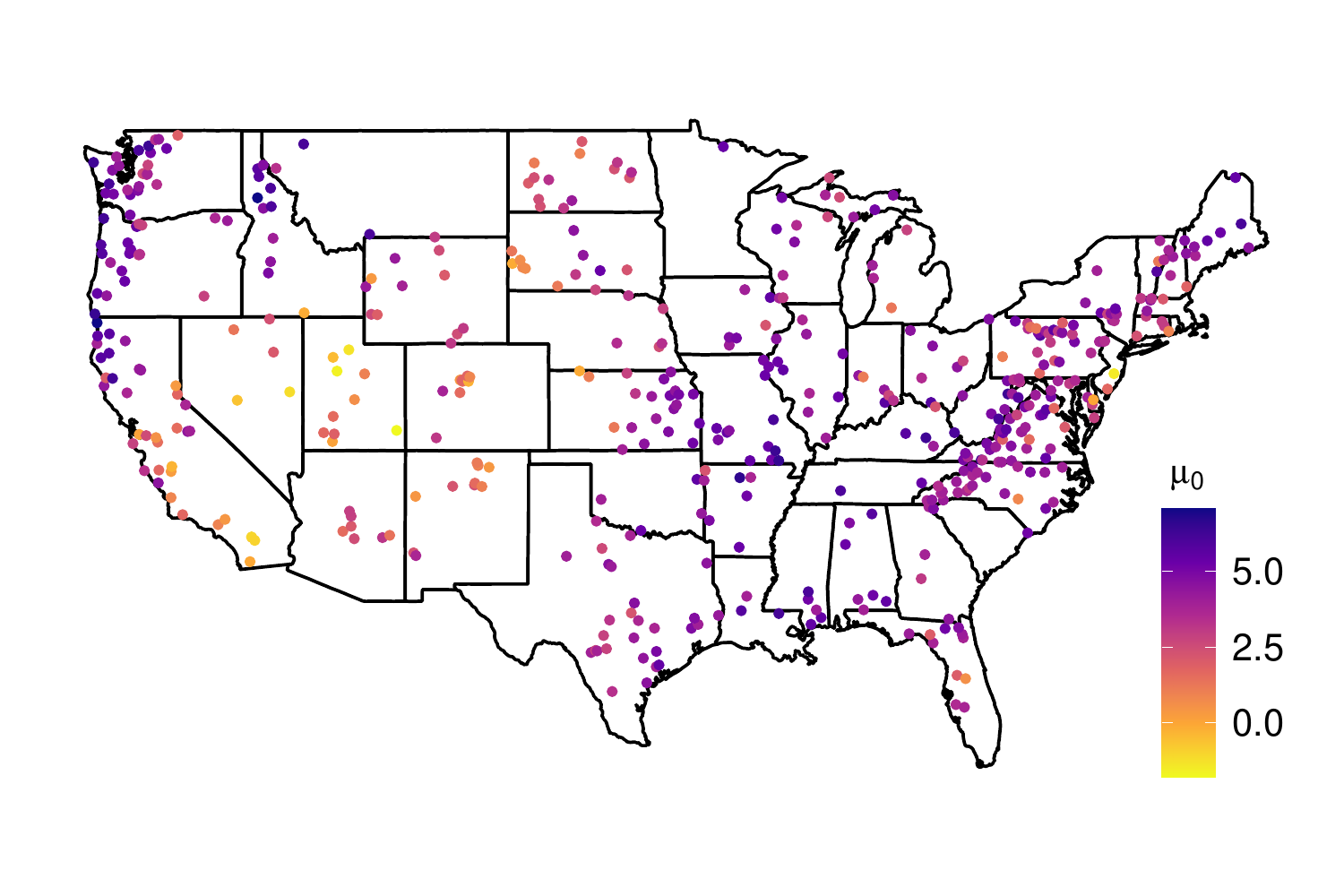}
\caption{Posterior mean of $\mu_0(\bs)$.}
    \label{fig:mu0_estimates}
    \end{subfigure}
    \begin{subfigure}[b]{0.49\linewidth}
        \centering
        \ \includegraphics[width=0.85\linewidth]{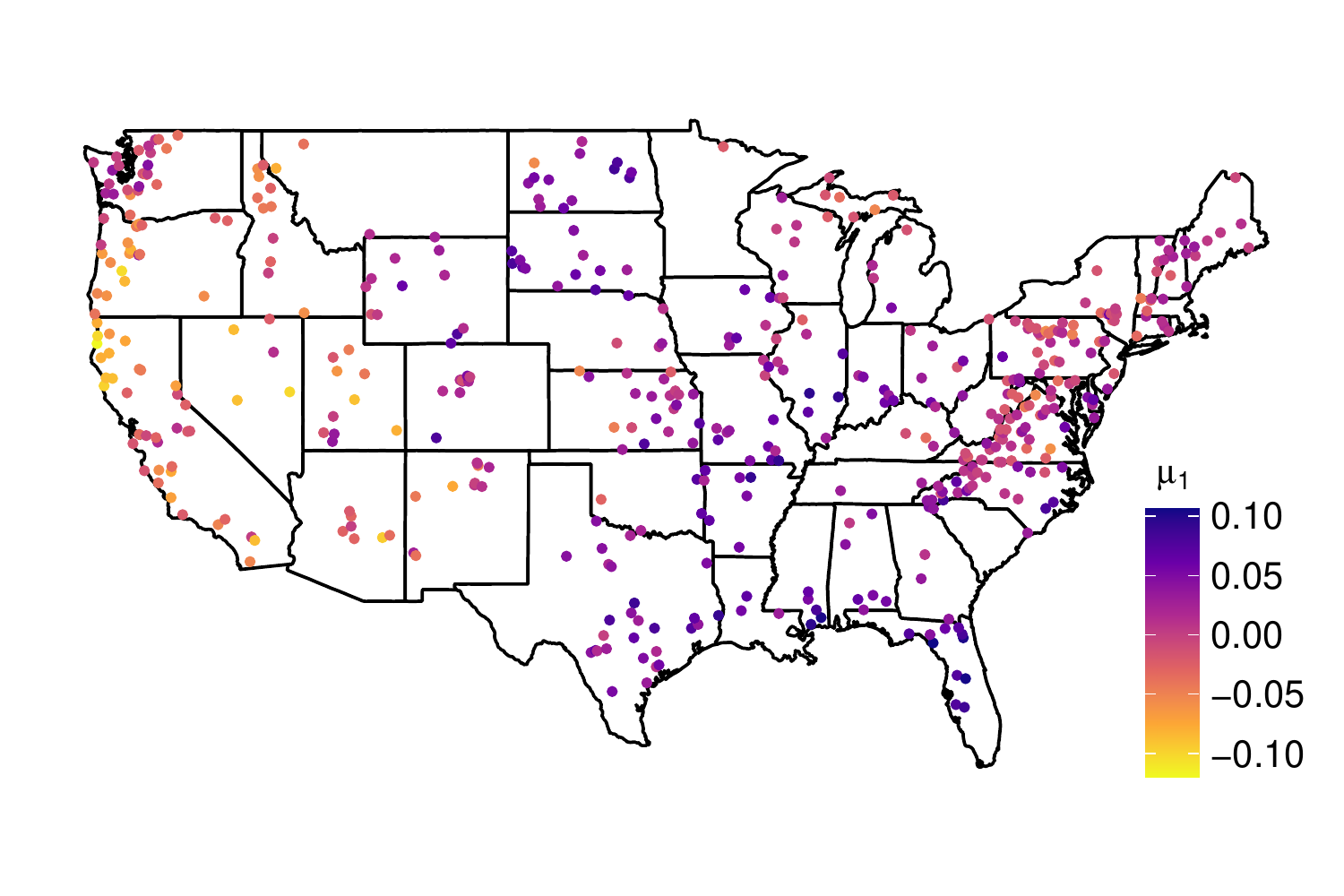}
\caption{Posterior mean of $\mu_1(\bs)$.}
    \label{fig:mu1_estimates}
    \end{subfigure}
    \begin{subfigure}[b]{0.49\linewidth}
        \centering
     \includegraphics[width=0.85\linewidth]{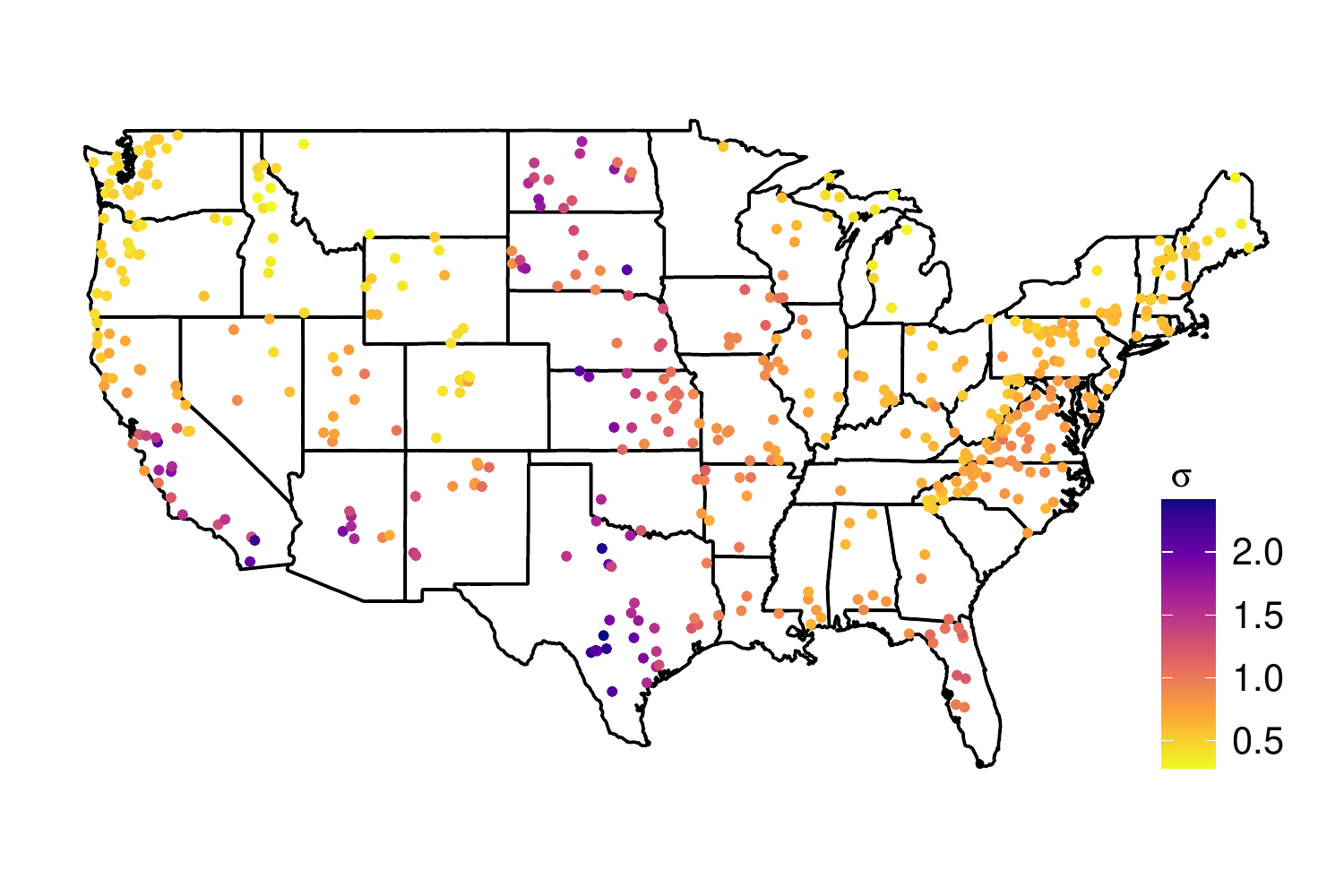}
\caption{Posterior mean of $\sigma(\bs)$.}
    \label{fig:sig_estimates}
    \end{subfigure}
    \begin{subfigure}[b]{0.49\linewidth}
        \centering
        \ \includegraphics[width=0.85\linewidth]{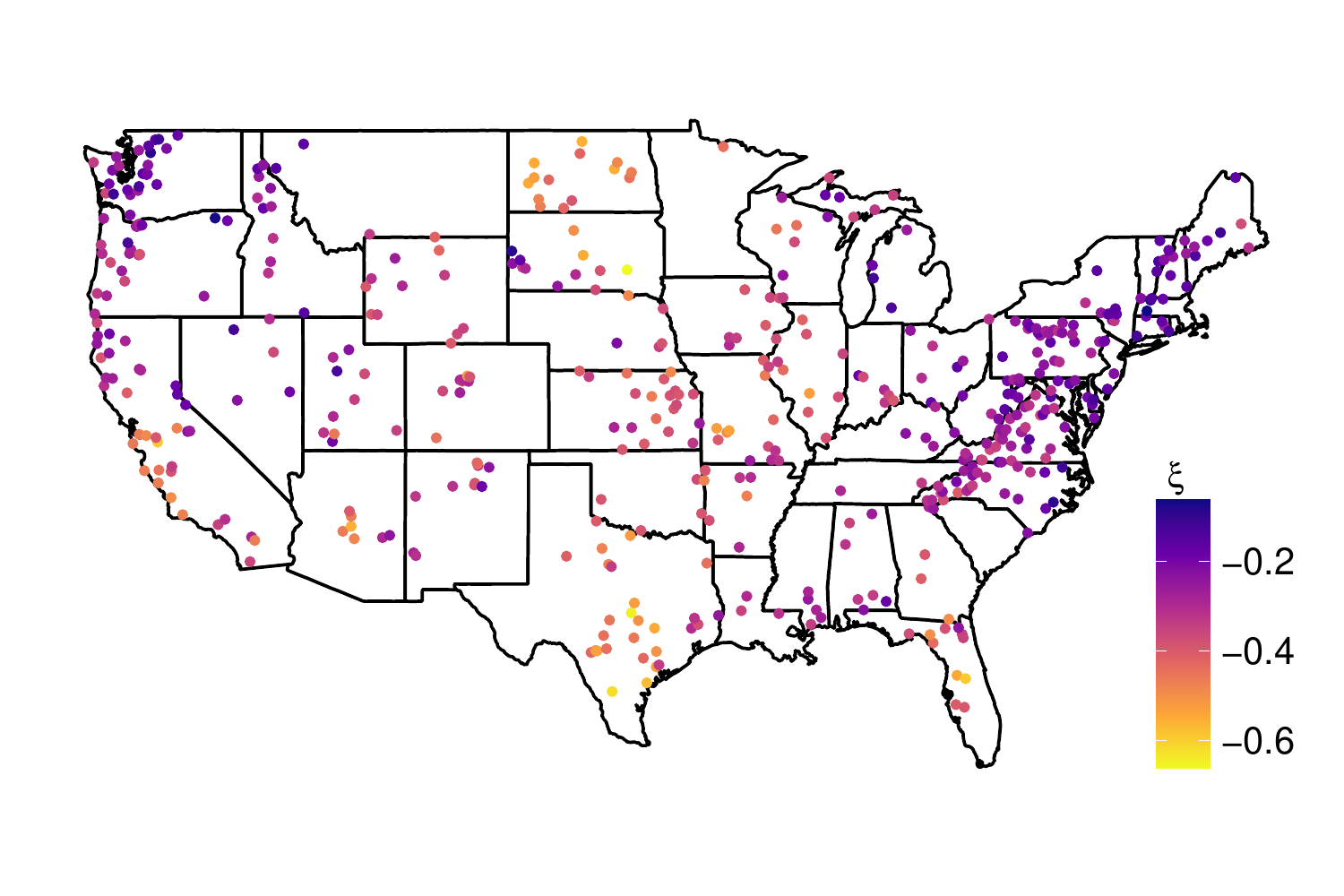}
\caption{Posterior mean of $\xi(\bs)$.}
    \label{fig:xi_estimates}
    \end{subfigure}
        \caption{{\bf HCDN GEV parameter estimates}: Posterior means for 487 stations in the USA based on log-transformed data from 1972--2021.}
    \label{f:cor2}
\end{figure}
\begin{figure}
    \centering
    \includegraphics[width=0.7\linewidth]{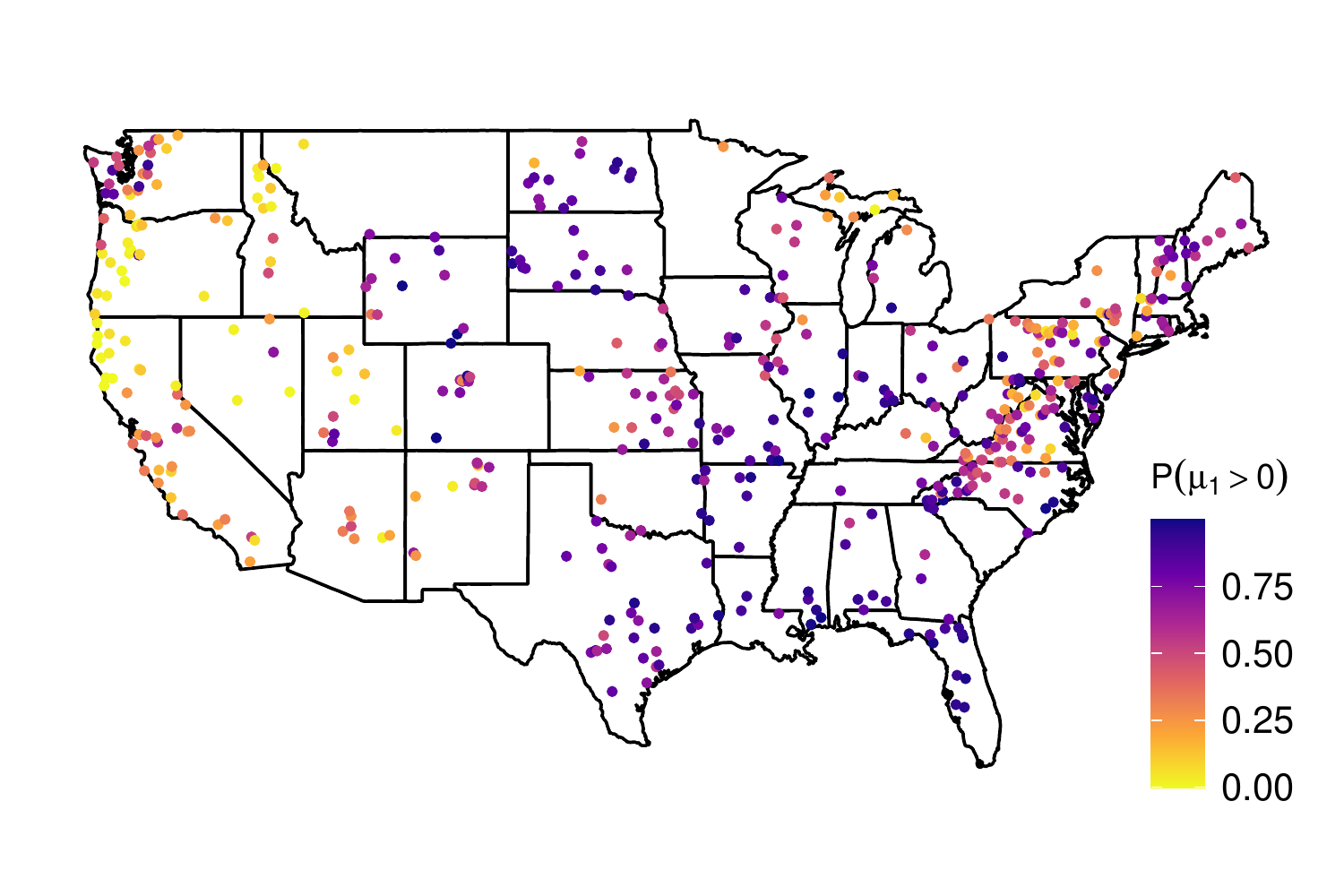}
\caption{Posterior of $Pr[\mu_1(\bs)>0]$ for GEV location parameters based on log-transformed data from 1972--2021.}
    \label{fig:map_mu1_positive}
\end{figure}

Figure \ref{f:cor2} plots the spatial distribution of posterior means for the GEV parameters. The shape and scale parameters are negatively associated for large parts of the country, possibly a consequence of the constraints on the GEV parameters. The scale parameter $\sigma(\bs)$ is highest in the Arkansas-Rio Grande-Texas Gulf and the Missouri basin regions\footnote{\href{https://www.doi.gov/employees/reorg/unified-regional-boundaries}{DOI unified regions}}, and the scale parameter is highest in the North Atlantic-Appalachian and the Columbia-Pacific Northwest regions. Areas with high estimates of $\mu_0(\bs)$ and $\xi(\bs)$ coincide with areas of high precipitation in the 1991--2020 US Climate Normals\footnote{\href{https://www.ncei.noaa.gov/products/land-based-station/us-climate-normals}{US Climate Normals}, \href{https://www.ncei.noaa.gov/access/climateatlas/}{US Climate Atlas}}, the current official baseline for describing average US climate. This suggests that an association between precipitation and streamflow maxima. In Figure \ref{fig:xi_estimates}, we note that the posterior means of the shape parameters $\xi(\bs)$ are negative at all 487 stations on the log transformed data. The use of the log transform for the streamflow data leads to negative GEV shape parameter estimates, imposing a finite upper bound on the distribution even on the original scale. However, we do not expect this to affect estimation of either the quantiles of the marginal distribution or the joint exceedance probabilities. To assess the effect of transforming the data, additional analysis was carried out on PMMs fitted to the original scale of the data, as well as on the square-root of streamflow. Results from both these analyses are provided in the Supplementary Material \citep[][Appendix C.4]{VecchiaDL_supplement}.

Our primary interest, however, are in estimates of the location parameters across the USA. Figure \ref{fig:mu0_estimates} plots the posterior mean of the intercept $\mu_0(\bs)$ of the location parameter, and has a spatial distribution similar to Figure \ref{fig:HCDN_quantiles}. 
 Figure \ref{fig:mu1_estimates} plots the posterior mean of the slope $\mu_1(\bs)$ of the location parameters with respect to time across the USA, and Figure \ref{fig:map_mu1_positive} plots the posterior probability of the slope parameter being positive, $Pr[\mu_1(\bs)>0]$. Positive slope estimates in Figure \ref{fig:mu1_estimates} indicate an increase in extreme streamflow over time, and high probabilities in Figure \ref{fig:map_mu1_positive} indicates stronger evidence for the increase being significant. The majority of the positive slope parameters are concentrated in the Mississippi and Missouri basins, and the Arkansas-Rio Grande-Texas Gulf regions. The North Atlantic-Appalachian region in the east has a large number of catchments with slope estimates near zero, but the majority of zero and negative slope estimates are concentrated around the Lower Colorado Basin, Columbia-Pacific Northwest, and the California-Great Basin regions. An exception is Washington, in the northwest, which has high estimates of the slope. The upper Colorado basin which includes Wyoming, Colorado, and New Mexico is of particular interest since HCDN stations in the region have relatively low 0.9 quantile values in Figure \ref{fig:HCDN_quantiles} as well as low $\mu_0(\bs)$ estimates in Figure~\ref{fig:mu0_estimates}, suggesting that extreme streamflow is starting to have large impacts in this region over time.
\subsubsection{Regional joint exceedance behavior}\label{s:jt-exceedance}
 \begin{figure}
    \centering
    \includegraphics[width=.4\linewidth]{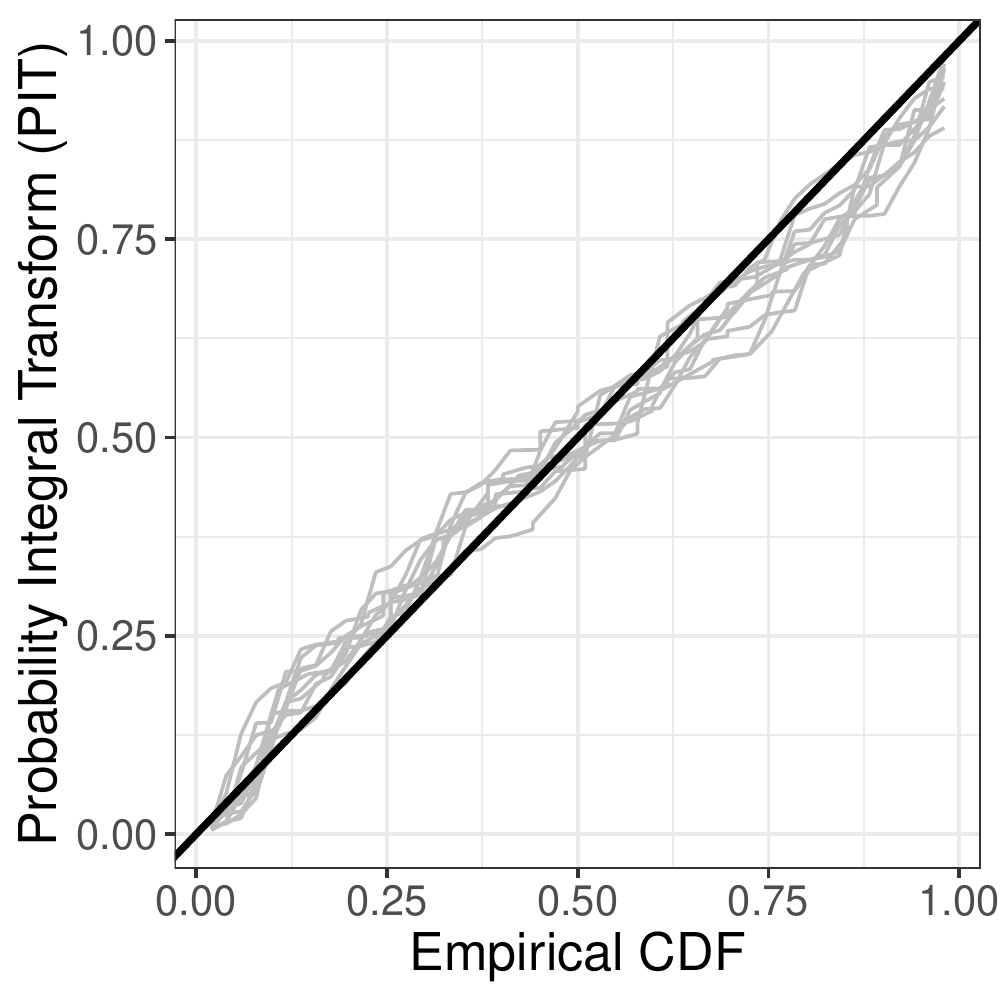}
    \includegraphics[width=.4\linewidth]{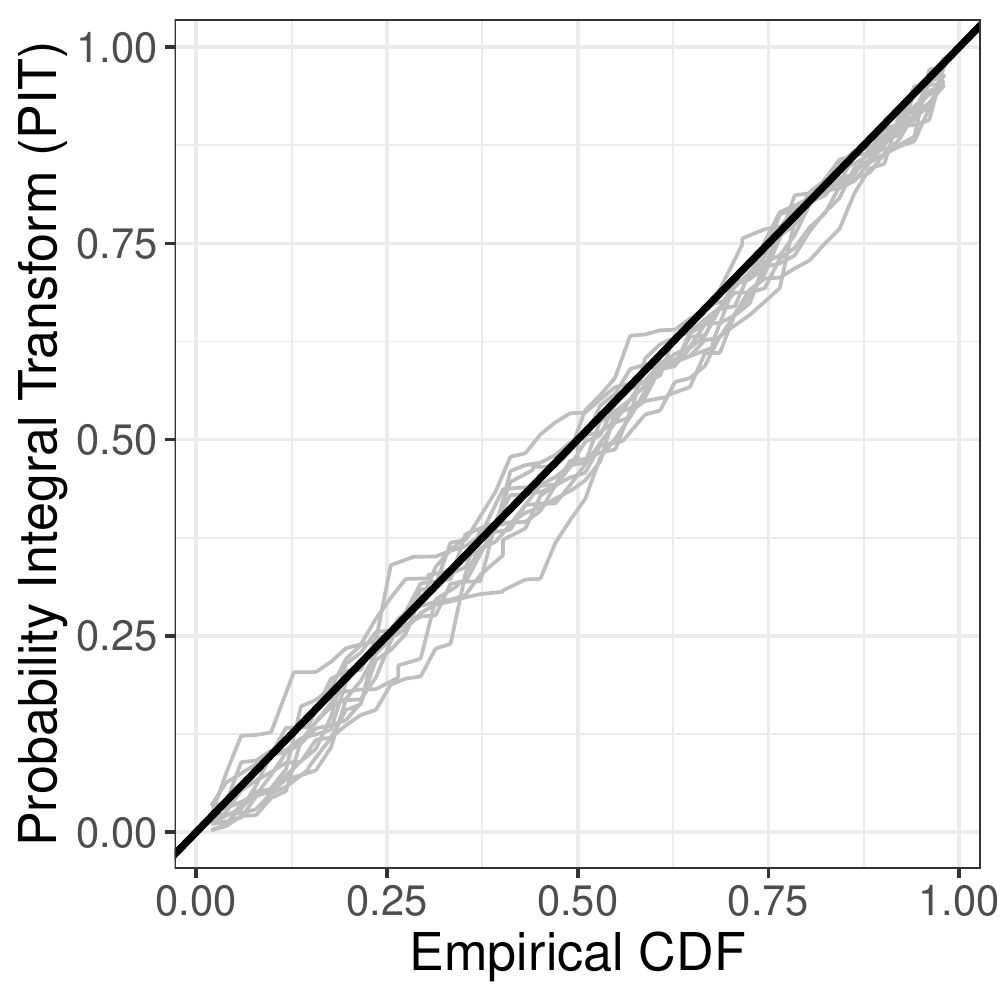}
\caption{\small PIT scores of marginal GEV fits for HCDN locations in Colorado (left) and New Mexico (right).}
    \label{fig:marginal_fits}
\end{figure}

To study the behavior of extreme streamflow jointly for multiple locations, we consider two clusters of HCDN stations in Colorado (CO) and in New Mexico (NM), comprising of 10 and 11 stations, respectively. Figure \ref{fig:marginal_fits} plots the probability integral transform (PIT) scores of marginal GEV fits for locations within each cluster based on posterior means of GEV parameters, which suggest adequate marginal fits for both sets of locations. To quantify the effect of changing climate on extreme streamflow for each cluster, we look at the joint posterior probability of streamflow maxima exceeding the observed 0.90 quantile values as shown in Figure \ref{fig:HCDN_quantiles}, i.e., Pr$[Y_t(\bs_i)>q_i,i=1,\ldots,n_i]$ for the sample 0.90 quantile $q_i$ at location $\bs_i$, where $n_i$ is the number of stations within each cluster. Since our marginal models have STVC, we are able to calculate posterior probabilities for both 1972 and 2021. The probabilities are calculated based on 200 post burn-in MCMC samples from the posterior distribution of the parameters. For each MCMC sample, 20,000 observations are generated from the fitted model; the 1972 and 2021 probabilities are calculated based on 10,000 observations each.

In Colorado, the 10 HCDN stations correspond to catchments with drainage ranging from $15.5\mbox{ km}^2$ to $432.9\mbox{ km}^2$. The cluster is well separated from other stations, and have high posterior estimates of $Pr[\mu_1(\bs)>0]$ (minimum = $0.10$, mean = $0.57$). The HCDN stations are all situated in the Upper Colorado Basin region, and the set of neighbors are spread across the Upper and Lower Colorado Basin regions.  The joint exceedance probability for 1972 has a mean of 0.075 with an SD of 0.04. The joint exceedance probability for 2021 has a mean of 0.17 with an SD of 0.046. This corresponds to an increase of around $125\%$ of the mean joint exceedance probability in the last 50 years. Note that if we assume independence across locations, the joint exceedance probability for the 10 locations would be approximately $10^{-10}$. Finally, the probability that the joint exceedance in 2021 is higher than in 1972 is 0.90, providing strong evidence in favor of increased extremal streamflow in the area, possibly due to changing climate.

In New Mexico, the 11 HCDN stations correspond to catchments with drainage ranging from $43.8 \mbox{ km}^2$ to $4804.9 \mbox{ km}^2$. The stations are all situated in the Upper Colorado Basin region, and the set of Vecchia neighbors are located across the Upper and Lower Colorado Basin as well as the California-Great Basin region. The catchments have a mix of high and low posterior estimates of $Pr[\mu_1(\bs)>0]$ (minimum = 0.07, mean = 0.48). The joint exceedance probability for 1972 has a mean of 0.045 with an SD of 0.012. The joint exceedance probability for 2021 has a mean of 0.053 with an SD of 0.017. This corresponds to an increase of around $18\%$ of the mean joint exceedance probability in the last 50 years. The probability that the the joint exceedance in 2021 is higher than in 1972 is 0.695, which is lower than the result for Colorado, but still significantly higher than the independent scenario. \citet[][Appendix C.2]{VecchiaDL_supplement} contains additional results for both clusters.

\subsubsection{Model comparison and model fit}
\begin{table}
\centering
\caption{Estimates and standard errors (in parentheses) from leave-one-out cross validation (LOO-CV) and the Watanabe-Akaike information criterion (WAIC) for comparing the process mixture model (PMM), the Huser-Wadsworth model (HW), the max-stable process model (MSP), and the Gaussian process model (GP). Lower values indicate a better fit.}
\label{t:waic}
\begin{tabular}{ccccc}
\toprule
 & \textbf{PMM} & \textbf{HW} & \textbf{MSP} & \textbf{GP} \\\midrule
\textbf{LOO-CV} & 29108 (540) & 29708 (544) & 32058 (583) & 33842 (561) \\
\textbf{WAIC} & 29559 (549) & 30193 (565) & 33441 (552) & 34440 (585)\\\bottomrule
\end{tabular}
\end{table}

We compared the PMM with three spatial processes - a GP, an MSP, and the Huser-Wadsworth (HW) process. In each case, inference is carried out by using SPQR for density estimation, and using MCMC afterwards for parameter estimation. This allows for a comparison of the appropriateness of different spatial processes for our application. For all three competing models, we used the same neural network architecture in our local SPQR fits as the PMM.

Table \ref{t:waic} lists estimates and standard errors from leave-one-out cross validation (LOO-CV) \citep{Vehtari_2016} and the Watanabe-Akaike information criterion (WAIC) \citep{waic} for the four models. Based on both metrics, the PMM has the lowest values and thus the best model fit, followed by the HW model. The MSP and the GP are noticeably worse than the PMM and HW models, but their LOO-CV and WAIC estimates are relatively close to each other. This suggests that models which assume asymptotic (in)dependence (e.g., the MSP and the GP) are likely to have worse fits when the dependence structure is complex. An inter-comparison of the posterior means of the slope parameter from each of these models is presented in the Supplementary Material \citep[][Appendix C.3]{VecchiaDL_supplement}.

\begin{figure}
    \centering
    \includegraphics[width=0.8\linewidth]{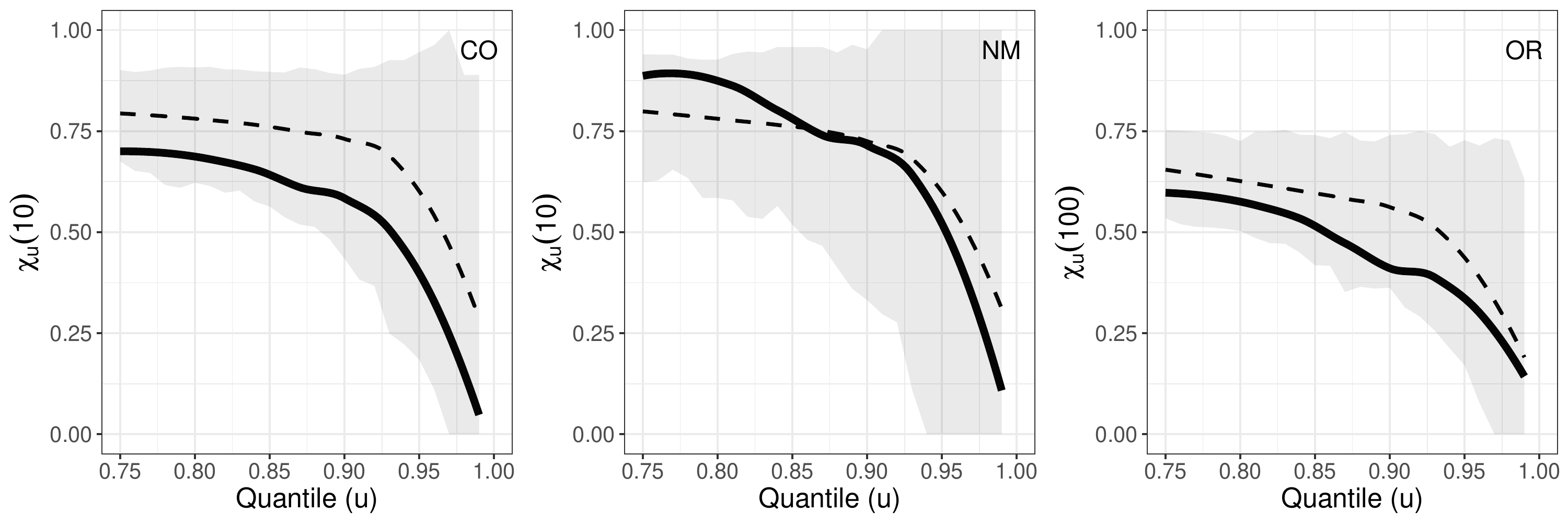}
    \includegraphics[width=0.8\linewidth]{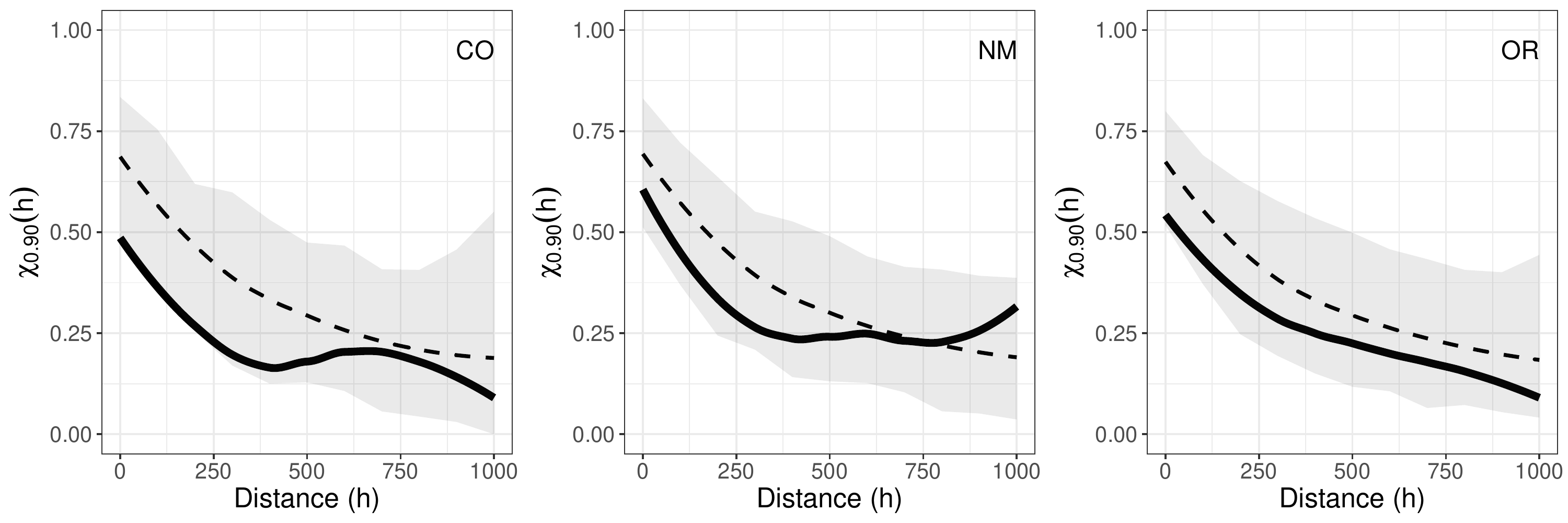}
    \caption{{\bf Estimates of $\chi_u(h)$ for three different regions}: Empirical estimates of the upper tail coefficient in bold, compared against estimates based on the posterior distribution. The dashed lines and the bands correspond to the mean and $95\%$ interval, respectively.}
    \label{f:chi_estimates}
\end{figure}

Finally, we evaluate the fit when using the PMM as the underlying spatial process by comparing estimates of $\chi_u(h)$ based on the posterior distribution with empirical estimates obtained from the rank-standardized observed data. For this, we choose 3 regions within the USA, the first two of which are the CO and NM clusters studied in Section \ref{s:jt-exceedance}. Both clusters have a total of 34 locations (including the Vecchia neighbor locations). The third area consists of HCDN locations in Oregon (OR) and with its Vecchia neighbors, accounting for a total of 56 locations in Oregon, Washington, Nevada, and California. This region provides a contrast to the CO and NM clusters; locations are farther apart, and have low estimates of the slope and scale but high estimates of the shape parameter. For each of the 3 clusters, we generated realizations of the spatial process based 200 samples from the posterior, and estimated $\chi_u(h)$ for each sample.

Figure \ref{f:chi_estimates} plots the empirical and posterior estimates of $\chi_u(h)$ for the three regions. The top row plots the upper tail coefficient for high quantiles, for locations $h=10 \mbox{ km}$ apart for CO and NM, and $h=100 \mbox{ km}$ apart for OR. The bottom row plots estimates of $\chi_u(h)$ for $u=0.90$. In each panel, the bold line represents the empirical estimate, and the dashed line and band represent the mean and $95\%$ interval of the posterior estimates respectively. We note that both the model-based and empirical estimates of the upper tail coefficient behaves similarly for the three regions, and are consistent with the empirical estimate that is obtained from data for the entire USA (Figure \ref{fig:chi_h}). This suggests that the stationarity assumption for the dependence structure in our model is appropriate. The $95\%$ intervals in the top row includes 0 for all 3 regions, reflecting the asymptotic independence corresponding to our estimate of $\delta$. Both the empirical and posterior estimates show behavior similar to Figure \ref{fig:empericalchi_a} for the asymptotic independent case. The tail coefficients can also been seen to decrease as distance increases, similar to Figure \ref{fig:empericalchi_b}. We noted similar behavior for other regions within the country. While the fitted model tends to overestimate $\chi_u(h)$ for several regions, it is able to capture the behavior of extremal dependence over large distances, and at high quantile levels.

\section{Discussion}\label{s:discussion}
In this paper, we proposed a process mixture model (PMM) for spatial extremes, where the marginal distributions at different spatial locations are GEV and their spatial dependence is captured using a convex combination of a GP and an MSP. The PMM extends \citet{Huser-Wadsworth}, and is flexible enough to accommodate missingness and censoring, as well as STVC for the marginal GEV distributions. We approximated the joint likelihood for the spatial model using a Vecchia approximation. We used the density regression model proposed in \citet{xu-reich-2021-biometrics} to approximate this likelihood, whose weights are modeled using a feed-forward neural network and learned using synthetic data generated from a design distribution. Parameter estimation for the model is carried out using MCMC.

We used the PMM to analyze changes in annual maximum streamflow within the US over the past 50 years. For this study, we used the annual maximum streamflow measured at 487 stations in the USGS Hydro-Climatic Data Network. The posterior means of the location parameter have non-zero slope estimates in several parts of the country. We noted a high concentration of positive slope estimates in the Mississipi and Missouri basins, and the Arkansas-Rio Grande-Texas Gulf regions, indicating that extremal streamflow has increased in those areas over the last 50 years.

Future work will focus on the theoretical properties of this model. While it is straightforward to derive an analytical expression for $\chi_u(h)$ for the trivial case of a shared MSP and has been provided in the Supplementary Material \citep[][Appendix A.4]{VecchiaDL_supplement}, i.e., for $R(\bs) = R$, obtaining an analytical expression for the general case is more challenging. This would also enable us to investigate the properties of $\chi_u(h)$ for the PMM for different values of $\delta$ and as $h \to \infty$. We would like to further investigate improvements to the computational aspects of this model and identify reasonable plug-and-play settings for local and global SPQR approximations. Finally, we would also like to extend this model to provide climate-informed estimates by regressing the spatiotemporal variability of the EVA parameters onto large-scale climate drivers from GCM output with spatially-varying regression coefficients for local calibration.  Another area of future work is to extend the model to accommodate more complex spatial dependence structures. Recent work on spatial extremes has incorporated graphical models as additional information for computing distances \citep{EngelkeHitz2020}. While our work considers only spatial coordinates, it is possible to incorporate additional distance measures in the form of river network information to reflect the physical structure of watersheds and the  so-called `river distance' between stations \citep{asadi2015extremes}. While the stream network information is not readily available for these data, incorporating this network structure might improve spatial modeling. Spatial models on stream networks have been developed for both max-stable \citep{asadi2015extremes} and Gaussian  \citep{santos2022bayesian} processes, and so it should be possible to incorporate these features into the PMM.

\section*{Acknowledgements}
This work was supported by grants from the Southeast National Synthesis Wildfire and the United States Geological Survey's National Climate Adaptation Science Center (G21AC10045), the National Science Foundation (CBET2151651, DMS2152887, and DMS2001433) and the National Institutes of Health (R01ES031651-01). The authors thank Prof. Sankarasubramanian Arumugam of North Carolina State University for discussion of the data and scope of the project.

\begin{singlespace}
	\bibliographystyle{rss}
	\bibliography{VecchiaDL}
\end{singlespace}
%\clearpage
\end{document}

\begin{center}
{\Large  Supplement to ``Modeling Extremal Streamflow using Deep Learning Approximations and a Flexible Spatial Process''}\\\vspace{6pt}
{\large Reetam Majumder\footnote[1]{North Carolina State University}, Brian J. Reich$^1$ and Benjamin A. Shaby\footnote[2]{Colorado State University}}\\
\today
\end{center}

\begin{appendix}
\section{Background and Properties of the PMM}\label{appA}
\singlespacing
\subsection{Connection to the main text}
This appendix supports the material in Section 3 of the main text. Appendix \ref{s:VI} provides an overview of the variable importance measure used in assessing the SPQR fits.
\subsection{Connection of the PMM to the Huser-Wadsworth model}\label{appA:HW}
We present an overview of the construction and interpretation of the Huser-Wadsworth (HW)  model \citep{Huser-Wadsworth} for spatial extremes, and how it can be generalized to develop the PMM.
Let $\{\tilde{W}(\bs):s\in\mathcal{S}\subset \mathbb{R}^2\}$ be a stationary spatial process with standard Pareto margins, and which has asymptotic independence with hidden regular variation. Further, let $\Tilde{R}$ be an independent standard Pareto random variable. \citet{Huser-Wadsworth} specify a spatial dependence model through the random field constructed as
\begin{equation}\label{e:HW1}
    \Tilde{X}(\bs) = \Tilde{R}^{\delta}\Tilde{W}(\bs)^{1-\delta}, \delta \in [0,1].
\end{equation}
Examples of $\Tilde{W}(\bs)$ include marginally transformed Gaussian processes, and inverted max-stable processes. \eqref{e:HW1} is used as a copula to model the extremal spatial dependence between locations. However, as the copula transformation is invariant to monotonically increasing marginal transformations, an alternative formulation with the same dependence structure is given by
\begin{equation}\label{e:HW2}
    X(\bs) := \delta R + (1-\delta)W(\bs),
\end{equation}
where $R =\log \Tilde{R} \sim \mbox{Exp}(1)$, independent of $W(\bs)\sim \log \Tilde{W}(\bs) \sim \mbox{ Exp}(1)$. The form of \eqref{e:HW2} implies that $X(\bs)$ can be interpreted as an interpolation of perfect dependence and asymptotic independence. 

The PMM can be constructed by replacing $\Tilde{R}$ in \eqref{e:HW1} by $\Tilde{R}(\bs)$, a max-stable process with asymptotic dependence, and specifying $\Tilde{W}(\bs)$ as a Gaussian process. The corresponding transformations to get the forms similar to \eqref{e:HW2} with exponential margins are provided in \ref{sec:margTransforms}.
Since we are interested in block maxima, the PMM has GEV margins. The spatial process can be interpreted as an interpolation of asymptotic dependence and asymptotic independence.
\subsection{Marginal transformations for the components of the PMM}\label{sec:margTransforms}
Let $\tilde{R}(\bs)$ be a max-stable process and $\tilde{W}(\bs)$ be a Gaussian process. Without loss of generality, we assume $\tilde{R}(\bs)$ has GEV(1,1,1) marginal distributions and $\tilde{W}(\bs)$ has standard normal marginal distributions. Consider the transformations 
\begin{align*}
 g_R(r) &=  -\log\{1-\exp(-1/r)\},\\   
 g_W(w) &= -\log\{1-\Phi(w)\},
\end{align*}
where $\Phi(w)$ is the standard Normal CDF. Then $R(\bs) = g_R(\tilde{R}(\bs))$ and $W(\bs) = g_R(\tilde{W}(\bs))$ have standard exponential margins. 

\subsection{Derivation of Conditional Exceedance for a common spatial process}
Denote $\vartheta(\bs_1,\bs_2)\in[1,2]$ as the extremal coefficient of the MSP so that for all $r>0$
$$
\mbox{Pr}\left\{\tilde{R}(\bs_1)<r,\tilde{R}(\bs_2)<r\right\} = \mbox{Pr}\left\{\tilde{R}(\bs_1)<r\right\}^{\vartheta(\bf s_1,s_2)}=\mbox{Pr}\left\{\tilde{R}(\bs_2)<r\right\}^{\vartheta(\bf s_1,s_2)},$$
where $\vartheta(\bs_1,\bs_2)$ is the extremal coefficient function \citep{SchlatherTawn2003}.
Therefore, small $\vartheta(\bs_1,\bs_2)$ indicates a strong dependence with $\vartheta(\bs_1,\bs_2)=1$ corresponding to complete dependence and $\vartheta(\bs_1,\bs_2)=2$ corresponding to independence.  
Extremal spatial dependence of the process between locations $\bs_1$ and $\bs_2$ is often measured in terms of the upper-tail coefficient \citep{Joe}, defined as the following conditional exceedance probability:
\begin{equation}\label{e:chi_defn}
    \chi_u(\bs_1,\bs_2) := \mbox{Prob}\{U(\bs_1) > u | U(\bs_2) > u\},
\end{equation}
where $u\in(0,1)$ is a threshold. The random variables $U(\bs_1)$ and $U(\bs_2)$ are defined as asymptotically dependent if the limit
\begin{equation}
    \chi(\bs_1,\bs_2) = \lim_{u\rightarrow 1}\chi_u(\bs_1,\bs_2)
\end{equation}
is positive and independent if $\chi(\bs_1,\bs_2) = 0$.  Since $R(\bs)$ and $W(\bs)$ are assumed to be isotropic processes, we can rewrite $\chi_u(h)$ and $\chi(h)$ as functions of the distance between locations.

Since the GP is asymptotically independent, for simplicity, we assume that $\tilde{W}(\bs)$ is independent at locations $(\bs_1,\bs_2)$, so that $$W(\bs_1),W(\bs_2)\iid\mbox{Exp}(1).$$
With these assumptions, we first find the joint survival probabilities of the variables $(Y_1,Y_2)^T \equiv \bigl(Y(\bs_1),Y(\bs_2)\bigr)^T$ arising from the process $Y(\bs)$ to derive the dependence measure $\chi_u(\bs_1,\bs_2)$. Note that $U_1 = F(Y_1)$ and $U_2 = F(Y_2)$ in \eqref{e:chi_defn} can be written as $G(V_1)$ and $G(V_2)$ using the definition of the PMM. In the simple case of a shared spatial process across all locations, i.e. $R(\bs_k) = R(\bs_l) = R(\bs)$, the joint survival probability is:
\begin{align*}
    & Pr[Y_1>y, Y_2>y] = \exp\bigl \{\frac{-2y}{1-\delta} \bigr \}\biggl[\frac{1-\delta}{3\delta - 1}\biggl(\exp\bigl \{\frac{3\delta-1}{\delta(1-\delta)}y\bigr \} - 1\biggr)\biggr] + \exp\bigl\{\frac{-y}{\delta}\bigr \},
\end{align*}
and the marginal corresponds to the survival function of the hypoexponential distribution with the CDF:
\begin{equation}
    \label{eq:hypo}
    Pr(Y_1>y) = \frac{1-\delta}{1-2\delta} \exp\{-\frac{y}{(1-\delta)}\} - \frac{\delta}{1-2\delta} \exp\{-\frac{y}{\delta}\}.
\end{equation}

This gives us
\begin{align*}
    \chi(\bs_1,\bs_2) &= \lim_{u \to 1} \chi_u(\bs_1,\bs_2)\\
    &= \lim_{y \to \infty}Pr[Y_1>y, Y_2>y]/Pr[Y_1>y]\\
    &= \begin{cases}
    0,\mbox{ for } \delta < 0.5,\\
    \frac{2(2\delta - 1)}{3\delta - 1},\mbox{ for } \delta>0.5.
    \end{cases}
\end{align*}

\subsection{Behavior of the conditional exceedance for the PMM}
\begin{figure}
    \centering
    \begin{subfigure}[b]{0.45\linewidth}
    \includegraphics[width=\linewidth]{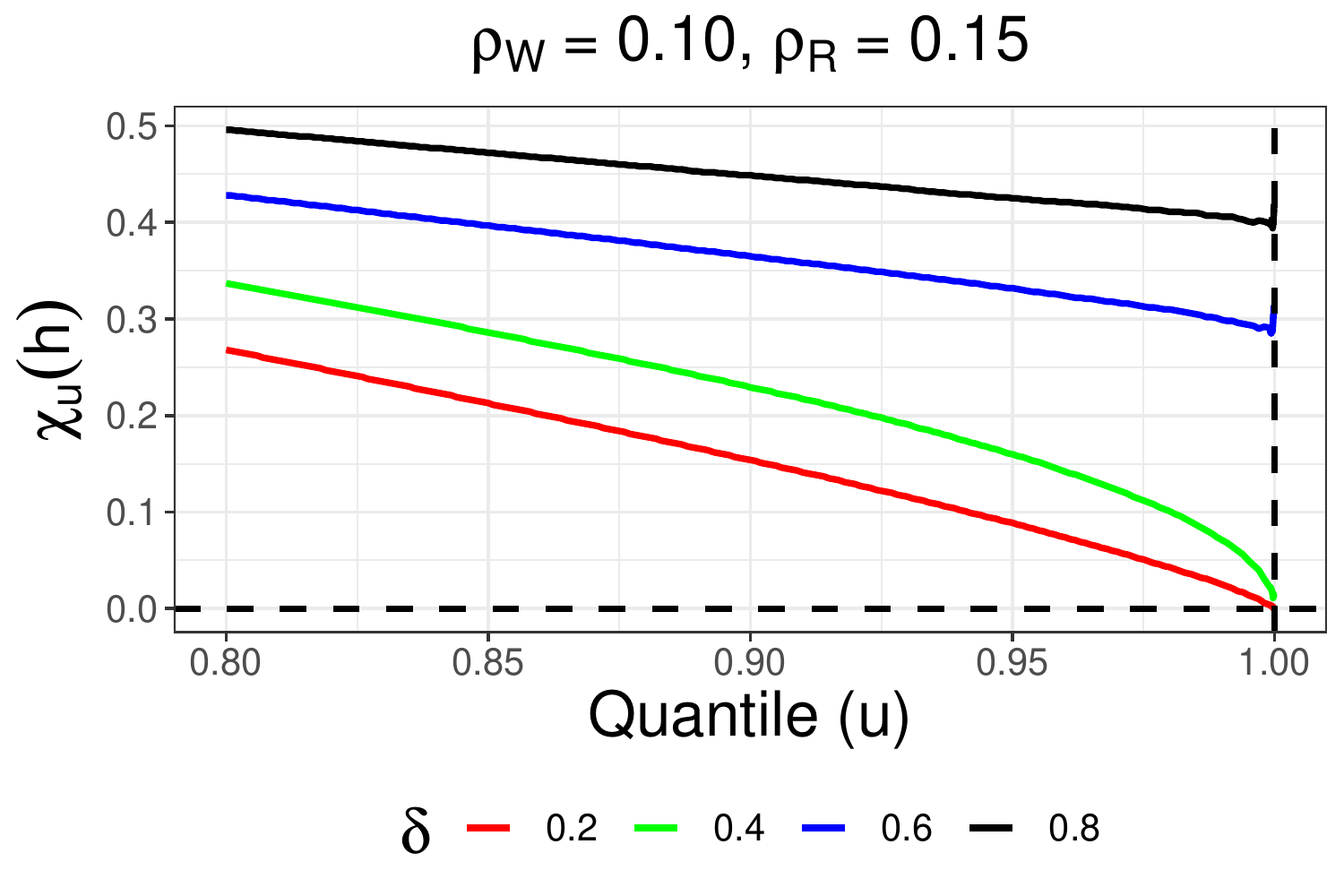}
    \includegraphics[width=\linewidth]{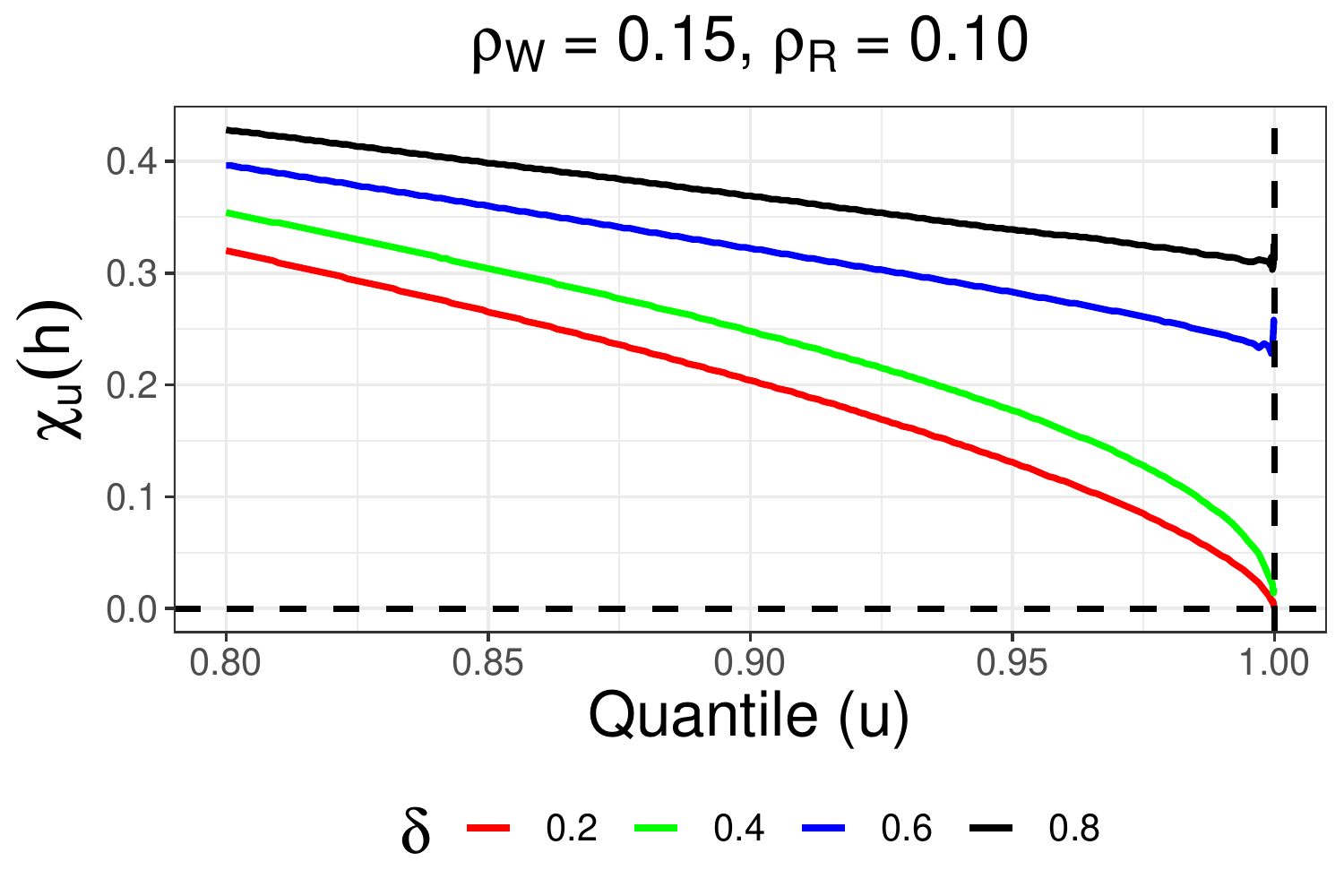}
    \includegraphics[width=\linewidth]{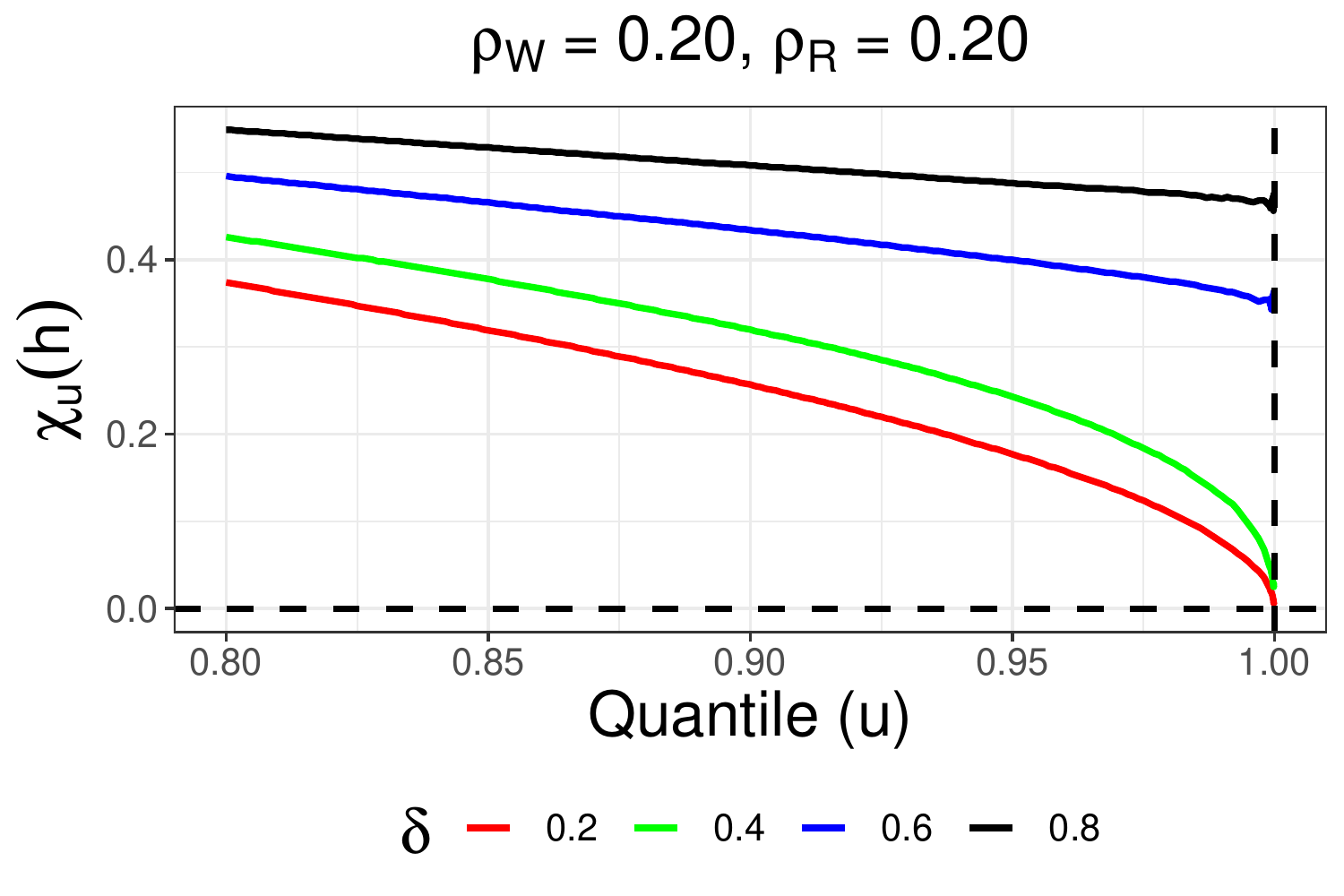}
    \includegraphics[width=\linewidth]{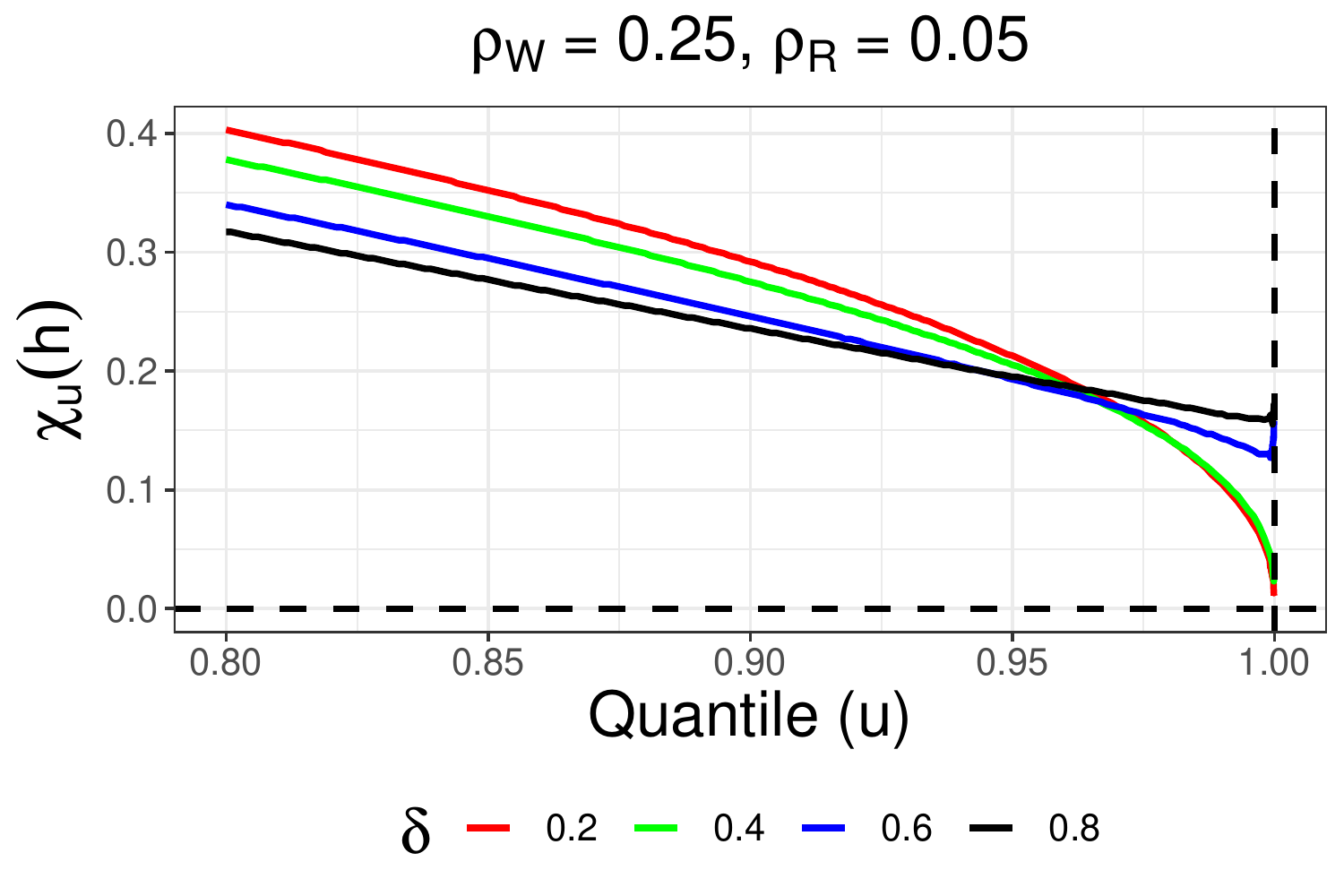}
    \includegraphics[width=\linewidth]{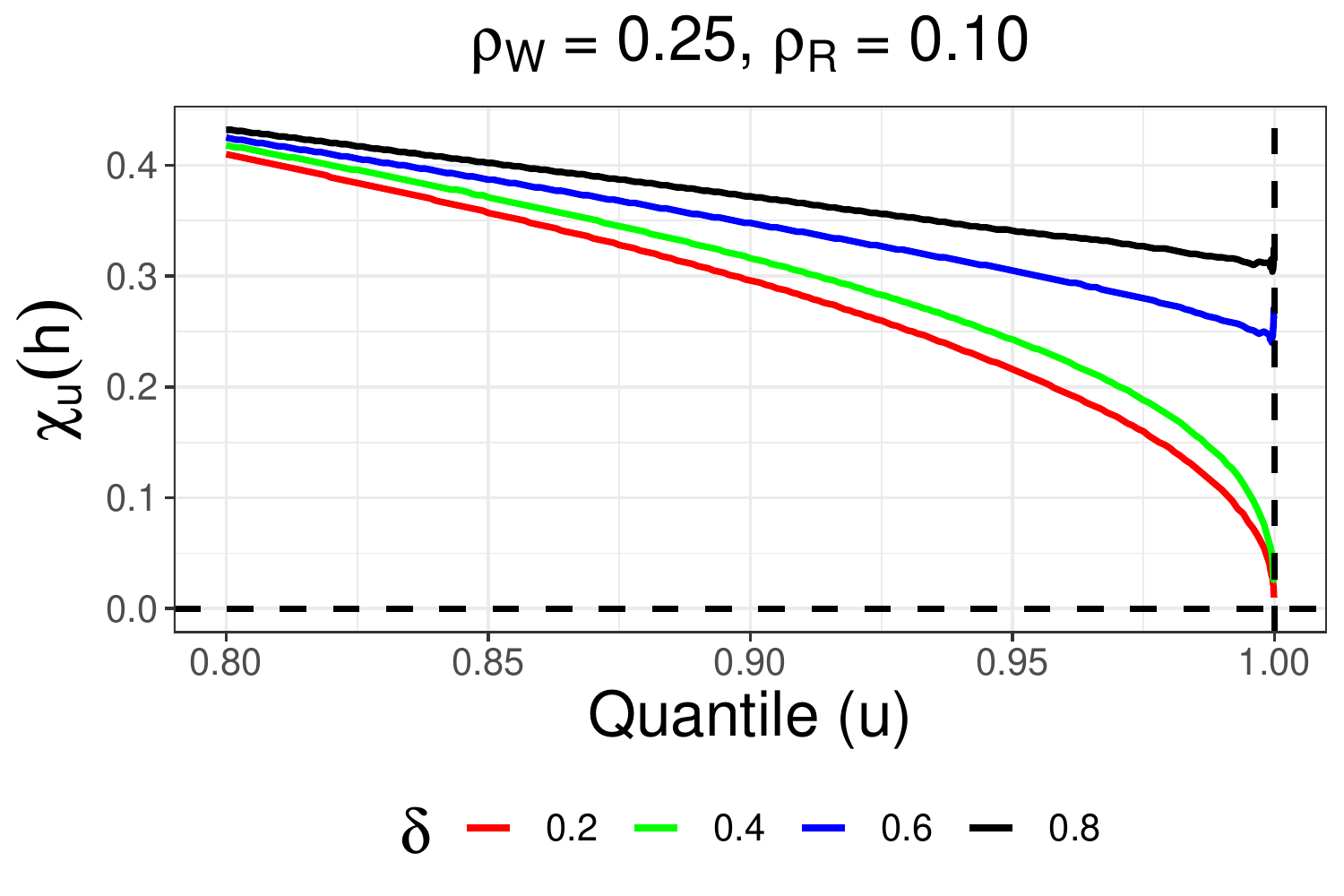}
\caption{$\chi_u(h)$ as a function of $u$ and $\delta$, at distance $h = 0.22$.}
    \label{fig:empericalchi_a}
    \end{subfigure}
    \hfill
    \begin{subfigure}[b]{0.45\linewidth}
    \includegraphics[width=\linewidth]{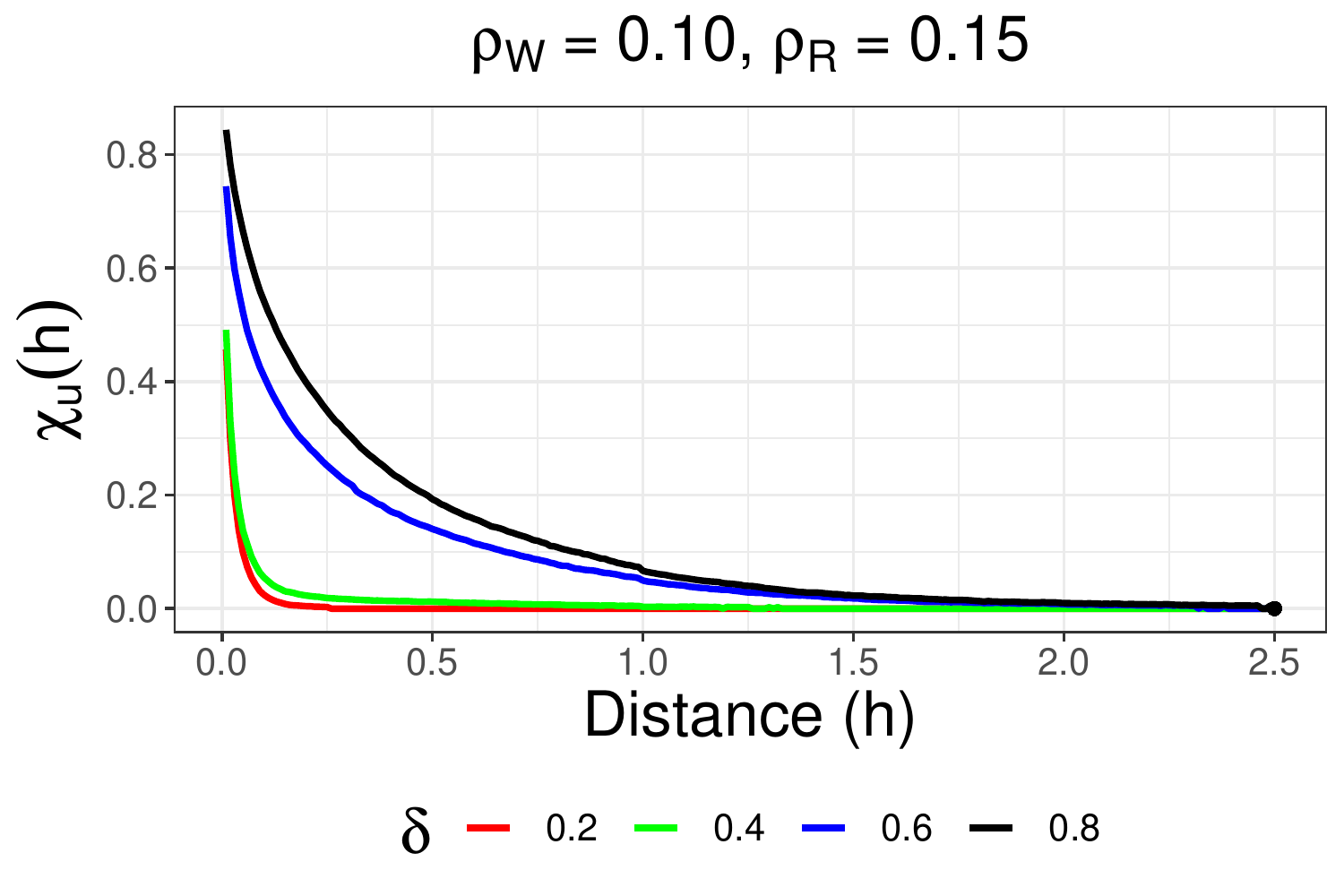}
    \includegraphics[width=\linewidth]{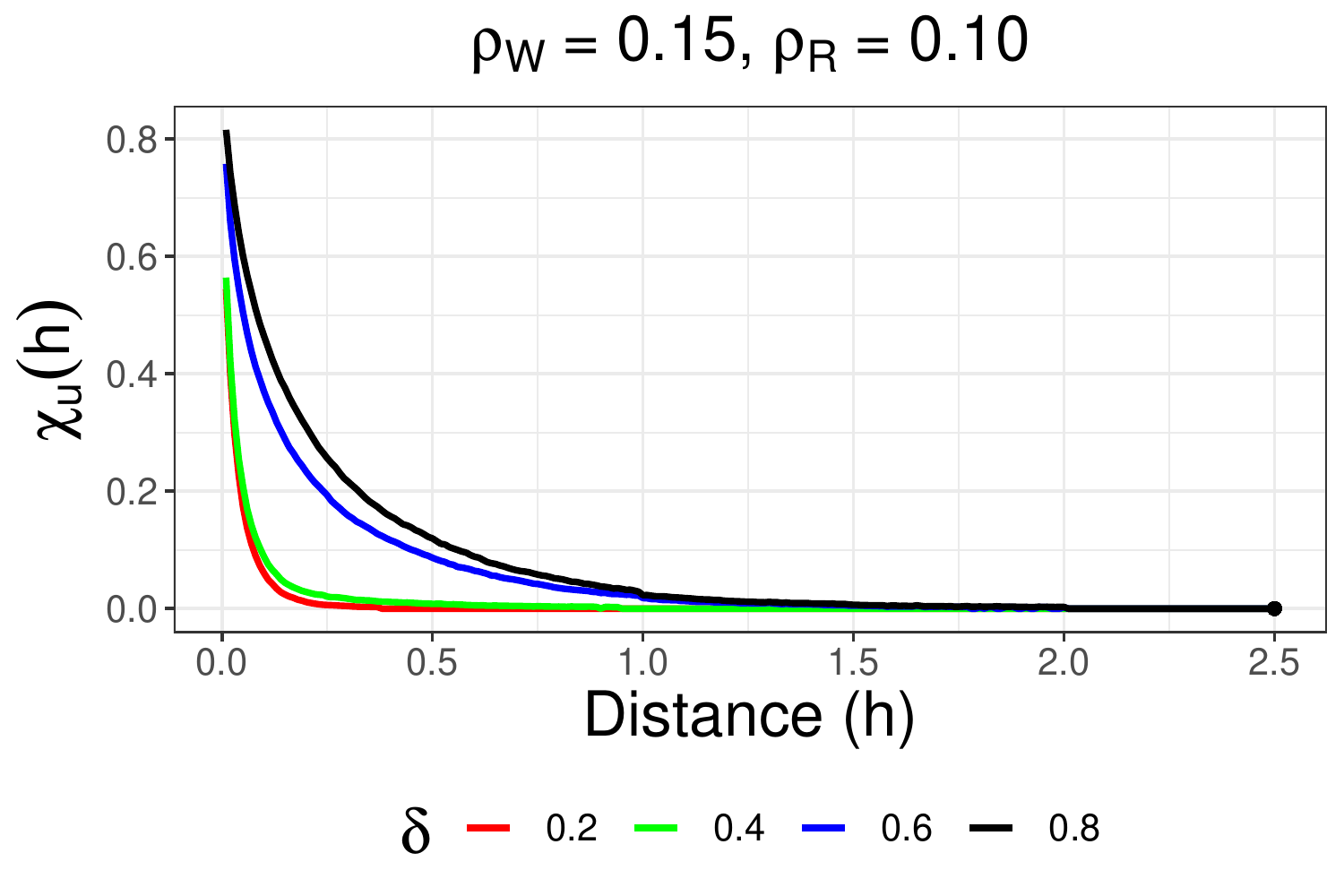}
    \includegraphics[width=\linewidth]{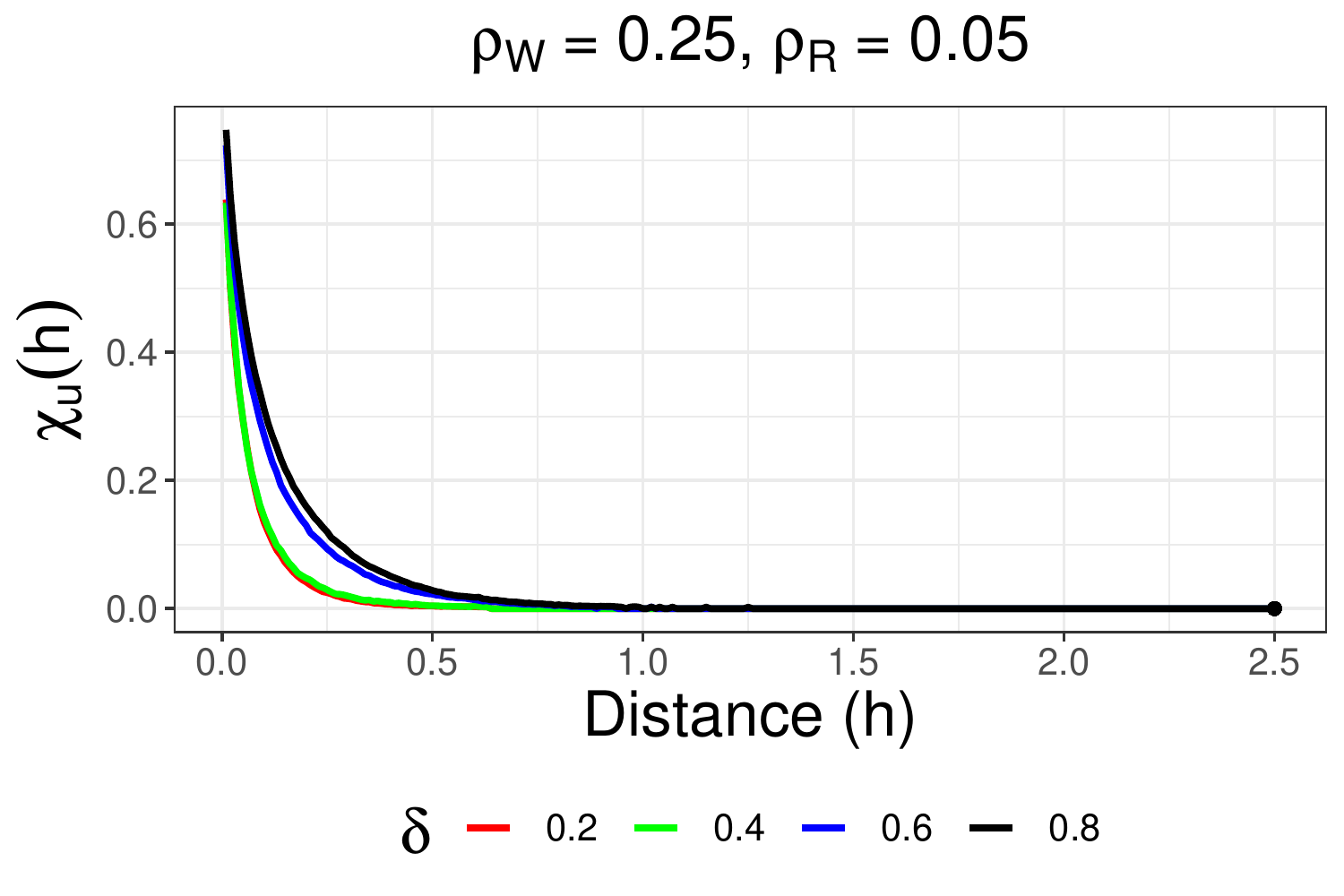}
    \includegraphics[width=\linewidth]{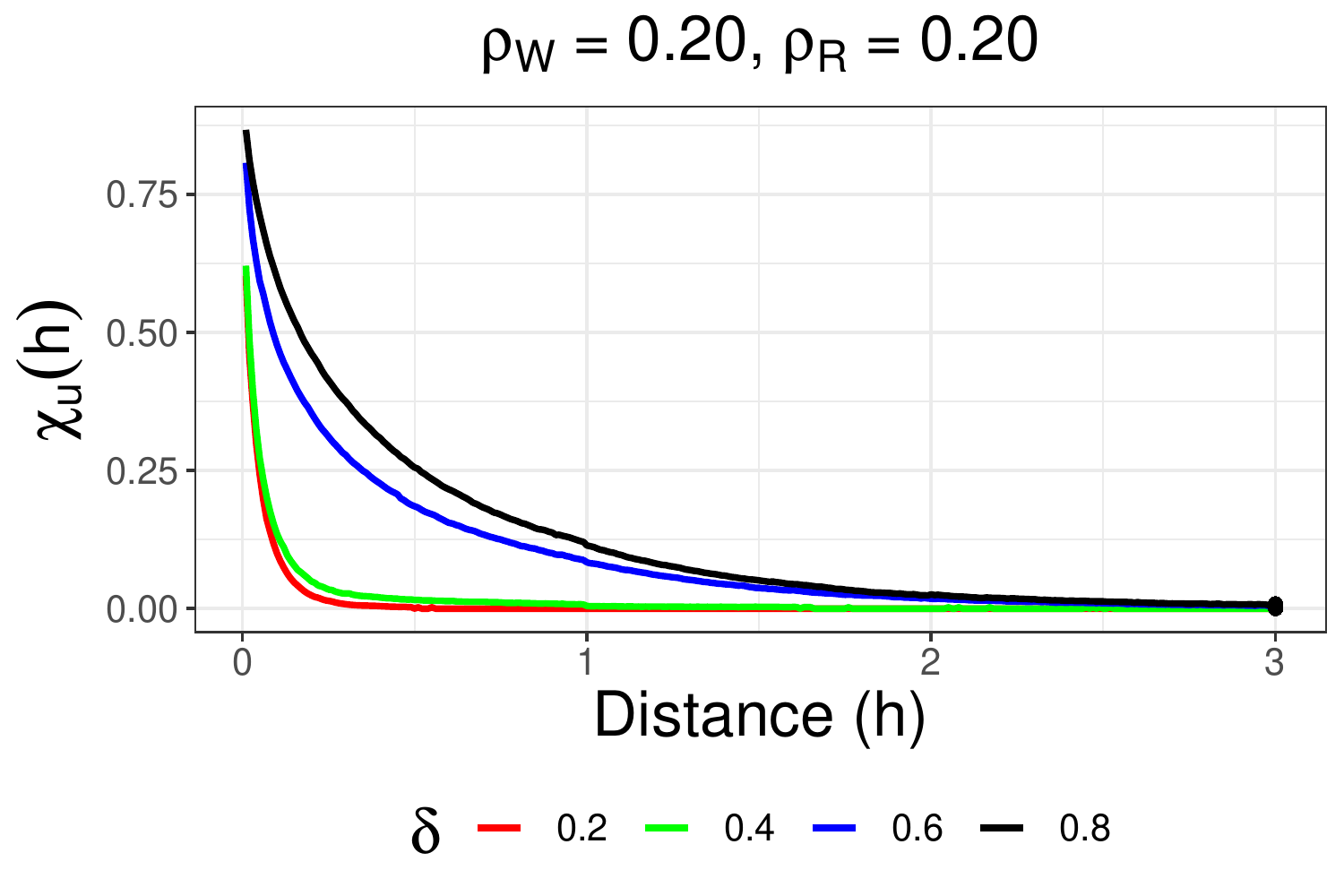}
    \includegraphics[width=\linewidth]{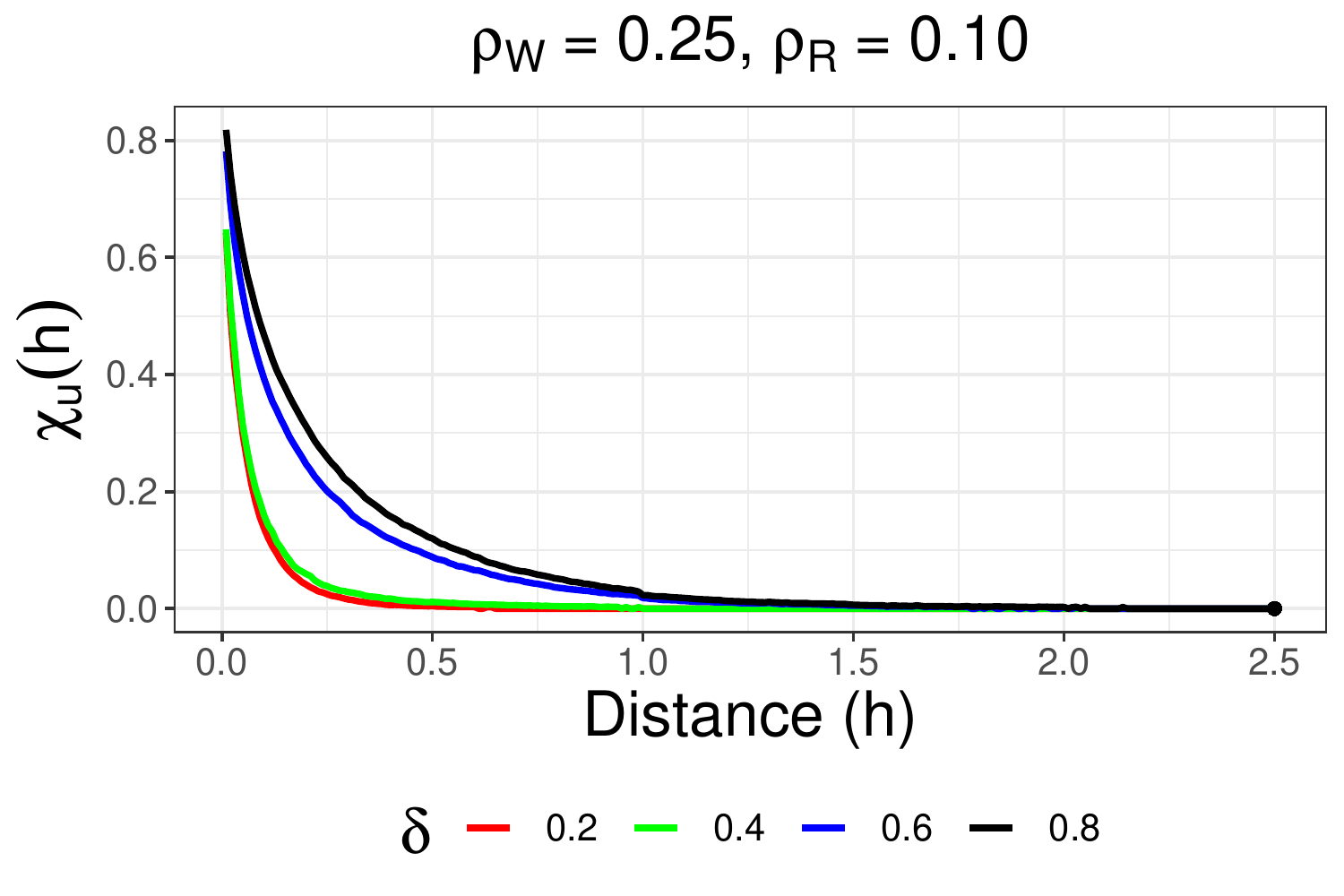}
\caption{$\chi_u(h)$ as a function of $h$ and $\delta$, for threshold $u = 0.999$.}
    \label{fig:empericalchi_b}
    \end{subfigure}

    \caption{{\bf Behavior of the empirical conditional exceedance:} Approximate $\chi_u(h)$ for the PMM plotted as a function of threshold $u$, distance $h$, asymptotic dependence parameter $\delta$, and GP range $\rho_W $, and MSP range $\rho_R$. Smoothness parameters $\alpha_W = \alpha_R = 1$ is fixed for all plots.}
    \label{fig:empericalchi}
\end{figure}

To understand the behavior of $\chi_u(h)$ for different values of $\rho_R$ and $\rho_W$, we computed $\chi_u(h)$ for different values of $\rho_W$ between 0.10 and 0.25, and $\rho_R$ between 0.05 and 0.20. The case $\rho_W = 0.20, \rho_R = 0.10$ is presented in the main text.

Figure \ref{fig:empericalchi} plots Monte Carlo approximations of $\chi_u(h)$ for the PMM as functions of $u$, $h$, $\rho_W$, and $\rho_R$. As in the main text, we fix $\alpha_R = \alpha_W = 1$, and choose $\delta\in\{0.2,0.4,0.6,0.8\}$. Figure \ref{fig:empericalchi_a} fixes $h=0.22$, and plots $\chi_u(h)$ as a function of the threshold $u$; the limit is 0 for $\delta<1/2$ and positive for $\delta>1/2$. Increasing $\rho_W$ or $\rho_R$ leads to a slower convergence to the limit in all the plots. Figure \ref{fig:empericalchi_b} sets $u=0.999$ and plots $\chi_u(h)$ as a function of the spatial lag $h$. While the limit is 0 in all cases, $\chi_u(h)$ approaches the limit much quicker when $\delta<1/2$ than when $\delta>1/2$.

\subsection{The global SPQR algorithm}
The global SPQR approximation sets $f_i(\cdot) = f(\cdot)$ in Eqn. (8) of the main text, thus pooling information over all locations. Instead of having separate models for each $f_i(\cdot)$, a single FFNN is used to model the SPQR weights for all the locations. We fit a density regression viewing $\bs_{(i)}$, $u_{(i)}$ and $\btheta^{SPAT}$ as the features ($\bx_i$). Since the process is assumed to be stationary in space, only the differences $\bs_j-\bs_i$ influence the regression model. To ensure that $\bx_{i}$ has the same length for all sites, we inflate the feature vector for sites $i \in \{1,\ldots, m\}$ with large values of $\bs_j - \bs_i$ and $u_j \sim \text{Uniform}(0,1)$. The feature set $\bx_i$ for modeling $u_i$ at location $\bs_i$ thus contains the spatial parameters $\btheta^{SPAT}$, process values at the neighboring locations $U(\bs_{(i)})$, as well as the spatial configuration of the neighboring set, $\{(\bs_{(i)} - \bs_i)\} \equiv \{(\bs_j - \bs_i); j \in \mathcal{N}_i\}$, where the sites in $\mathcal{N}_i$ are ordered by the distances to $\bs_i$. 

\begin{algorithm}
\caption{Global SPQR approximation}\label{a:global}
\begin{algorithmic}
\Require Locations $\bs_1, \ldots, \bs_n$ and corresponding sets of neighboring locations $\bs_{(1)}, \ldots, \bs_{(n)}$
\Require Design distribution $p^*$, sample size $N$
\State $k \gets 1$
\While{$k \leq N$}
    \State Draw sample location $\bs_{l_k}$, where $l_k \in \{2, \ldots, n\}$
    \State Draw values of $\mathbf{\btheta}_{k}^{SPAT} \sim p^*$
    \State Generate $U(\bs) = G\{V(\bs)\}$ at $\bs \in \{\bs_{l_k}, \bs_{(l_k)}\}$
    \State Define features $\bx_{l_k} = ( \mathbf{\btheta}_{k}^{SPAT},u_{(l_k)},\bs_{(l_k)} - \bs_{l_k})$, where $u_{(l_k)} = \{U_{l_k}(\bs); \bs \in \bs_{l_k}\}$
    \State $k \gets k + 1$
\EndWhile
\State solve $\hat{\cal W} \gets \arg_{\cal W}\max \prod_{k=1}^N f(u_{l_k}|\bx_{l_k})$, for $f(u|\bx, \cal W)$ using \texttt{SPQR}
\end{algorithmic}
\end{algorithm}
Algorithm \ref{a:global} details the global SPQR approximation. Both approximations (local and global) have their advantages. Each location has a unique spatial configuration of its neighbors for ungridded data, but the local approximation is not affected by these differences. The local SPQR models requires fewer features for training and relatively shallow networks tend to be sufficient. The global approximation requires more features and, therefore, benefits from a deeper network. It can however be computationally more attractive, as a single FFNN will require significantly fewer computational resources than $n-1$ shallower FFNNs for most real life examples. In Section \ref{s:sim:GP}, we compare both the local and global approximations, and provide our reasoning for favoring the local approximation.

\subsection{Variable importance measures used in this study}\label{s:VI}
In most applications where quantile regression is used, understanding the covariate effect on different quantiles is of paramount interest. Therefore, we seek to understand the full conditional distributions and spatial dependence structure by measured the important of these covariates on specific aspects of the response distribution as measured by the quantile function $Q(\tau|\bx)$, where $\tau$ is the quantile level of interest. In our application, the covariates are the conditioning set of observations and the spatial dependence parameters. While SPQR can capture complex non-linear covariate effects on the entire response distribution, it is difficult to interpret the effect of individual covariates on different values of $\tau$ as the weights $\mathcal{W}$ are not individually identified and do not correspond to meaningful quantities. 

The \texttt{SPQR} package quantifies covariate quantile effects using the accumulative local effects (ALEs) of \cite{ApleyZhu2020}. The sensitivity of $Q(\tau|\mathcal{W},\bx)$ to covariate $j$ is naturally quantified by the partial derivative $$q_j(\tau|\mathcal{W},\bx) = \frac{\partial Q(\tau|\mathcal{W},\bx)}{\partial x_j}.$$ The ALE begins by averaging $q_j(\tau|\mathcal{W},\bx)$ over $\bx$ conditioned on $x_j = u$,  i.e.,
\begin{equation*}
    {\bar q}_j(\tau|\mathcal{W},u) = \mathbb{E}_{\bx}\{ q_j(\tau|\mathcal{W},\bx)|x_j=u\}.
\end{equation*}
The ALE main effect function of $x_j$ is then defined as 
\begin{equation*}
    \mbox{ALE}_j(\tau|\mathcal{W},x) =\int_{0}^{x}{\bar q}_j(\tau|\mathcal{W},u)du.
\end{equation*}
Second-order ALE interaction effect for $x_j$ and $x_l$ can be defined analogously by taking the partial derivative with respect to both $x_j$ and $x_l$. These functions can be plotted by $\tau$ to summarize how the predicted quantile changes with respect to change in the covariate values. The ALE function is then distilled to one-number summaries following \cite{greenwell2018} to compare variable importance by quantile level. The variable importance (VI) for continuous covariates are characterized by the standard deviation of the ALE with respect to the marginal distribution of $\bx$, i.e., $$\mbox{VI}_j(\tau|\mathcal{W}) = \mbox{SD}\{ \mbox{ALE}_j(\tau|\mathcal{W},x_j)\}$$.

The ALE and VI summaries depend on the model parameters, $\mathcal{W}$. In our work, we evaluate them using point estimates of $\widehat{\mathcal{W}}$ to give a point-estimate of the summaries, $ALE_j(\tau|\widehat{\mathcal{W}},x)$ and $\mbox{VI}_j(\tau|\widehat{\mathcal{W}})$. If a Bayesian neural network is used instead, posterior samples can be used to quantify uncertainty of the summaries such as the posterior probability that variable $j$ is more important than variable $l$ for predicting conditional quantile at $\tau$.

\section{Additional simulation studies}
\subsection{Connection to the main text}
This appendix supports Section 5 of the main text in the form of 3 additional simulation studies.
\subsection{GP with fixed margins}\label{s:sim:GP}
This study considers a GP as the underlying spatial process, a special case of the PMM corresponding to $\delta = 0$. Of course, the conditional distributions of a GP are univariate Gaussian and so the SPQR approximation is unnecessary. This simple case, however, will allow comparisons to the exact conditional distribution, which is not available for conditional densities associated with the general form of the PMM. The local and global SPQR approximations of $f_u$ are trained at 100 locations chosen randomly on the unit square, ordered by their distance from the origin. For the local SPQR approximation, we simulate $10^6$ independent realizations of a GP at each of the 100 locations. For the global SPQR approximation, we simulate $10^8$ independent GP realizations at the 100 locations. The data in each case are generated from a GP with mean $0$, variance $1$, and exponential correlation $\mbox{Cor}\{Y_t(\bs_i),Y_t(\bs_j)\}=(1-r)I(i=j) + r\exp(-||\bs_i-\bs_j||/\rho)$ for $r \in (0,1)$ and $\rho>0$. The proportion of the variance explained by the spatial error, $r$, and the spatial range $\rho$ make up $\btheta^{SPAT}$. After drawing $\btheta^{SPAT} \sim p^*$ from the distributions given in Table \ref{tab:hyperparams}, the $t^{th}$ realization $Y_{t1},...,Y_{t100}$ is generated from a multivariate normal distribution with correlation defined by $\btheta_{t}^{SPAT}=(r_t,\rho_t)$. For each $Y_{tj}$, $t>1$, its up to $m = 10$ nearest neighbors are identified from the Vecchia neighboring set; $u_{t1}=\Phi(Y_{t1})$ is selected as the response, and $\btheta_{t}^{SPAT}$ and the $m$ remaining $Y_{tj}$ constitute $\bX_t$. In practice, using $Y_{tj}$ as features instead of $u_{tj} = \Phi(Y_{tj})$ provided a better model fit in this simulation study. For the local SPQR approximation, the feature set consists of $Y_{tj}$ along with $\log\rho_t$ and $\log( r_t/(1-r_t))$. For the global SPQR model, the feature set consists of $Y_{tj}, \frac{||s_{j1} - s_{i1}||}{\rho_t}, \frac{||s_{j2} - s_{i2}||}{\rho_t}$, and $\log\frac{r_t}{1-r_t}$, where $\bs_j = (s_{j1},s_{j2})$ for $j = 2,\ldots,11$.

\begin{table}
\centering
\caption{Design distribution $p^*$ (top), and FFNN hyperparameters (bottom) for the global and local SPQR approximations.}
\label{tab:hyperparams}
\begin{tabular}{lcc}
\toprule
\textbf{Hyperparameter} & \textbf{Global SPQR} & \textbf{Local SPQR}  \\\midrule
$r$ & Uniform$(0,1)$ & Uniform$(0,1)$ \\
$\rho$ & Uniform$(0.1,2)$& Uniform$(0.1,1.23)$ \\\midrule
Number of features & 31 & 12 \\
Hidden layer neurons & (60, 40, 30) & (25, 15)\\
Output knots & 15 & 10 \\
Activation function & sigmoid & sigmoid  \\
Learning Rate & 0.001 & 0.005  \\
Batch size & 1000 & 1000  \\
Epochs & 20 & 20 \\\bottomrule
\end{tabular}
\end{table}

Hyperparameter tuning for the global and local SPQR approximations was carried out by comparing fitted models on a validation data set of $10^6$ observations. Models were compared on the basis of the log-score and the Kullback-Leibler (KL) divergence between the estimated and true densities. The lower section of Table \ref{tab:hyperparams} lists the hyperparameter configuration chosen for the global and local SPQR approximations. The global model has more layers and higher complexity than the local models, since it contains more features and more variability in the data. Increasing the complexity further in either model results in diminishing improvements in log-scores and KL divergences, and our hyperparameter choices reflect a balance of computational cost and goodness-of-fit.

\begin{figure}
    \centering
            \includegraphics[width=0.8\linewidth]{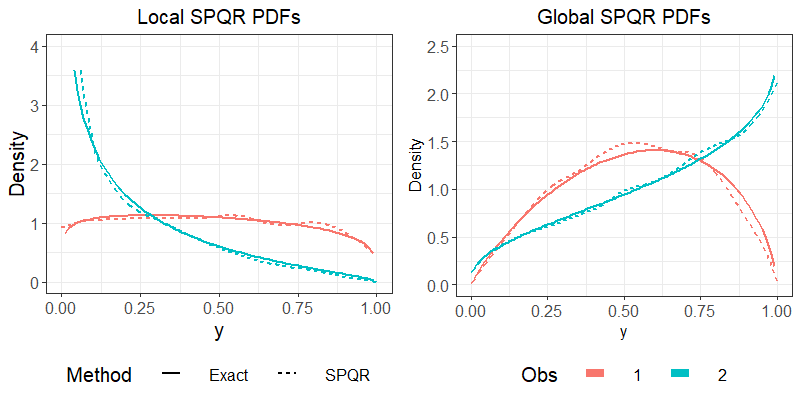}
\caption{{\bf SPQR fit for simulated data from a GP}: True and estimated PDFs for two out-of-sample observations fitted using the local and global SPQR approximations.}
    \label{fig:GP_PDF_plots}
    \includegraphics[width=0.8\linewidth]{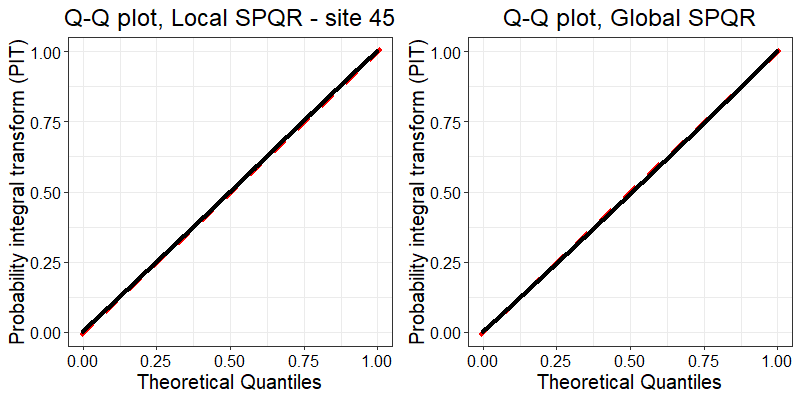}
\caption{{\bf Goodness-of-fit for SPQR fits on GP data}: Q-Q plots on the uniform scale based on the local and global SPQR models.}
    \label{fig:qqplot}
    \includegraphics[width=0.4\linewidth]{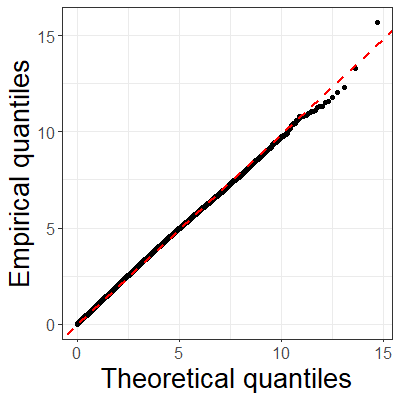}
        \includegraphics[width=0.4\linewidth]{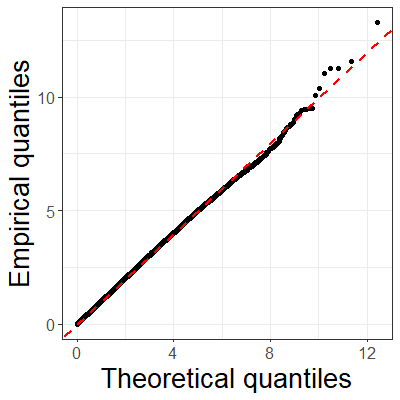}
\caption{{\bf Goodness-of-fit for SPQR fits on GP data}: Q-Q plots on the exponential scale based on the local and global SPQR models.}
    \label{fig:qqplot2}
\end{figure}

We estimate the model weights ${\cal W}$ using the global and local SPQR approximations described in Algorithms 1 and 2 of the main text. To improve the stability of the global SPQR, we split the training data into ten training sets each of size $10^7$ and obtain $R=10$ estimates of ${\cal W}$, denoted ${\hat {\cal W}}^1,...,{\hat {\cal W}}^R$. The averaged probabilities $\pi_k(\bX) = \sum_{r=1}^{R}\pi_k^r(\bX)/R$, where $\pi_k^r(\bX)$ is evaluated using ${\hat {\cal W}}^r$, are used to evaluate the approximate densities. Each global SPQR model takes approximately 232 minutes to fit, while each local SPQR with all neighbors takes around 21 minutes of computation time.

\begin{figure}
    \centering
\includegraphics[width=\linewidth]{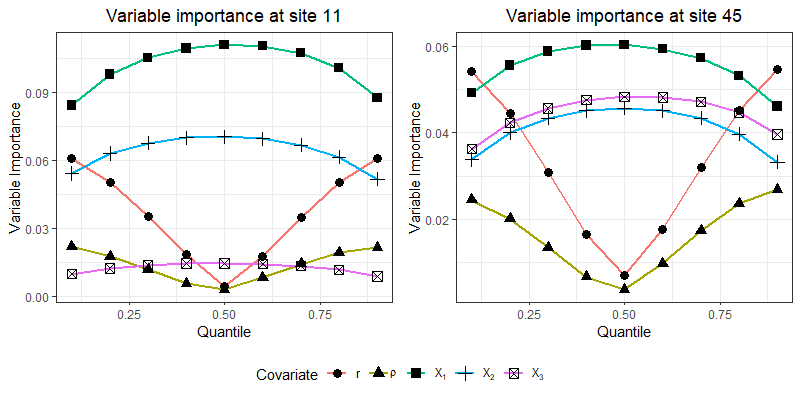}
\caption{{\bf Variable importance for the local SPQR approximation on GP data}: VI for sites 11 (left) and 45 (right) for the 5 most important variables, including parameters $r$ and $\rho$ as well as the three nearest neighbors in the Vecchia approximation, across quantiles between (0.05,0.95).}
    \label{fig:VI}
\end{figure}

Figure \ref{fig:GP_PDF_plots} plots the true and estimated PDFs for two randomly selected test set observations for location 45 from the local (left) and global (right) SPQR approximations and shows that the model fits well. In both cases, $r=1$ and $\rho = 0.2$. Figure \ref{fig:qqplot} plots the PIT scores for the two approximations. The PIT score for the true model is $F(Y_{tj})$ where $F$ is the true Gaussian conditional distribution of $Y_{tj}$ given its neighbors, and this is plotted against the same measure for the fitted CDF $\hat{F}(Y_{tj})$ obtained from the local and global SPQR fits. The PIT statistics falling on the $Y=X$ line shows that the models fit well. While they look identical on the uniform scale which provides equal weight to the entire distribution, Figure \ref{fig:qqplot2} which presents the same data on the exponential scale shows differences in their tail behavior. This is to be expected, since the global SPQR tries to capture the distribution of all the locations, and the local SPQRs model the spatial structure of individual locations. 

Finally, Figure \ref{fig:VI} plots the variable importances (VI) of the five most important variables across all quantiles of the local SPQR model. These include $\btheta^{SPAT}$, as well as the three nearest neighbors based on the Vecchia approximation. For both locations 11 (left) and 45 (right), we note that the nearest neighbors have lower variable importance for the extreme quantiles at either end and higher variable importance in the middle. While they are ordered by their importance (the nearest neighbor has highest importance and so on), their positions relative to each other are different for the two locations in Figure \ref{fig:VI} and likely depend on the spatial configuration of the neighbors. The opposite behavior is seen for the spatial parameters $r$ and $\rho$, which have highest variable importance for the extreme quantiles.

For parameter estimation, we simulate 200 independent datasets. Each dataset consists of 5 independent realizations of a GP at the 100 locations. The GP has mean $\mu$, variance $\sigma^2$ and exponential correlation function with range $\rho$ and variance parameter $r$.  Therefore, there are four parameters to be estimated: $\btheta^{MARG}=(\mu,\sigma)$ and $\btheta^{SPAT}=(\rho,r)$. For priors, we assume that $\mu,\log(\sigma)\iid\mbox{Normal}(0,10^2)$, $\log(\rho)\sim\mbox{Normal}(-2,1)$, and $\log\{r/(1-r)\}\sim \mbox{Normal}(0,1.5^2)$. Three scenarios are considered with the true values of $\btheta \equiv(\mu,\sigma,\rho,r)$ set to:
\begin{enumerate}
    \item $\mu = 3, \sigma = 2, r = 0.7, \rho = 0.2$
    \item $\mu = 2, \sigma = 2, r = 0.5, \rho = 0.2$
    \item $\mu = 1, \sigma = 3, r = 0.7, \rho = 0.3$.
\end{enumerate}
For each scenario, we use 10,000 MCMC samples after a burn-in of 1,000 iterations. For local SPQR models, runtimes were approximately 13 seconds per 1,000 MCMC iterations; the global SPQR takes approximately 27 seconds per 1,000 MCMC iterations.

\begin{figure}
    \centering
    \includegraphics[width=\linewidth]{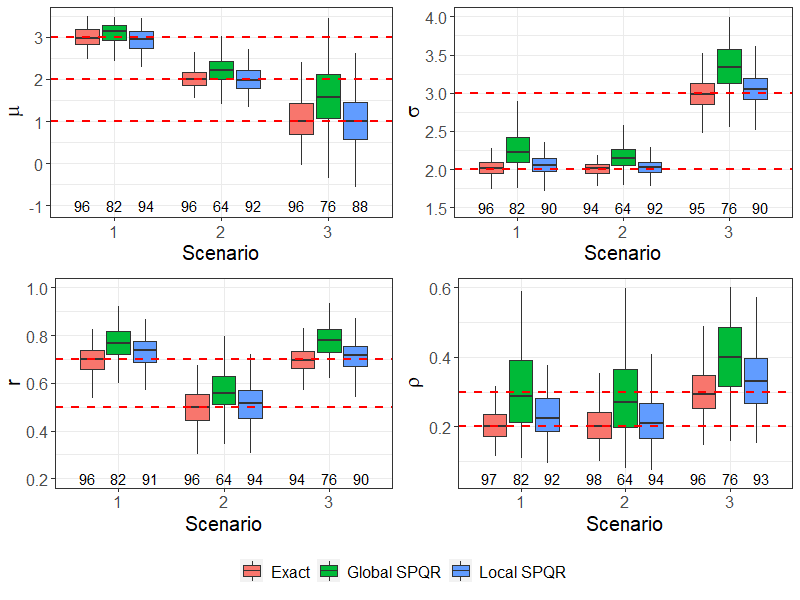}
\caption{{\bf Parameter estimation for the GP simulation study}: Sampling distribution of the posterior mean for the GP parameters for three different scenarios.  The boxplots compare the exact Gaussian full conditional distributions (red) versus approximate full conditionals obtained via the global (green) and local SPQR (blue) algorithms. The horizontal dashed lines are true values; empirical coverage of the 95\% intervals are provided below each scenario.}
    \label{fig:sim}
\end{figure}

Figure \ref{fig:sim} shows the results for the Vecchia approximation alongside the exact Gaussian conditional distribution and the global and local SPQR approximations based on density regression. The local SPQR performs significantly better than the global SPQR, and its parameter estimates have lower bias and coverage close to the nominal level. The local SPQR also has less sampling variance than the global approximation, and its parameter estimates are comparable to those generated using the exact conditionals. This is especially noticeable in estimates for the range $\rho$, where the global SPQR has significantly higher bias and variablility than the other methods. Since neighbor configurations vary for each location, local SPQRs where each model is already conditioned on a specific spatial configuration seem to be better at estimating the spatial range. 

\subsection{PMM with fixed margins}\label{s:sim:PMM} 

\begin{figure}
    \centering
    \includegraphics[width=.6\linewidth]{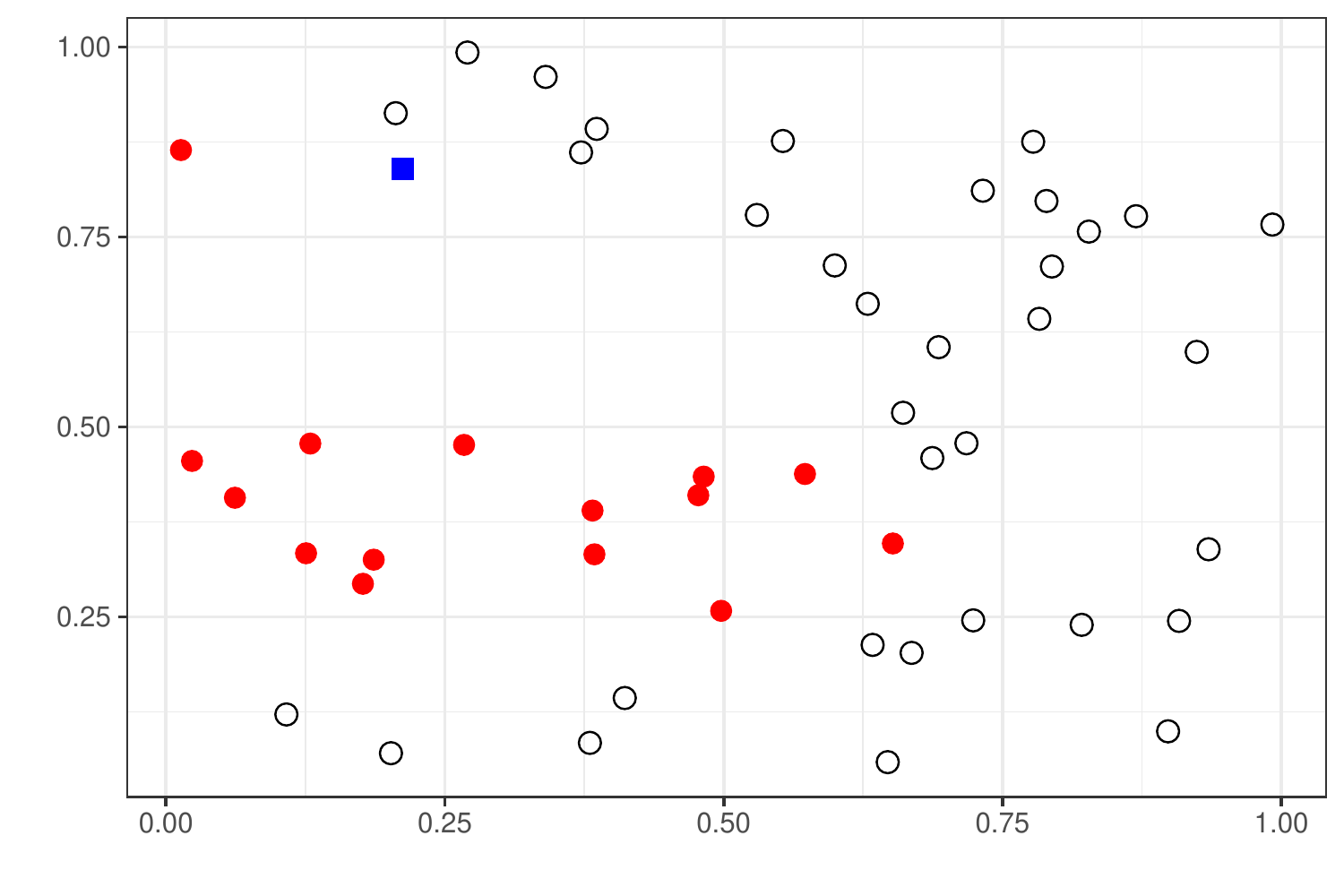}
\caption{{\bf Locations used in the PMM simulation study:} 50 randomly generated locations on a unit square. The blue square corresponds to site 25, and the red circles are its nearest neighbors from the Vecchia conditioning set.}
    \label{fig:sim:pmmsvc_coverage}
\end{figure}

We consider the spatial process trained in Section 4 of the main text, but with the same marginal parameters across all 50 locations.
For all simulations, the GEV location and scale are $\mu=2$ and $\sigma=1$ and the spatial dependence parameters are $\alpha=1$ and $\rho=0.15$. The simulation study scenarios for parameter estimation vary based on the GEV shape $\xi\in\{-0.1,0.1\}$ and asymptotic dependence parameter $\delta\in\{0.2,0.8\}$. We also add a fifth scenario with observations set to be missing (completely at random over space and time) with probability $\pi_M=0.05$ and censored below the threshold $T$, set to the sample median ${\hat q}_{0.5}$ (over space and time).  The scenarios are:
\begin{enumerate}
    \item $\xi=0.1$, $\delta=0.2$
    \item $\xi=0.1$, $\delta=0.8$
    \item $\xi=-0.1$, $\delta=0.2$
    \item $\xi=-0.1$, $\delta=0.8$
    \item $\xi=0.1$, $\delta=0.2$, $\pi_M=0.05$, $T={\hat q}_{0.5}$.
\end{enumerate}

For priors, we select $\mu,\log(\sigma)\sim\mbox{Normal}(0,10^2)$, $\xi\sim\mbox{Normal}(0,0.25^2)$ for $\btheta^{MARG}$, and $\delta \sim \mbox{Uniform}(0,1)$, $\rho\sim \mbox{Uniform}(0.0,0.5)$ for $\btheta^{SPAT}$. The posterior distribution is approximated using MCMC with 11,000 (21,000 for scenario 5) iterations and Metropolis candidate distributions tuned to have acceptance probability near 0.4.  
%Generating 11,000 MCMC samples takes around seven minutes on standard PC.  
After discarding the first 1,000 iterations as burn-in, the remaining samples are used to compute the posterior mean and 95\% interval for each parameter. For the first four scenarios, runtimes were approximately 1 minute per 1,000 MCMC iterations for the first 4 scenarios and about 6 minutes for scenario~5.

\begin{figure}
    \centering
    \includegraphics[width=.8\linewidth]{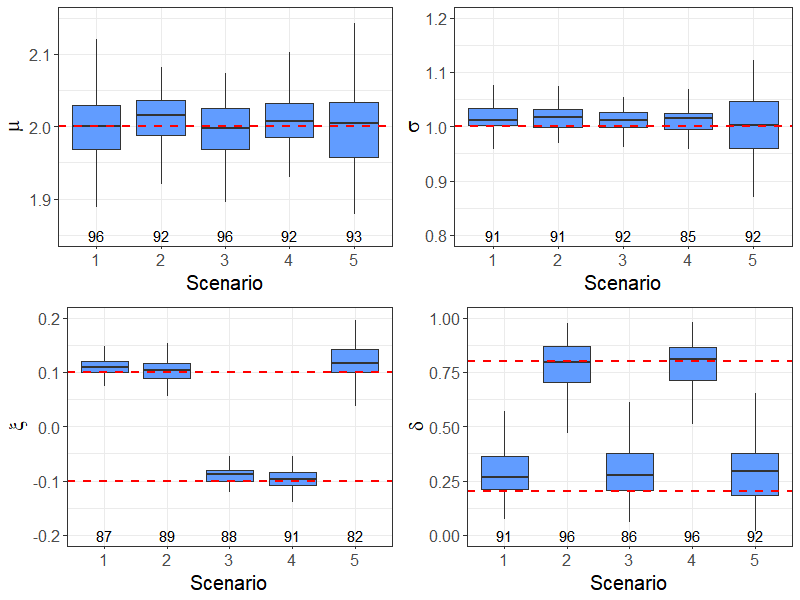}
\caption{Sampling distribution of the posterior mean for the GEV parameters and asymptotic dependence parameter $\delta$ for the five simulation scenarios. The horizontal dashed lines are true values and the numbers along the bottom give the empirical coverage of the 95\% intervals. }
    \label{fig:sim:MCMC_posterior_mean_EVP}
\end{figure}
Figure \ref{fig:sim:MCMC_posterior_mean_EVP} plots the sampling distribution of the posterior mean estimator of the model parameters of interest and gives empirical coverage of the 95\% posterior interval. The posterior mean estimator for the GEV parameters generally has low bias and coverage near the nominal level. While the sampling variance of the posterior mean estimator of $\delta$ is high, the method is clearly able to distinguish between the two asymptotic regimes with expected value near 0.30 for the asymptotic independence Scenarios 1 and 3 compared to roughly 0.78 for the asymptotic dependence Scenarios 2 and 4. As expected, the sampling variance increases in Scenario 5 with missing data and censoring, but the method is still able to reliably estimate the model parameters. Finally, we see that coverage of $\delta$ is higher for the asymptotic dependence scenarios. We believe this to be caused by model assumptions made for the components of $\theta^{SPAT}$, and that it can likely be alleviated by relaxing some assumptions.

\subsection{PMM with linear models for SPQR}\label{s:sim:LIN}
\begin{comment}
\begin{figure}
    \centering
    \includegraphics[width=.9\linewidth]{figs/svc_params.png}
\caption{True values of the marginal GEV parameters for the 50 locations used in the PMM simulation study.}
    \label{fig:sim:svc_params}
\end{figure}
\end{comment}
To understand the need for a neural network that underlies SPQR, we conduct a simulation study whose setup is identical to the one presented in Section 4 of the main text, but with the weights $\pi_k(\bx,\mathcal{W})$ arising from a linear model instead of a non-linear one. In particular, we consider a NN without any hidden layers, with everything else kept unchanged. Two scenarios are considered corresponding to $\delta \in (0.2,0.8)$. The marginals have STVC model specifications for the GEV parameters, and the same priors are used for the MCMC simulations as before. The simulations are carried out on 50 independent datasets, each with 50 replications. Each MCMC chain is run for 10,000 iterations after discarding 1,000 burn-in iterations.

\begin{table}
\centering
\caption{Coverage (in $\%$) for marginal GEV parameters under 2 scenarios based on MCMC simulations over 50 datasets. The 3 values represent the minimum, mean, and maximum coverage across the 50 study locations.}
\label{t:coverage}
\begin{tabular}{ccccc}
\toprule
 & $\mu_0$ & $\mu_1$ & $\sigma$ & $\xi$ \\\midrule
$\delta = 0.2$ & (72, 86, 98) & (86, 95, 100) & \multicolumn{1}{c}{(28, 51, 80)} & (22, 60, 98) \\
$\delta = 0.8$ & (70, 84, 92) & (86, 95, 100) & (0, 18, 70) & (0, 23, 88)\\\bottomrule
\end{tabular}
\end{table}

\begin{figure}
    \centering
    \includegraphics[width=.45\linewidth]{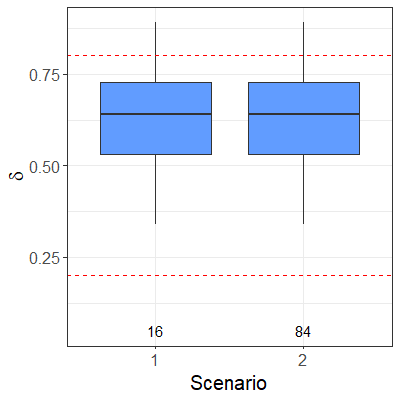}
\caption{Sampling distribution of the posterior mean for the asymptotic dependence parameter $\delta$ for two simulation scenarios. The horizontal dashed lines are true values and the numbers along the bottom give the empirical coverage of the 95\% intervals. }
    \label{fig:sim:linsvc_coverage}
\end{figure}

Table \ref{t:coverage} details coverage of the empirical $95\%$ intervals for the posterior distribution of the marginal GEV parameters. Mean coverage across locations is decent for the location parameters, but based on the SPQR likelihood, the scale and shape parameters cannot be estimated reliably. Figure \ref{fig:sim:linsvc_coverage} plots the sampling distribution of the posterior mean estimator of $\delta$ for the 2 scenarios, and provides empirical coverage of the $95\%$ posterior interval. The distribution is nearly identical for the two scenarios, and the SPQR likelihood is unable to distinguish between an asymptotic dependence and asymptotic independence scenario. This indicates that the non-linear nature of the NN is necessary to capture the spatial distribution of the model. While alternative optimization routines not considered here might give better performance than backpropagation using an Adam optimizer, they are likely to suffer from problems related to the dimension of the covariate vector or in the presence of more complex features, e.g., gridded covariates. Using simple NNs which are not individually computationally intensive provides a balance of computational cost and predictive power, and datasets with more complex features can take advantage of sophisticated NN architectures like convolutional neural networks (CNNs) within the SPQR framework presented in this work.

\section{Additional results from HCDN data analysis}\label{s:fitdetails}
\subsection{Connection to the main text}
    This appendix supports Section 6 of the main text, and provides additional output from the fitted model. Section \ref{s:originalscale} provides selected results from analyzing the HCDN data in its original scale.
\subsection{Parameter estimation using MCMC}
\begin{figure}
    \centering
    \begin{subfigure}[b]{0.45\linewidth}
    \includegraphics[width=\linewidth]{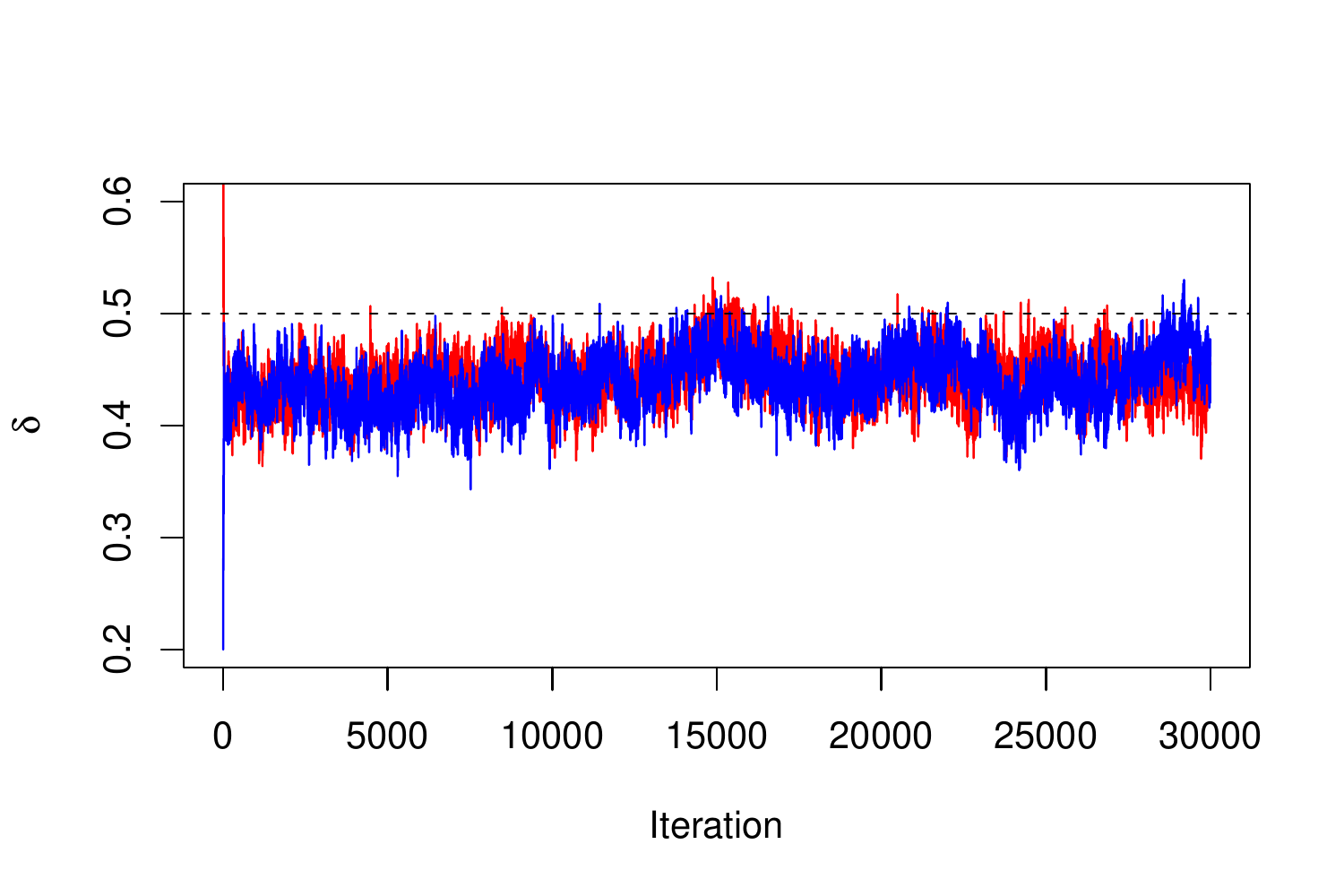}
\caption{Trace plot of 2 MCMC chains with different starting values.}
    \label{fig:tracePlot_delta}
    \end{subfigure}
    \hfill
    \begin{subfigure}[b]{0.45\linewidth}
    \includegraphics[width=\linewidth]{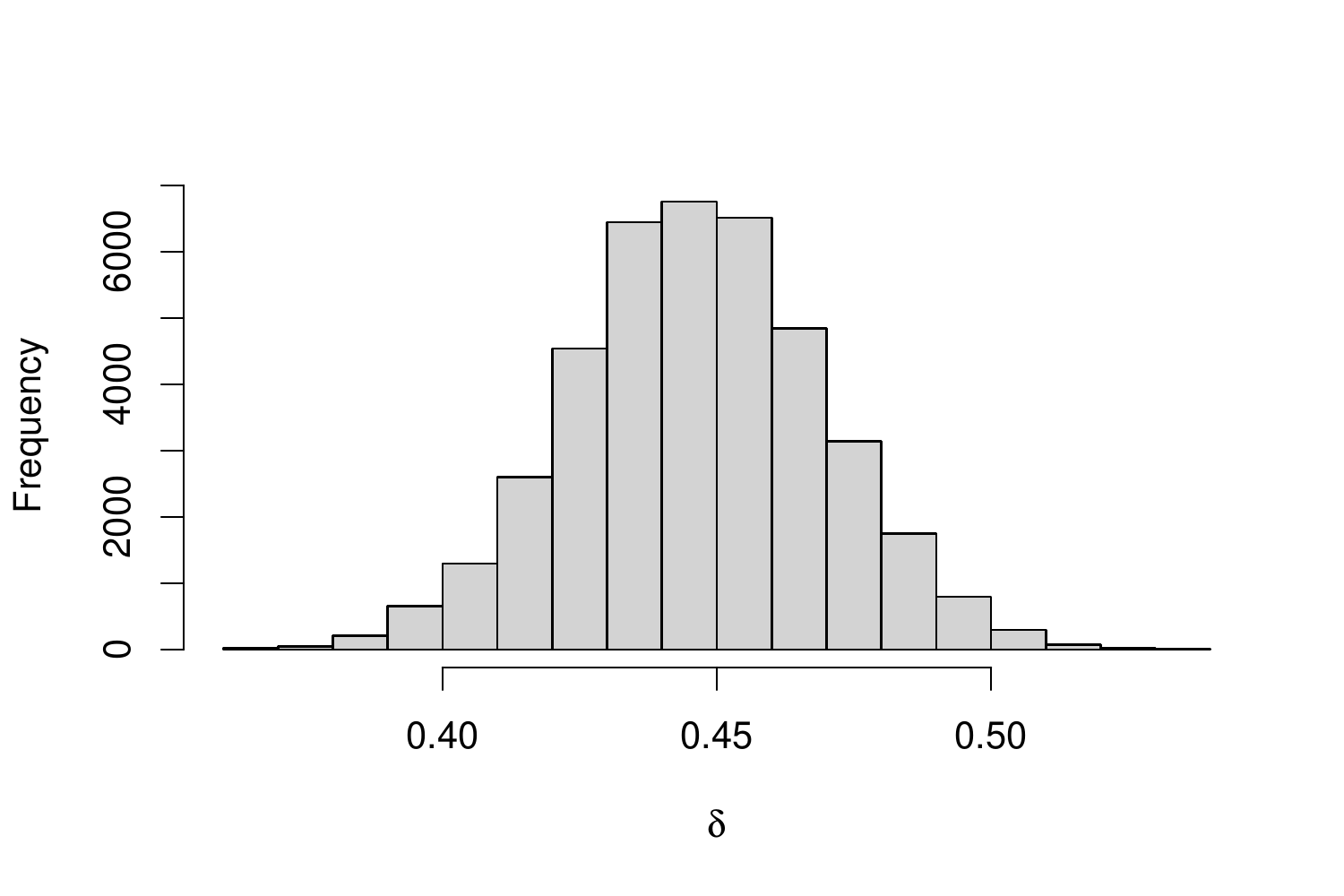}
\caption{Posterior distribution based on 20,000 post burn-in samples.}
    \label{fig:hist_delta}
    \end{subfigure}
    \caption{{\bf Estimation of $\delta$:} MCMC trace plots and the posterior distribution of $\delta$.}
    \label{fig:posterior_delta}
\end{figure}

The local SPQR models for extreme streamflow data was fitted using the \texttt{SPQR} package on \texttt{R 4.2.1}. Each model had 2 hidden layers with 30 and 20 neurons respectively, 15 output knots, a batch size of 1000, learning rate of 0.01, and up to 15 neighbors. Of the 200,000 synthetic observations used to fit the models, 80\% was used for training and 20\% for validation. The model was run for 200 epochs, and the local SPQR takes approximately 14 minutes for locations with a full feature set (i.e., all 15 neighbors).

For the MCMC, we ran the 2 chains in parallel for 30,000 iterations. It takes us approximately 8 minutes per 100 MCMC iterations. Figure \ref{fig:tracePlot_delta} overlays the trace plots for $\delta$ from the two chains, and we see that they are well mixed. The first 10,000 iterates from each chain are discarded as burn-in, and the remaining samples from both chains are used to obtain our posterior estimates. Figure \ref{fig:hist_delta} plots a histogram of the posterior distribution of $\delta$, which indicates that the spatial process is in the asymptotic independence regime.
\subsection{Regional joint exceedance behavior}
 \begin{comment}
 \begin{figure}
    \centering
    \includegraphics[width=.4\linewidth]{figs/CO_cluster.pdf}
    \includegraphics[width=.4\linewidth]{figs/NM_cluster.pdf}
\caption{\small HCDN stations in blue and their Vecchia neighbors in red for studying joint exceedance in Colorado (left) and New Mexico (right).}
    \label{fig:CO_cluster}
\end{figure}
\end{comment}
Table \ref{t:exceedance} provides further details of the joint exceedances for the two clusters. Beyond the comparison for the $0.9$ quantile that has been presented in the main text, we also compared the median of the distribution for 1972 and 2021. The mean probability for joint exceedance is over 10 times higher for the median than for the 0.90 quantile. The probability of joint exceedance is also higher for 2021 compared to 1972 for both clusters of locations.
\begin{table}
\centering
\caption{Joint exceedance probabilities of streamflow maxima between 1927--2021 for two HCDN location clusters.}
\label{t:exceedance}
\begin{tabular}{cccccc}
\textbf{} & \textbf{Quantile} & \multicolumn{2}{c}{\textbf{0.50}} & \multicolumn{2}{c}{\textbf{0.90}} \\\midrule
\textbf{Cluster} & \textbf{Year} & Mean & SD & Mean & SD \\\midrule
\multirow{2}{*}{\textbf{CO}} & 1972 & 0.404 & 0.102 & 0.075 & 0.040 \\
 & 2021 & 0.617 & 0.074 & 0.169 & 0.046 \\\midrule
\multirow{2}{*}{\textbf{NM}} & 1972 & 0.451 & 0.054 & 0.045 & 0.012 \\
 & 2021 & 0.482 & 0.054 & 0.053 & 0.017\\\bottomrule
\end{tabular}
\end{table}

\subsection{Additional model comparison and model fit results}
\begin{figure}
    \centering
        \begin{subfigure}[b]{\linewidth}
     \includegraphics[width=0.5\linewidth]{figs/map_mu1.pdf}
     \includegraphics[width=0.5\linewidth]{figs/map_mu1_positive.pdf}
\caption{Posterior means of $\mu_1(\bs)$ and $Pr[\mu_1(\bs)>0]$ based on the process mixture model.}
    \label{fig:map_mu1_PMM}
    \end{subfigure}
    \begin{subfigure}[b]{\linewidth}
     \includegraphics[width=0.5\linewidth]{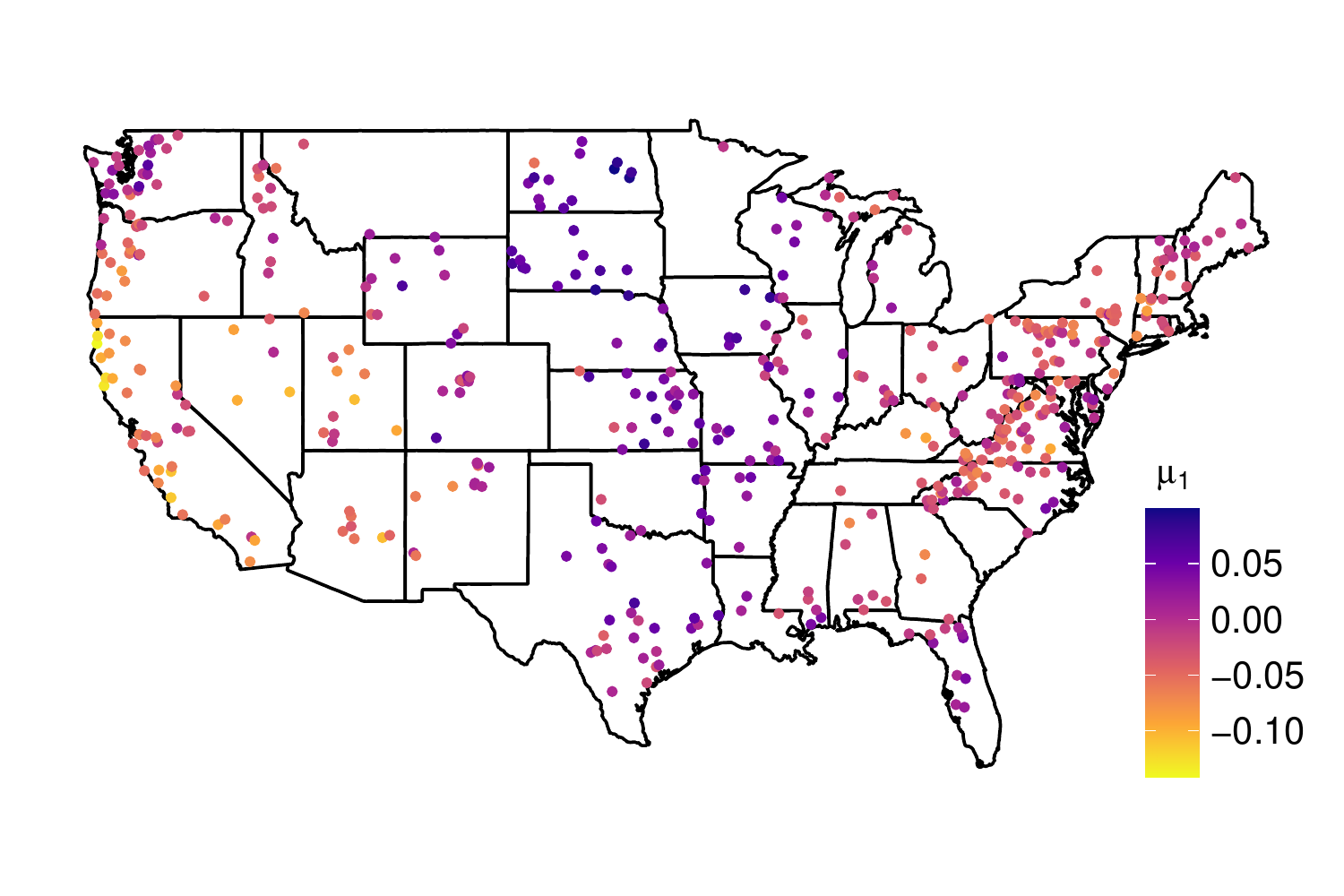}
     \includegraphics[width=0.5\linewidth]{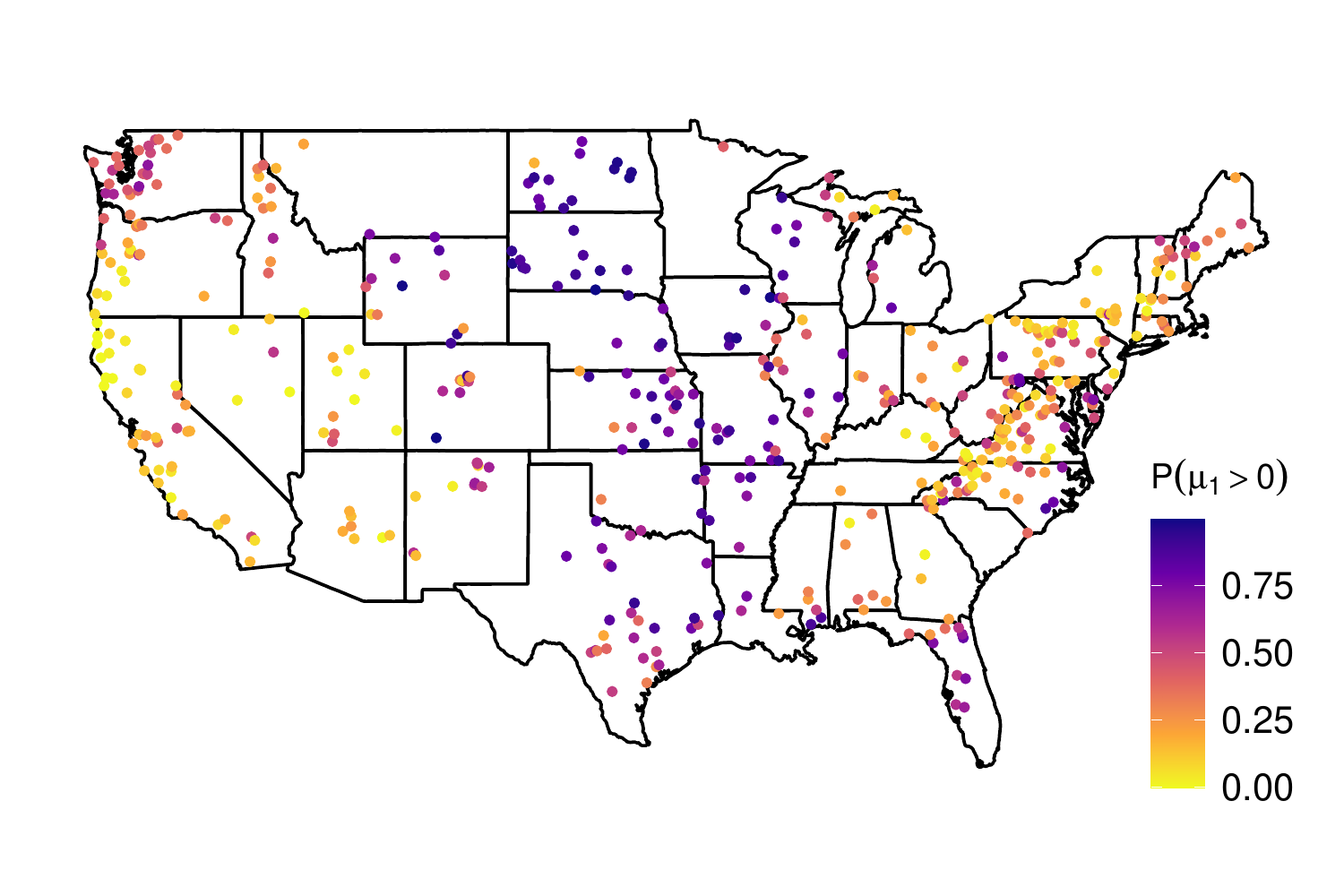}
\caption{Posterior means of $\mu_1(\bs)$ and $Pr[\mu_1(\bs)>0]$ based on the Huser-Wadsworth process.}
    \label{fig:map_mu1_HW}
    \end{subfigure}
    \begin{subfigure}[b]{\linewidth}
        \includegraphics[width=0.5\linewidth]{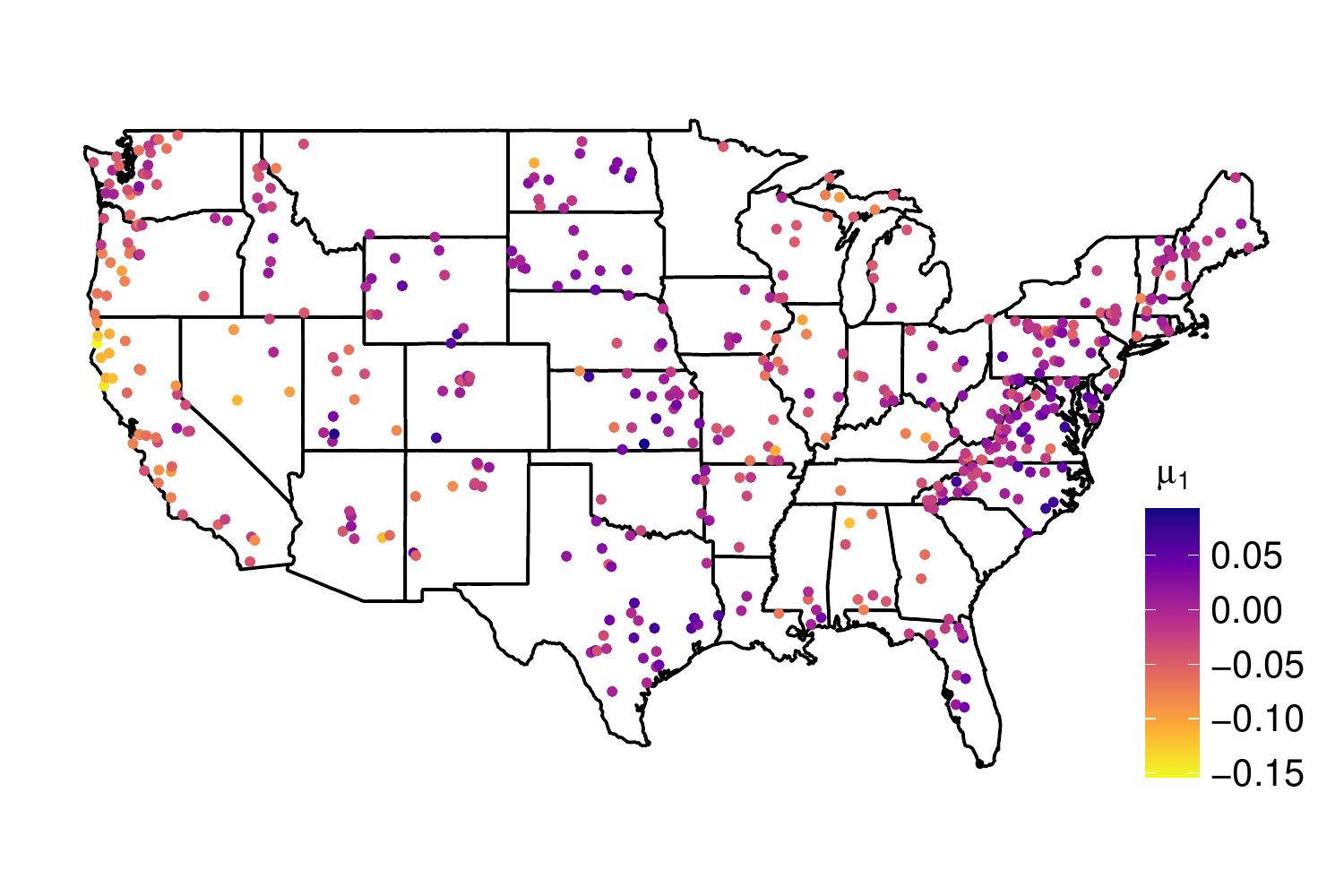}
        \includegraphics[width=0.5\linewidth]{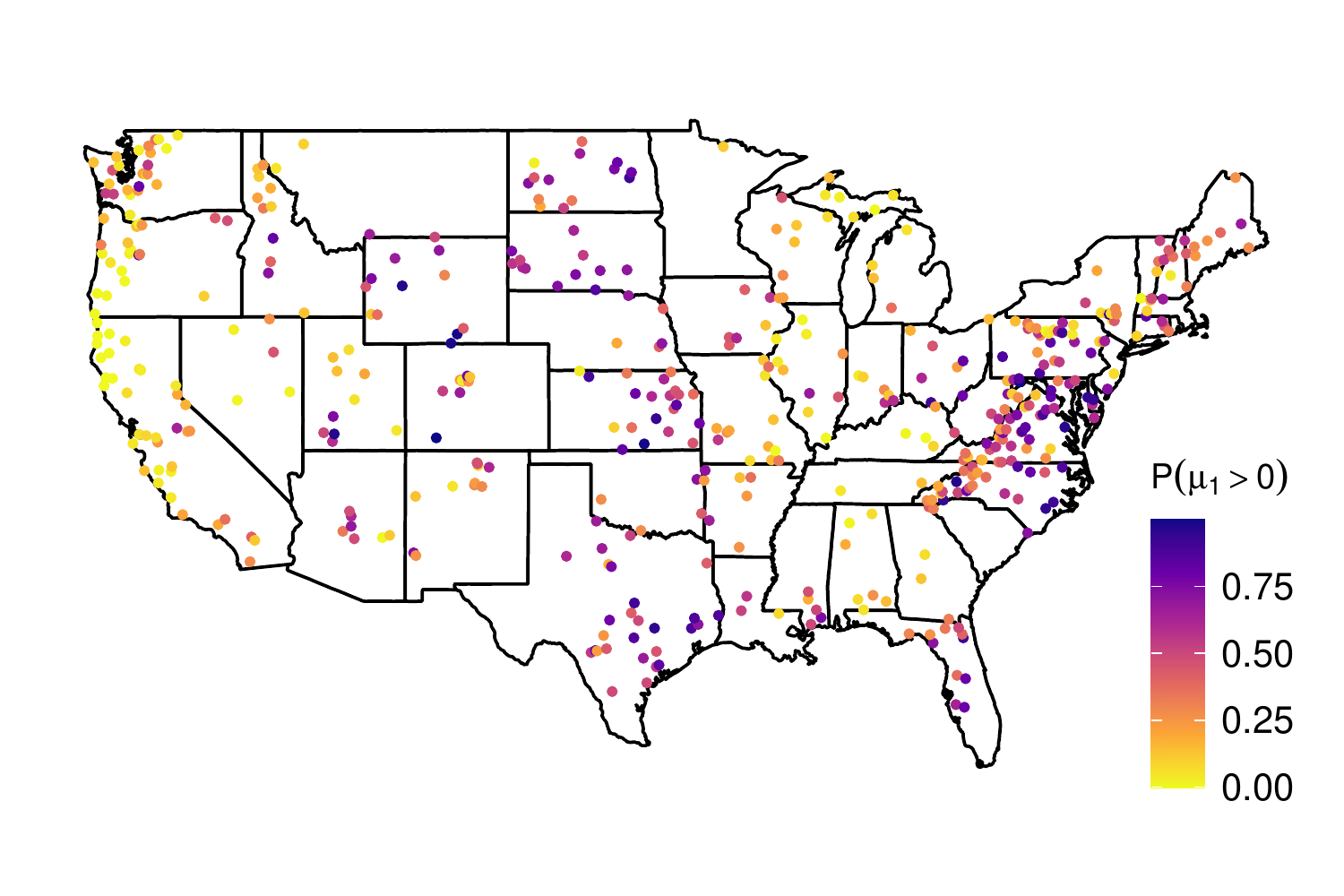}
\caption{Posterior means of $\mu_1(\bs)$ and $Pr[\mu_1(\bs)>0]$ based on a max-stable process process.}
    \label{fig:map_mu1_MSP}
    \end{subfigure}
        \begin{subfigure}[b]{\linewidth}
        \includegraphics[width=0.5\linewidth]{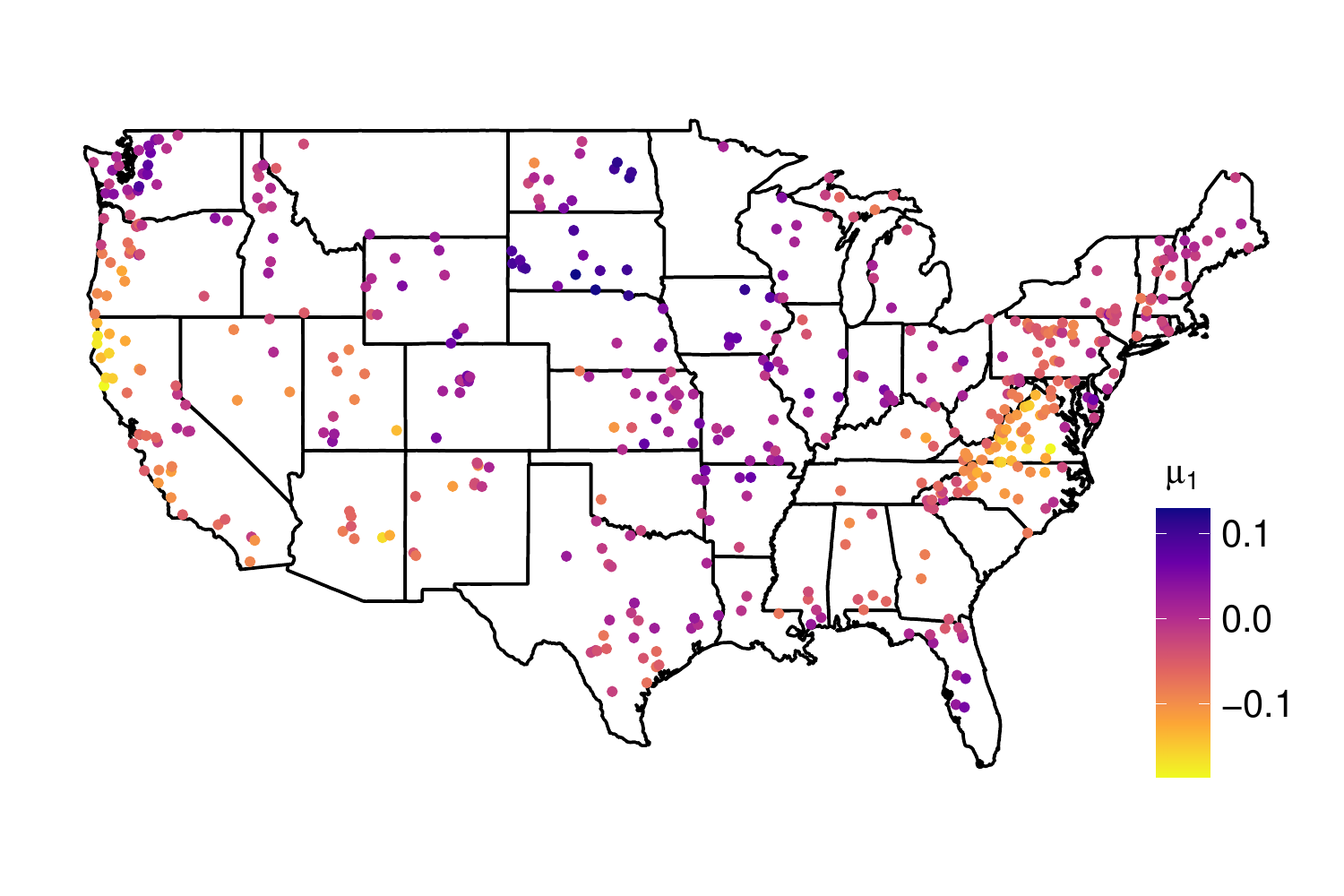}
        \includegraphics[width=0.5\linewidth]{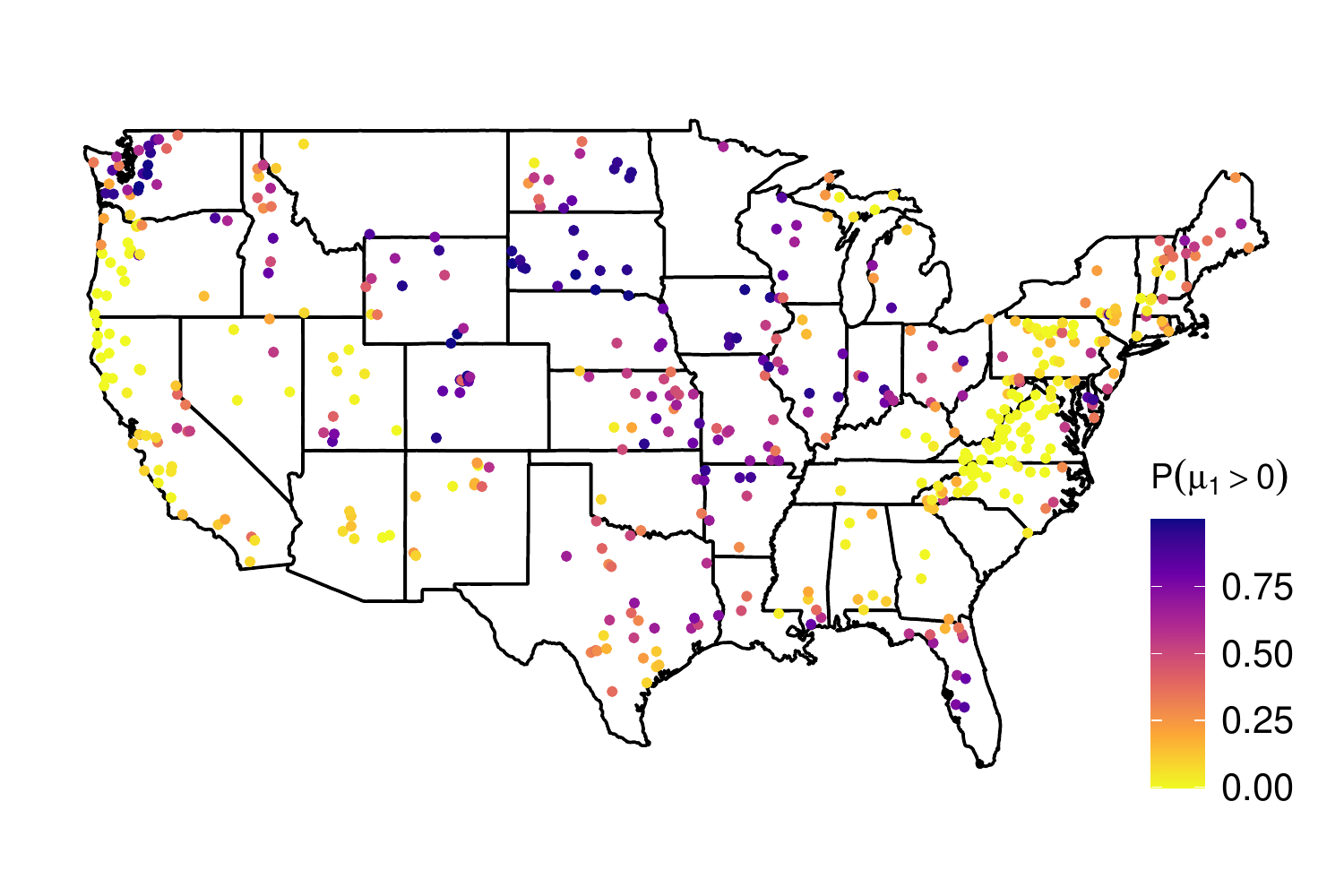}
\caption{Posterior means of $\mu_1(\bs)$ and $Pr[\mu_1(\bs)>0]$ based on a Gaussian process.}
    \label{fig:map_mu1_GP}
    \end{subfigure}
        \caption{{\bf Slope parameter estimates for competing models}: Posterior means of the slope and its probability of being positive for the 487 HCDN stations based on models fitted using a PMM, an HW process, an MSP, and a GP.}
    \label{f:map_mu1_competing}
\end{figure}
Figure \ref{f:map_mu1_competing} plots posterior means of $\mu_1(\bs)$ and $Pr[\mu_1(\bs)>0]$ based on four competing models with different spatial processes - the process mixture model (PMM), the Huser-Wadsworth (HW) process, a max-stable process (MSP), and a Gaussian process (GP). The HW model is the closest to the PMM in terms of the spatial distribution of $\mu_1(\bs)$. However, both the HW process and the MSP have lower estimates of the posterior mean of the slope compared to the PMM. The GP, on the other hand, has a higher range of $\mu_1(\bs)$ estimates across the country. Comparing $Pr[\mu_1(\bs)>0]$ for the 3 models, we see each model have slight differences among each other, and with the PMM. The HW model and the GP have lower probabilities in eastern USA compared to the MSP and the PMM. The MSP has low probabilities for Washington, which is the only region in the west coast with high slope probabilities as per the remaining 3 models. The MSP also estimates lower probabilities in the Mississippi Basin and Great Lakes region compared to the remaining models. Finally, the GP shows less long range dependence compared to the other models, with smaller clusters and several areas with high and low estimates adjoining each other.

\subsection{Sketch of analysis in alternative data scales}\label{s:originalscale}
\begin{figure}
    \centering
    \begin{subfigure}[b]{0.45\linewidth}
    \includegraphics[width=\linewidth]{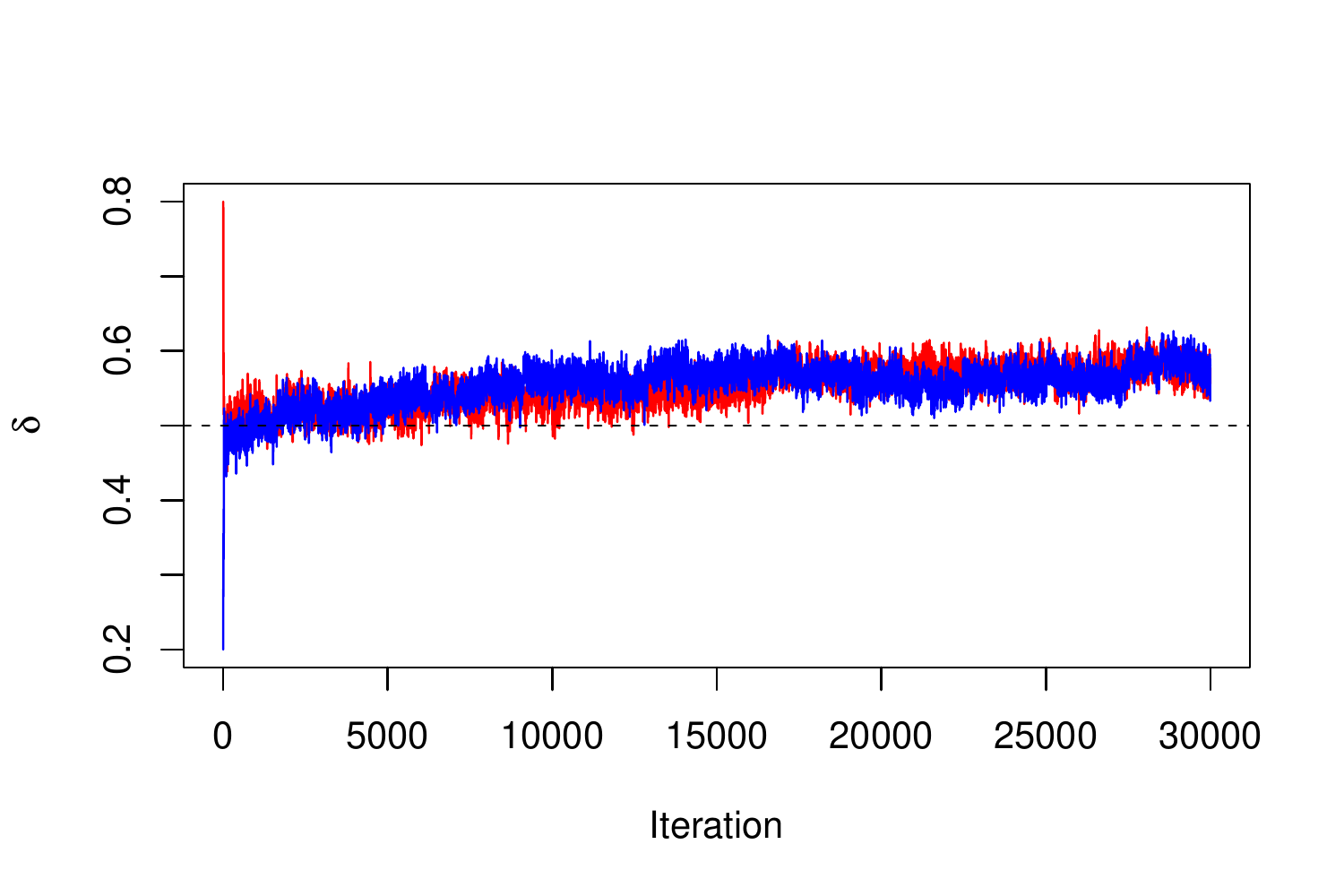}
\caption{Trace plot of 2 MCMC chains with different starting values.}
    \label{fig:tracePlot_delta_og}
    \end{subfigure}
    \hfill
    \begin{subfigure}[b]{0.45\linewidth}
    \includegraphics[width=\linewidth]{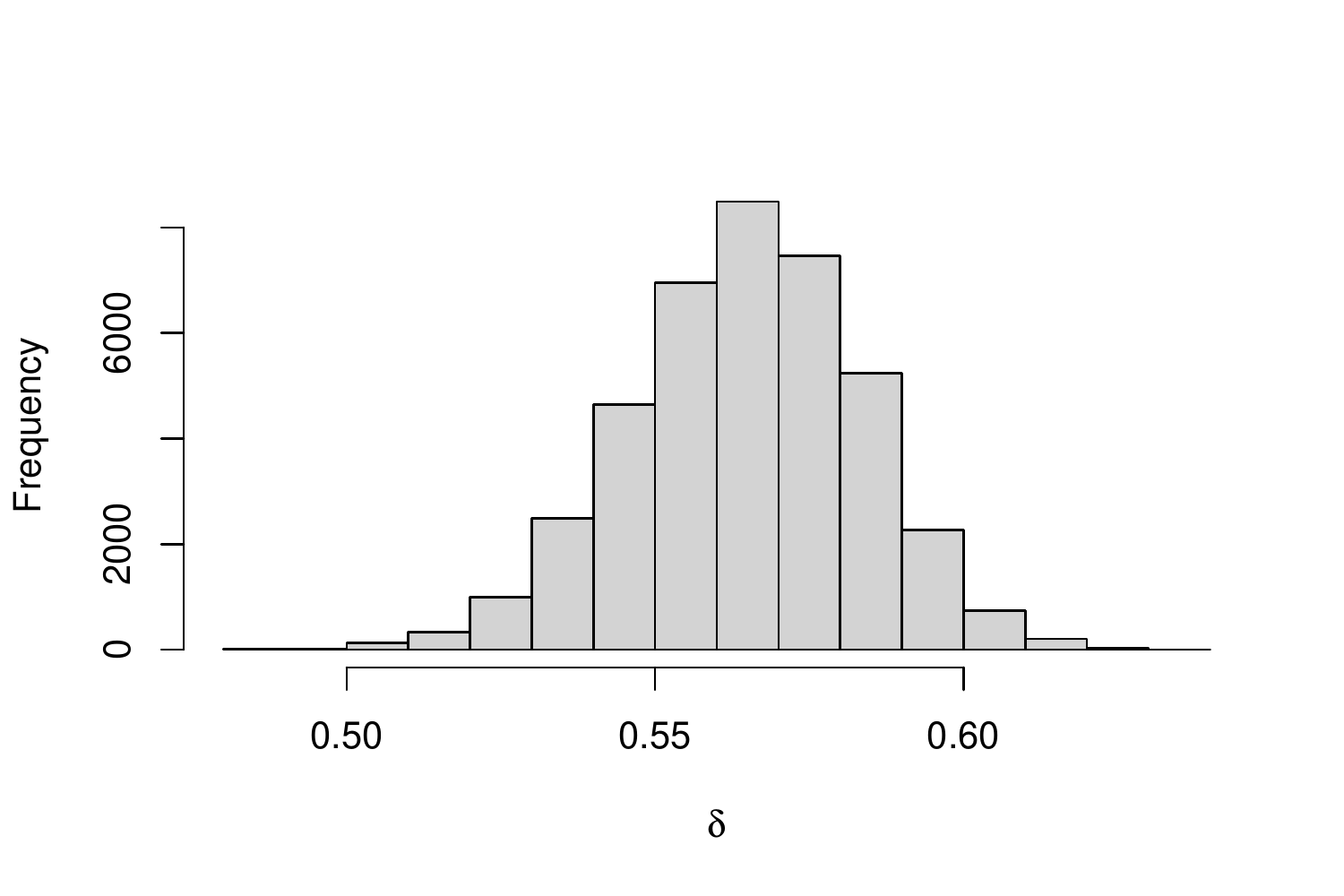}
\caption{Posterior distribution based on 20,000 post burn-in samples.}
    \label{fig:hist_delta_og}
    \end{subfigure}
    \caption{{\bf Estimation of $\delta$ from a PMM fitted to the original streamflow data}:} MCMC trace plots and the posterior distribution of $\delta$.
    \label{fig:posterior_delta_og}
\end{figure}

\begin{figure}
    \centering
    \begin{subfigure}[b]{0.45\linewidth}
    \includegraphics[width=\linewidth]{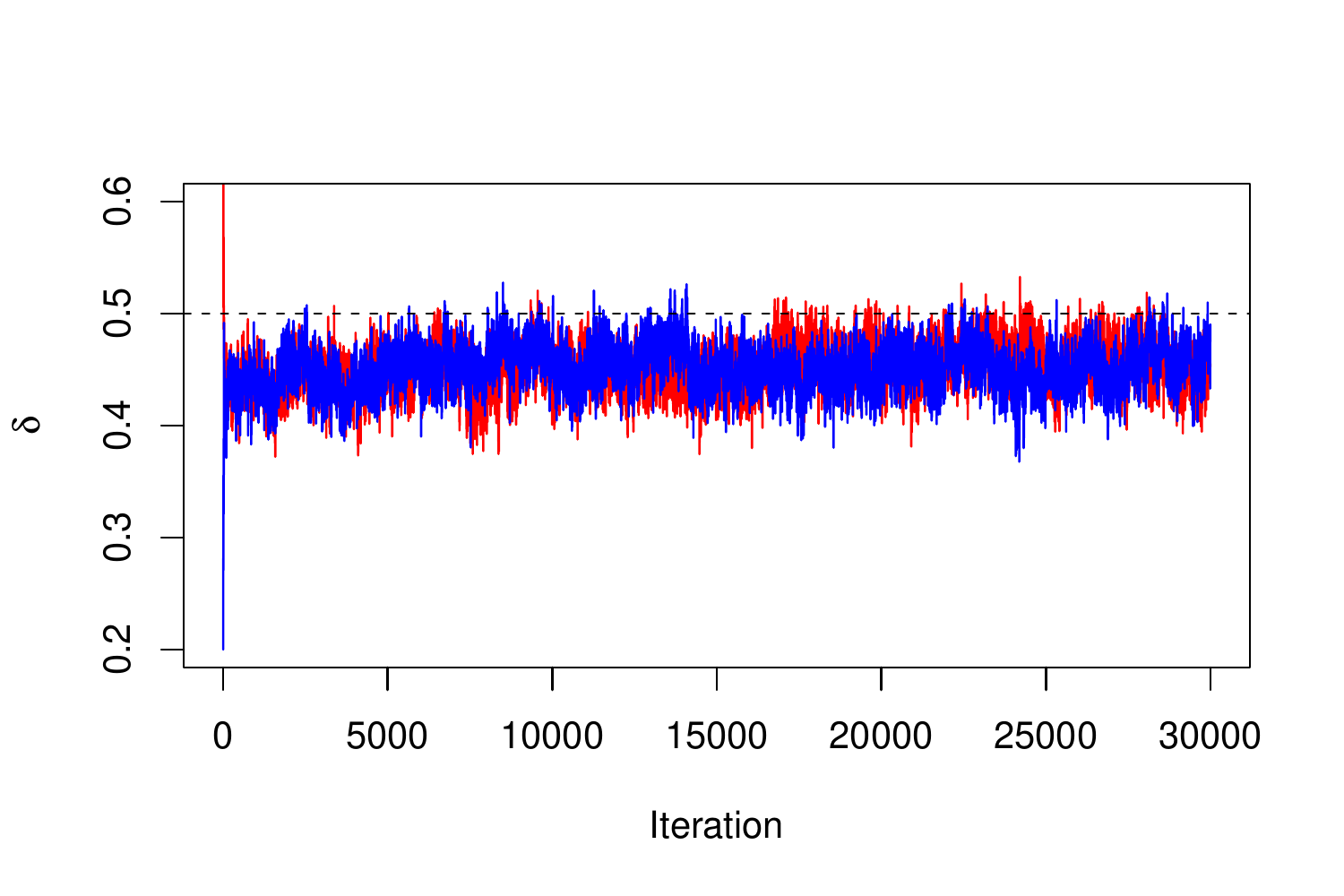}
\caption{Trace plot of 2 MCMC chains with different starting values.}
    \label{fig:tracePlot_delta_sqrt}
    \end{subfigure}
    \hfill
    \begin{subfigure}[b]{0.45\linewidth}
    \includegraphics[width=\linewidth]{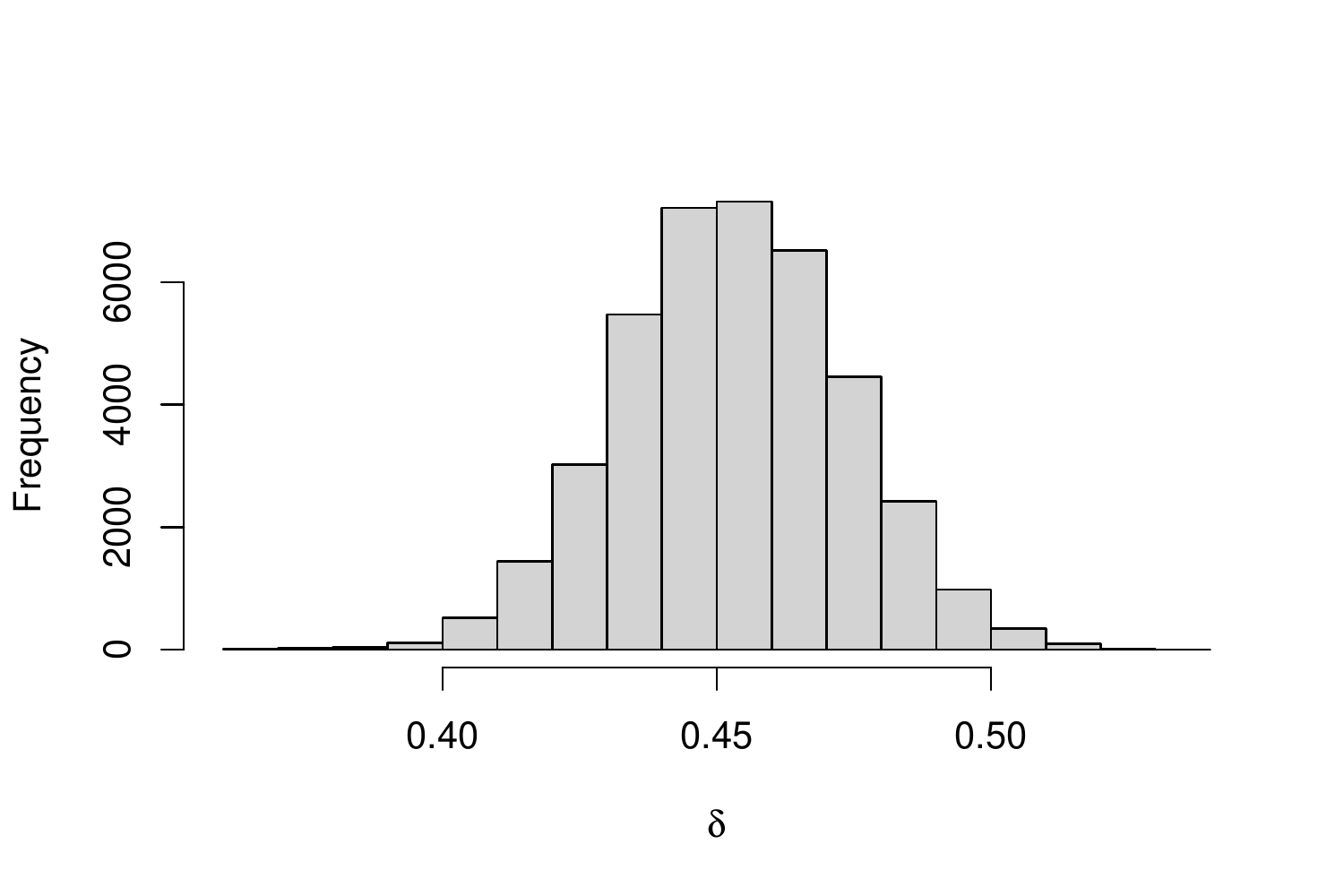}
\caption{Posterior distribution based on 20,000 post burn-in samples.}
    \label{fig:hist_delta_sqrt}
    \end{subfigure}
    \caption{{\bf Estimation of $\delta$ from a PMM fitted to the square root of the streamflow data}:} MCMC trace plots and the posterior distribution of $\delta$.
    \label{fig:posterior_delta_sqrt}
\end{figure}
The log-transform that is carried out on the data leads to negative GEV shape parameter estimates and that imposes a finite upper bound on the distribution even on the original scale. To better understand the properties of the marginal distribution, we fitted the PMM to data on the original scale, as well to the square root of the streamflow.

Data fitted to the original scale, i.e., without taking the log-transformation, had convergence issues for several of the parameters. However, it is still possible to use the results to help interpret posterior estimates from the log-transformed streamflow data. Figure \ref{fig:posterior_delta_og} plots the posterior of $\delta$ for the PMM fitted to the original data, which has a mean of $0.57$ and a standard deviation of 0.02. The 95\% interval for the posterior is (0.53,0.60). This suggests that the PMM is able to distinguish between asymptotic dependence and asymptotic independence regimes.

We also investigated the distribution of shape parameter in this model, since estimates of $\xi(\bs)$ in the log-transformed scale are negative implying a finite upper bound to the log-transformed data. On the original scale, we found that the posterior of the shape parameter was positive at 486 out of 487 locations, with a mean of 0.48. The range of the estimates was (-0.42,3.02), and indicates that the original data does not have a finite upper bound.

Data fitted to the square root of streamflow avoids most of these convergence issues. Figure \ref{fig:posterior_delta_sqrt} plots the posterior of $\delta$ in this case, which has a mean of 0.4531 and a standard deviation of 0.02. These are consistent with the estimates of $\delta$ obtained when the PMM is fitted to log-transformed data. The $95\%$ posterior interval is $(0.41,0.49)$. Finally, the estimates of $\xi(\bs)$ are between (-0.39, 0.53). The estimates of the shape parameter in all cases suggest that a spatially varying model for $\xi(\bs)$ is more appropriate than fixing it across the entire country. While the square root of streamflow is in some ways a more appropriate transformation than the log of streamflow from a theoretical perspective, we prefer the use of the log-transform in our work due to ease of interpretation and its frequent use in weather and climate literature.

\end{appendix}

\begin{singlespace}
	\bibliographystyle{rss}
	\bibliography{VecchiaDL}
\end{singlespace}

%% file: commands.tex
\newcommand{\btheta}{ \mbox{\boldmath $\theta$}}

\newcommand{\bx}{ \mbox{\bf x}}
\newcommand{\bX}{ \mbox{\bf X}}

\newcommand{\bs}{ \mbox{\bf s}}

\newcommand{\iid}{\stackrel{iid}{\sim}}

\newcommand{\calN}{{\cal N}}

\newcommand{\beq}{ \begin{equation}}
\newcommand{\eeq}{ \end{equation}}
\newcommand{\beqn}{ \begin{eqnarray}}
\newcommand{\eeqn}{ \end{eqnarray}}

\usepackage{caption}